\documentclass[aps,pra,preprint,groupedaddress,showpacs]{revtex4-1}

\usepackage{amsfonts,amssymb,amsmath}

\usepackage{amstext}
\usepackage{graphicx}

\usepackage{paralist}

\newcommand{\trans}{\mathsf{T}}
\newcommand{\SDGEP}{SDGEP}
\newcommand{\HDGEP}{HDGEP}
\newcommand{\Lanczos}{Lanczos}
\newcommand{\Krylov}{Krylov}

\newcommand{\Neumann}{Neumann}
\newcommand{\DM}{Dzyaloshinskii--Moriya}

\newcommand{\Brillouin}{Brillouin}

\newcommand{\varmat}[1]{\mathsf{[#1]}}
\newcommand{\rank}{\mathop{\mathrm{rank}}\nolimits}

\newcommand{\Or}{\mathcal{O}}

\newcommand{\secref}[1]{Sec.~\ref{#1}}
\newcommand{\Secref}[1]{Section~\ref{#1}}
\newcommand{\ssecref}[1]{Subsec.~\ref{#1}}
\newcommand{\Ssecref}[1]{Subsection~\ref{#1}}
\newcommand{\ssecrefs}[1]{Subsecs.~\ref{#1}}

\newcommand{\equaref}[1]{Eq.~\eqref{#1}}
\newcommand{\Equaref}[1]{Equation~\eqref{#1}}
\newcommand{\equarefs}[1]{Eqs.~\eqref{#1}}

\newcommand{\explcite}[1]{Ref.~[\onlinecite{#1}]}

\newcommand{\explcites}[1]{Refs.~[\onlinecite{#1}]}

\newcommand{\explcitepart}[1]{[\onlinecite{#1}]}
\newcommand{\extequaref}[1]{Eq.~#1}
\newcommand{\extequarefs}[1]{Eqs.~#1}
\newcommand{\extequarefpart}[1]{#1}

\newcommand{\figref}[1]{Fig.~\ref{#1}}
\newcommand{\Figref}[1]{Figure~\ref{#1}}

\newcommand{\figrefsub}[2]{Fig.~\ref{#1}(#2)}
\newcommand{\Figrefsub}[2]{Figure~\ref{#1}(#2)}
\newcommand{\figrefssub}[2]{Figs.~\ref{#1}(#2)}

\newcommand{\figrefsubpart}[2]{\ref{#1}(#2)}
\newcommand{\figsub}[1]{(#1)}

\newcommand{\refappendix}{Appendix}
\newcommand{\refsubappendix}{Subsection}
\newcommand{\sappref}[1]{Subsec.~A.\ref{#1}}

\newcommand{\refsection}{Section}
\newcommand{\refsections}{Sections}

\begin{document}


\title{Zero modes in magnetic systems: general theory and an efficient computational scheme}


\author{F. J. Buijnsters}
\email[]{F.Buijnsters@science.ru.nl}
\author{A. Fasolino}
\author{M. I. Katsnelson}
\affiliation{Institute for Molecules and Materials, Radboud University Nijmegen, Heyendaalseweg~135, 6525~AJ Nijmegen, Netherlands}


\date{\today}

\begin{abstract}
The presence of topological defects in magnetic media often leads to normal modes with zero frequency (zero modes). Such modes are crucial
for long-time behavior, describing, for example, the motion of a domain wall as a whole.
Conventional numerical methods to calculate the spin-wave spectrum in magnetic media are either inefficient or they fail for systems with zero modes.
We present a new efficient computational scheme that reduces the magnetic normal-mode problem to a generalized Hermitian eigenvalue problem also in the presence of zero modes.
We apply our scheme to several examples, including two-dimensional domain walls and Skyrmions, and show how the effective masses that determine the dynamics can be calculated directly.
These systems highlight the fundamental distinction between the two types of zero modes that can occur in spin systems, which we call \emph{special} and \emph{inertial} zero modes.
Our method is suitable for both conservative and dissipative systems. For the latter case, we present a perturbative scheme to take into account damping, which can also be used to calculate dynamical susceptibilities.
\end{abstract}


\maketitle

\newlength{\figwidthhalf}
\setlength{\figwidthhalf}{246.0pt}   
\newlength{\figwidthfull}
\setlength{\figwidthfull}{1.8\figwidthhalf}

\section{Introduction}

Many properties of magnetic systems can be understood at the classical level by studying their magnetic structure and behavior on the sub-micron lengthscale (micromagnetics
\cite{Brown1963,Aharoni2001,*[{For a review see, for example, }] [{}] Fidler2001}) or atomistically (atomistic spin dynamics \cite{Chen1994,Skubic2008}). In these approaches, the dynamics of the microscopic magnetic moments is described by the Landau--Lifshitz--Gilbert (LLG) equation \cite{Landau1935,Gilbert2004}. The various competing interactions (exchange, anisotropy, dipolar, Zeeman,~\ldots)\ in micromagnetic models often result in a rich energy landscape with multiple local energy minima and hysteresis \cite{Brown1963,Skomski2008}. Nontrivial magnetic configurations may be very stable, for instance if they contain topological defects such as domain walls or magnetic Skyrmion bubbles \cite{Skomski2008,Dell1986}.

It is often useful to study the dynamics of small-amplitude deviations from a given magnetic equilibrium configuration (linearization). The eigenmodes of the linearized LLG equation are known as magnetic normal modes. In homogeneous systems, the magnetic normal modes are spin waves, which propagate through the material \cite{Akhiezer1968,Vonsovsky1974}. The presence of inhomogeneities, whether intrinsic (lattice defects, boundaries) or configurational (domain walls, Skyrmions), changes this picture. Such defects do not only affect the dynamics of the spin waves; they also often give rise to special low-energy normal modes that are \emph{localized} near the defect \cite{Thiele1973B,Helman1991,Makhfudz2012}. The modes localized on configurational defects are particularly interesting. They provide valuable insight into the dynamics of domain walls \cite{Tatara2008} and other topological defects, a sound understanding of which will be important for the development of novel magnetic-storage technologies such as racetrack memory \cite{Parkin2008}. The low-energy modes also provide a channel for dissipation \cite{Wieser2010,Wang2012}. Microscopic magnetic elements, such as ferromagnetic rings, are another class of systems with potential for technological application \cite{Giesen2007}. The spin-wave mode spectrum of these elements can be determined experimentally using magnetic-response measurements or \Brillouin{} light scattering, providing a very direct test of micromagnetic models \cite{Giesen2007,Montoncello2008B,Gubbiotti2006,Talbi2010}.

While exact or approximate analytical solutions of the magnetic normal-mode problem do exist in certain special cases \cite{Thiele1973B,Helman1991,Giesen2007}, in general it can be solved only numerically.
In some cases, the magnetic normal modes can be obtained by a `brute-force' method: numerically integrating the LLG equation over a certain time interval and performing a fast Fourier transformation (FFT) in the time domain \cite{Bolte2006,Giesen2007,Makhfudz2012,Mochizuki2012}. While in principle effective, this approach is limited to relatively small systems by the large amounts of CPU power and memory storage it requires, especially if a good frequency resolution is to be achieved (long simulation times). Moreover, it requires some manual tuning (reasonable settings for the initial amplitudes and sampling frequencies) and it fails to detect zero-frequency and degenerate modes.
In this work, we present a direct numerical procedure that can be used to find the magnetic normal modes of any spin system near any given equilibrium configuration (more precisely, near any local energy minimum). It can deal efficiently and scalably with any type of interaction, including long-range interactions, and does not assume that the material is homogeneous or that the equilibrium configuration is collinear \cite{Arias2001}.

An efficient approach should somehow be based on a direct calculation of the eigenvectors and eigenvalues of the dynamical matrix that results from linearization of the LLG equation \cite{Grimsditch2004}.
However, we shall see that this dynamical matrix is not necessarily diagonalizable, so that eigenvectors in the usual sense may not even exist.
Diagonalizability can only be guaranteed if no zero-frequency modes (zero modes) are present.
To the best of our knowledge, this fact has been overlooked in all previous works describing general methods for the magnetic normal-mode problem \cite{Grimsditch2004,Zivieri2012}.
While there certainly are many cases in which this issue does not occur
\cite{Grimsditch2004,Tacchi2011,Tacchi2010,Giovannini2004,Montoncello2008A,Montoncello2008B,Gubbiotti2006,Zivieri2011,Giovannini2007,Zivieri2013},
we shall see that zero modes appear in many relevant physical systems.
Indeed, precisely these zero modes are often the most important for the dynamics of topological defects. For example, we shall see that it is the zero modes that determine whether the dynamics of a topological defect is inertial, and if so, with what effective mass.

Our method has a firm basis in the general theory of Hamiltonian systems \cite{Arnold1989}. We shall show that the normal-mode problem of an \emph{arbitrary} (conservative) Hamiltonian system at a local energy minimum can be cast in the form of a \emph{Hermitian definite generalized eigenvalue problem} (\HDGEP{}) \cite{Demmel2000}, $Dx = \lambda Sx$, where the matrices $D$ and $S$ are Hermitian and $S$ is positive definite, which can be solved particularly efficiently. The most popular methods for large eigenvalue problems (\Lanczos{}, conjugate-gradient nonlinear optimization,~\ldots)\ require the problem to be of this form. Important features of these methods are that they operate in an incremental fashion (the lowest modes are calculated first) and that they can be implemented in a matrix-free manner \cite{Knyazev2000} (they are \Krylov{}-subspace methods \cite{Saad2003}). These features make the \HDGEP{} methods considerably scalable.
First, the low modes of a very large system, which are often the most physically relevant, may be obtained without solving the full eigenvalue problem for all eigenvectors, which would obviously take at least $\Or (N^2)$ time.
Second, it is not necessary to store the interaction matrix in explicit form, which will contain $\Or (N^2)$ nonzero values if the long-range dipolar interactions are taken into account. It is sufficient to provide a routine that evaluates the forces or torques for any given specific configuration. When implemented using FFT or multigrid techniques, such a routine can run in $\Or (N \log N)$ instead of $\Or (N^2)$ time \cite{Labbe1999}.

We obtain a solution method for the normal-mode problem of the conservative (zero damping) spin system as an immediate special case of our method for general Hamiltonian systems. A similar reduction of the conservative magnetic normal-mode problem to the \HDGEP{} was proposed in \explcite{Zivieri2012} by assuming, wrongly, that the Hessian matrix of a function is always positive definite at a local minimum.
A particular strong point of our approach is that it also works if the Hessian matrix of the Hamiltonian at the equilibrium configuration is not positive definite but merely positive semidefinite (also called nonnegative definite), as it is in the presence of zero modes.
An additional advantage of our method is that it may be used directly in Cartesian coordinates, in which the micromagnetic Hamiltonians normally take a very simple form (often quadratic). We do not need to go over to spherical coordinates, which are more computationally expensive and have singularities at certain points.

For the spin system with damping, we derive explicit expressions for the normal modes by treating the damping term of the LLG equation as a perturbation.
In this way we can obtain the damped modes and decay rates to a good approximation without the need for solving non-Hermitian eigenvalue problems.

This paper is organized as follows.
In \secref{sec:hamsysbrief}, we state some general properties of the normal modes of linearized Hamiltonian systems that are essential for what follows.
Here we introduce the nomenclature of \emph{special} and \emph{inertial} zero modes and specify their distinct dynamics.
A more detailed discussion is provided in the \refappendix{}.
In \secref{sec:spincons}, we make the definitions of \secref{sec:hamsysbrief} explicit for the conservative spin system.
\Secref{sec:redHDGEP} then shows how the normal-mode problem of a Hamiltonian system, such as the conservative spin system, near a local energy minimum can be reduced to the \HDGEP{}. We specifically show how to deal with zero modes in a robust way.
We present perturbative expressions for the spin system with damping in \secref{sec:spindamped}.
\Secref{sec:implementation} explains how the method can be efficiently implemented in a computer code.
\Secref{sec:examples} provides examples of magnetic normal modes in various spin systems, highlighting some key features of magnetic normal modes.
In \ssecref{sec:inertial}, we focus on the two qualitatively different types of effective dynamical behavior (inertial and noninertial) that may be found when a magnetic equilibrium configuration containing some (topological) defect is perturbed by an external force.
We show how a normal-mode analysis that includes zero modes immediately provides the equations of motion and effective masses of such magnetic structures.
\Secref{sec:summary} summarizes our results.

\section{\label{sec:hamsysbrief}Normal modes of Hamiltonian systems}

This \refsection{} states some results from the theory of Hamiltonian systems that are essential for the following \refsections{}.
In particular, we introduce our nomenclature for the three types of normal modes (positive, special zero, inertial zero) that may appear in systems with a positive semidefinite Hamiltonian.
A more thorough discussion with explanations and references is provided in the \refappendix{}.

Let us consider a time-invariant dynamical system near an equilibrium point, which we take to lie at $x=0$.
Its equation of motion is given by
\begin{equation}\label{eq:geneom}
\dot x^i = {M^i}_j x^j + \Or({\| x \|}^2)\text{,}
\end{equation}
where $x^1,\ldots,x^{m}$ represent a nonsingular system of coordinates and the dot denotes the time derivative.
Our goal is to find the eigenvalues and eigenvectors of $M$. This cannot normally be accomplished by a diagonalization of $M$, because
\begin{inparaenum}[\itshape a\upshape)]
\item in general, $M$ is very large but not symmetric, so that the efficient iterative methods for the \HDGEP{} cannot be used; and
\item $M$ might not be diagonalizable at all (it may be \emph{defective}).
\end{inparaenum}
However, if the dynamical system \eqref{eq:geneom} is a linear or nonlinear Hamiltonian system, we shall see that we can bypass these problems by introducing a certain antisymmetric matrix $\Omega$. The elements of $\Omega$ are given by
\begin{equation}\label{eq:genomega}
\Omega^{ij} = - {\lbrace x^i,x^j\rbrace|}_{x=0} = {\lbrace x^j,x^i\rbrace|}_{x=0}\text{,}
\end{equation}
the value at the equilibrium point of the Poisson bracket between the coordinates $x^j$ and $x^i$.
It can be shown (see \refappendix{}) that for a Hamiltonian system, the matrix $M$ is such that $M\Omega$ is symmetric.

For certain physical systems, Hamiltonian dynamics takes place only on a subspace of the space where the coordinates are defined. An example is the spin system: while a magnetic moment $\mathbf m$ is defined on $\mathbb{R}^{3}$, its dynamics is restricted to a subset of the form $\{ \mathbf m \in \mathbb{R}^3 : \| \mathbf m \|=c \}$ for some $c \ge 0$.
The dimension of this `accessible subspace' (\emph{symplectic leaf} \cite{Weinstein1983}) is always even. For a system of $n$ spins in Cartesian coordinates, we have $m = 3n$, while the dimension of the symplectic leaf is only $2n$.
We remind the reader that the \emph{image space} of a matrix $A$ consists of all vectors $x$ that can be written as $x = A y$ for some vector $y$; the dimension of this linear subspace is denoted by $\rank A$.
The image space of $\Omega$, which has dimension $2n = \rank\Omega$, is identical to the tangent space of the symplectic leaf at $x=0$.
Vectors that are not contained in the image space of $\Omega$ correspond to an infinitesimal displacement of the system out of the symplectic leaf and are unphysical. We may thus restrict the matrices $\Omega$ and $M\Omega$ to the image space of $\Omega$. We shall denote these restricted matrices by $\langle \Omega \rangle$ and $\langle M\Omega \rangle$; that is, we define
\begin{equation*}
\langle \Omega \rangle \equiv F^\trans \Omega F \hspace{1.3em}\text{ and }\hspace{1.3em} \langle M\Omega \rangle \equiv F^\trans M\Omega F\text{,}
\end{equation*}
where $F$ is an $m \times 2n$ matrix whose columns form an orthonormal basis of the image space of $\Omega$. Since the image space of $M\Omega$ is contained in the image space of $\Omega$, these restrictions are well defined and without loss. Notice that the matrix $\langle \Omega \rangle$ is invertible by definition. In this paper, we shall implicitly convert between vectors in $\mathbb{R}^{2n}$ and vectors in the image space of $\Omega$ without writing $F$. It is unnecessary to explicitly construct $F$ in a computer code (see \secref{sec:implementation}).

It can be shown (see \refappendix{}) that the $2n \times 2n$ matrix $\langle M\Omega \rangle$ is the Hessian matrix (the matrix of second-order partial derivatives) at $x=0$ of the restriction of the Hamiltonian $\mathcal H$ to the symplectic leaf (for a certain parametrization of the symplectic leaf).
Therefore, if $x=0$ is a constrained local minimum of $\mathcal H$ on the symplectic leaf, the Hessian matrix $\langle M\Omega \rangle$ is guaranteed to be positive \emph{semi}definite. However, it may not be assumed (compare \explcite{Zivieri2012}) that $\langle M\Omega \rangle$ is also positive definite. To see this, consider the following simple counterexamples with $m=2n=2$: $\mathcal{H}(p,q)=0$, $\mathcal{H}(p,q)=p^2$ and $\mathcal{H}(p,q)=p^4+q^4$ all have minima at $p=q=0$ but not positive-definite Hessians at that point.

\begin{figure}
  \includegraphics[width=\figwidthhalf]{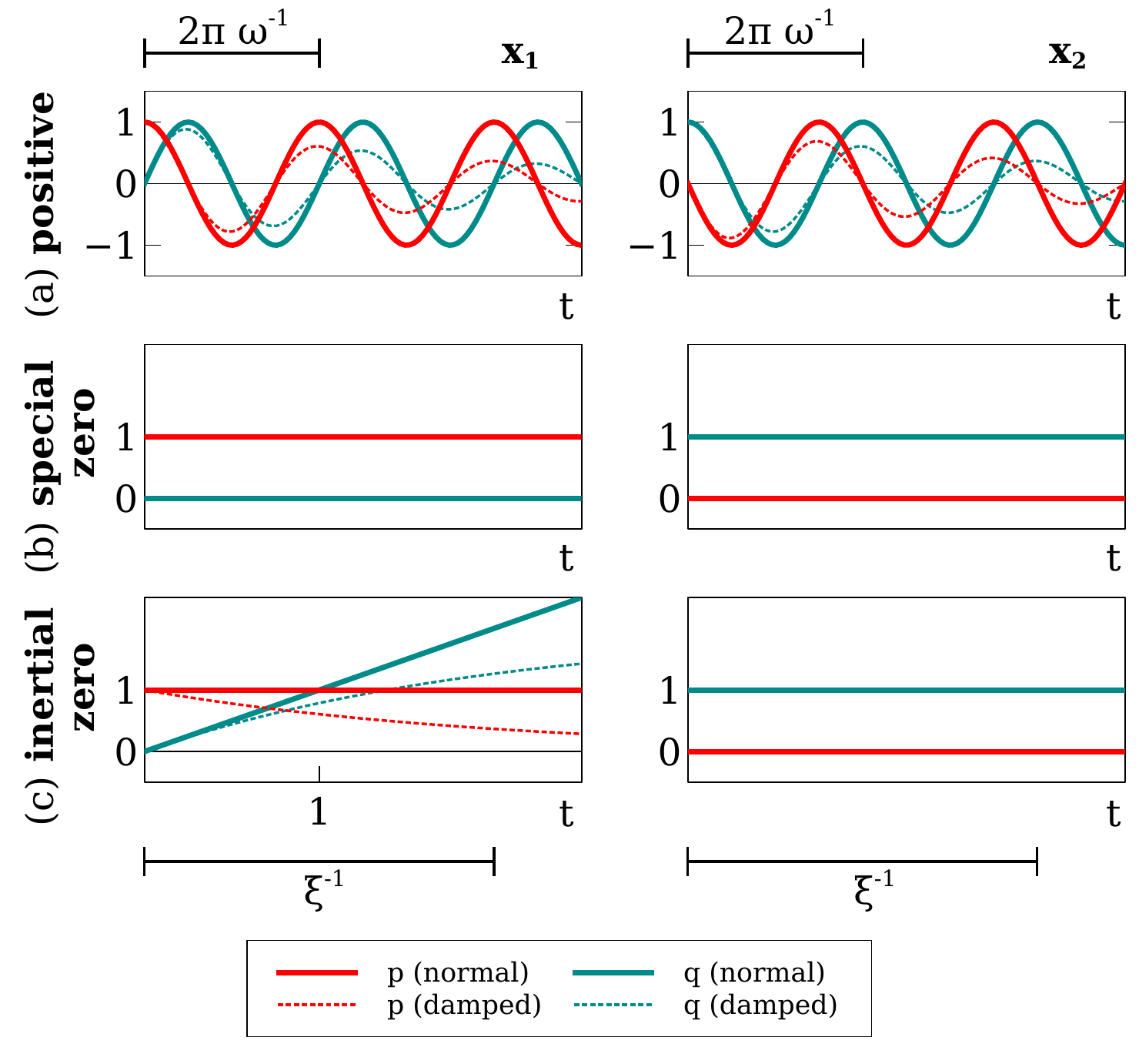}
  \caption{\label{fig:fundsols}Color) Fundamental solutions $x_1$ and $x_2$ of the linearized equation of motion \eqref{eq:geneom} corresponding to the three types of normal modes of a Hamiltonian system:
    \figsub{a}~positive \eqref{eq:posnormmode},
    \figsub{b}~special zero \eqref{eq:ordzeronormmode} and
    \figsub{c}~inertial zero \eqref{eq:defzeronormmode} modes.
    The dynamical variables $p$ and $q$ are the amplitudes of the vectors $u_1$ and $u_2$ respectively, as defined in \equaref{eq:decomppq}. Dashed lines: effect of damping with the indicated decay time $\xi^{-1}$ (see \secref{sec:spindamped}).}
\end{figure}
If $\langle M\Omega \rangle$ is positive semidefinite, the normal modes of $M$ may be of three distinct types (see \refappendix{}).
We introduce the following names for these three types of modes.
\begin{enumerate}
\item A \emph{positive normal mode} of $M$ is a pair $(u_1,u_2)$ of vectors in the image space of $\Omega$ that satisfy
\begin{equation}\label{eq:posnormmode}
\left\{
\begin{array}{l}
M u_1 = \phantom{-}\omega u_2 \\
M u_2 = -\omega u_1
\end{array}
\right.
\end{equation}
for some $\omega>0$. The corresponding fundamental solutions of the linearization of \equaref{eq:geneom} are (see \figrefsub{fig:fundsols}{a})
\begin{equation}\label{eq:fundsolconspos}
\begin{array}{l}
x_1(t) = \phantom{-}\cos(\omega t)u_1 + \sin(\omega t)u_2\text{,}\\
x_2(t) =          - \sin(\omega t)u_1 + \cos(\omega t)u_2\text{.}
\end{array}
\end{equation}
Each positive normal mode corresponds to a pair of eigenvectors of $M$. The eigenvectors are $u_1 - iu_2$ (eigenvalue $i\omega$) and $u_1 + iu_2$ (eigenvalue $-i\omega$).
\item A \emph{special zero normal mode} is a pair $(u_1,u_2)$ of vectors in the image space of $\Omega$ that satisfy
\begin{equation}\label{eq:ordzeronormmode}
\left\{
\begin{array}{l}
M u_1 = 0 \\
M u_2 = 0
\end{array}
\right.\text{.}
\end{equation}
The corresponding fundamental solutions are (see \figrefsub{fig:fundsols}{b})
\begin{equation}\label{eq:fundsolconsord}
\begin{array}{l}
x_1(t) = u_1\text{,}\\
x_2(t) = u_2
\end{array}
\end{equation}
(constant functions).
A special zero normal mode also corresponds to a pair of linearly independent eigenvectors of $M$ ($u_1$ and $u_2$).
\item An \emph{inertial zero normal mode} is a pair $(u_1, u_2)$ of vectors in the image space of $\Omega$ that satisfy
\begin{equation}\label{eq:defzeronormmode}
\left\{
\begin{array}{l}
M u_1 = u_2 \\
M u_2 = 0
\end{array}
\right.\text{.}
\end{equation}
The corresponding fundamental solutions are (see \figrefsub{fig:fundsols}{c})
\begin{equation}\label{eq:fundsolconsdef}
\begin{array}{l}
x_1(t) = u_1 + t u_2\text{,}\\
x_2(t) = u_2\text{.}
\end{array}
\end{equation}
This type of mode results from a nondiagonalizable (defective) matrix $M$.
Technically, an inertial zero mode corresponds to a Jordan block of size $2$ in the Jordan normal form of $M$.
\end{enumerate}
The nomenclature chosen for the three types of modes (positive, special and inertial) is explained below.
Notice that different types of modes may have different units: for an inertial zero normal mode $\|u_1\| / \|u_2\|$ has units of time, while for a positive normal mode $\|u_1\| / \|u_2\|$ is dimensionless.
Since each mode contains two vectors, the total number of independent modes $n$ is one half of the dimension of the symplectic leaf.
If $\langle M\Omega \rangle$ is positive definite, all normal modes are positive normal modes.

We may write the vectors that make up a normal mode as
\begin{equation}\label{eq:defw}
\left\{
\begin{array}{l}
u_1=\Omega w_1 \\
u_2=\Omega w_2
\end{array}
\right.
\end{equation}
for certain vectors $w_1$ and $w_2$ in the image space of $\Omega$.
\Secref{sec:redHDGEP} presents an efficient procedure by which suitable vector pairs $w_1,w_2$ may be found.
All normal modes can and should be chosen to satisfy the relations
\begin{subequations}
\begin{align}
\label{eq:orthopq}                               w_{1k}^\trans \Omega w_{2l}& = \delta_{kl} \\
\label{eq:orthopp} w_{1k}^\trans \Omega w_{1l} = w_{2k}^\trans \Omega w_{2l}& = 0\text{,}
\end{align}
\end{subequations}
where $k,l=1,\ldots,n$ index the modes.
As a result, we may decompose an arbitrary vector $x$ in the image space of $\Omega$ in terms of the normal modes as
\begin{equation}\label{eq:decompx}
x 
  = \sum_{k=1}^n \left[ -(w_{2k}^\trans x) \, u_{1k} + (w_{1k}^\trans x) \, u_{2k} \right] \text{.}
\end{equation}
Using the fundamental solutions \eqref{eq:fundsolconspos}, \eqref{eq:fundsolconsord} and \eqref{eq:fundsolconsdef}, such a decomposition immediately yields a solution of the initial-value problem for \equaref{eq:geneom} in the linear approximation.

Given a state vector
\begin{equation}\label{eq:decomppq}
x = \sum_{k=1}^n \left( p_{k} \, u_{1k} + q_{k} \, u_{2k} \right) + \Or(p_k^2+q_k^2)\text{,}
\end{equation}
the quadratic part of the Hamiltonian is given by
\begin{equation}\label{eq:Hquadratic}
\mathcal H = \sum_{k\text{ pos.}} \frac{1}{2} \omega_k \left( p_{k}^2 + q_{k}^2 \right) + \sum_{k\text{ def.}} \frac{1}{2} p_{k}^2\text{,}
\end{equation}
where the first sum is taken over the positive normal modes and the second sum over the inertial zero normal modes. Special zero modes do not contribute to \equaref{eq:Hquadratic}.
The variables $p_k$ and $q_k$ in \equaref{eq:decomppq} are canonically conjugate momenta and coordinates (see \refappendix{}).
Notice that for a given configuration $m=m_0+x$, the values of these momenta and coordinates can be determined, to first order, using \equaref{eq:decompx}.
We find, in the linear limit, that for a special zero normal mode
\begin{equation}
\label{eq:specialhamilt}
\left\lbrace
\begin{array}{l}
\dot{p}_k =          - \frac{\partial \mathcal H}{\partial q_k} = 0 \\
\dot{q}_k = \phantom{-}\frac{\partial \mathcal H}{\partial p_k} = 0
\end{array}
\right.\text{,}
\end{equation}
while for an inertial zero normal mode
\begin{equation}
\label{eq:inertialhamilt}
\left\lbrace
\begin{array}{l}
\dot{p}_k =          - \frac{\partial \mathcal H}{\partial q_k} = 0 \\
\dot{q}_k = \phantom{-}\frac{\partial \mathcal H}{\partial p_k} = p_k
\end{array}
\right.\text{.}
\end{equation}
The latter type of dynamics \eqref{eq:inertialhamilt} corresponds (after a suitable scaling of $p_k$ and $q_k$) to the dynamics of a free massive particle, which explains our choice of the name `inertial zero normal mode'. The former type of dynamics \eqref{eq:specialhamilt} does not occur in conventional Newtonian systems such as systems of coupled oscillators (see \refappendix{}), whence the name `special zero normal mode'.

\section{\label{sec:spincons}Conservative spin systems}

The conservative dynamics of a spin system is described by the LLG equation without damping,
\begin{equation}\label{eq:consLLG}
\dot{\mathbf{m}}_i = \tilde\gamma \mathbf{m}_i \times \nabla_{\mathbf{m}_i}\mathcal H\text{,}
\end{equation}
where $\mathbf{m}_i \in \mathbb{R}^3$ represents the magnetic moment with position index $i=1,\ldots,n$, $\mathcal H$ is the Hamiltonian, and $\tilde\gamma$ is a physical constant.
Notice that the magnitude $\| \mathbf{m}_i \|$ of each magnetic moment is constant in time. These magnitudes are fixed by the physics of the system.
\Equaref{eq:consLLG} is equivalent to
\begin{equation}
\dot{m}_{i\alpha} = \lbrace m_{i\alpha}, \mathcal H \rbrace\text{,}
\end{equation}
the generalized form of Hamilton's equations applied to the (time-invariant) variables $m_{i\alpha}$,
for the Poisson bracket
\begin{equation}\label{eq:spinpoissonfull}
\lbrace m_{i\alpha}, m_{j\beta} \rbrace =
\left\lbrace
\begin{array}{l l}
-\tilde\gamma \varepsilon_{\alpha\beta\gamma} m_{i\gamma} & \text{for $i=j$} \\
0 &\text{for $i \neq j$}
\end{array}
\right.\text{,}
\end{equation}
where Greek indices represent Cartesian coordinates $x,y,z$ and $\varepsilon_{\alpha\beta\gamma}$ is the Levi-Civita symbol.
Thus, the dynamics of the conservative spin system is Hamiltonian.

For convenience, we shall write \equaref{eq:consLLG} as
\begin{equation}\label{eq:LLGfcp}
\dot{m} = \tilde \gamma [ m , \nabla \mathcal H ]\text{.}
\end{equation}
The variable $m \in \mathbb{R}^{3n}$ can be seen as a compound vector that assigns to every position $i=1,\ldots,n$ a vector $\mathbf{m}_i \in \mathbb{R}^3$.
The square brackets in \equaref{eq:LLGfcp} denote an elementwise cross product:
given $x,y \in \mathbb{R}^{3n}$, we define $z = [x,y] \in \mathbb{R}^{3n}$ such that $\mathbf{z}_i = \mathbf{x}_i \times \mathbf{y}_i$ for each position $i$. In other words, it is just the ordinary cross product (vector product) performed $n$ times.
For small deviations $x=m-m_0 \in \mathbb{R}^{3n}$ from some fixed configuration $m_0$, \equaref{eq:LLGfcp} becomes
\begin{equation}\label{eq:linLLG}
\dot{x} = -\tilde\gamma [ m_0,h ] + Mx + \Or({\| x \|}^2)\text{,}
\end{equation}
where $h_{(i\alpha)}=-{\partial \mathcal H / \partial m_{i\alpha} |}_{m=m_0}$ is the effective field at $m_0$. The matrix $M$ is given by
\begin{equation}\label{eq:spinM}
Mx =   \tilde\gamma[ m_0,Ax ] + \tilde\gamma[ h,x ] \text{,}
\end{equation}
where $A_{(i\alpha)(j\beta)} = {\partial^2 \mathcal H / (\partial m_{i\alpha} \partial m_{j\beta})|}_{m=m_0}$ is the $3n\times 3n$ Hessian matrix of $\mathcal H$ at $m_0$.
To be explicit, let us mention that the elements of $M$ are given by
\begin{equation}
M_{(i\alpha)(j\beta)} =
\left\lbrace
\begin{array}{l l}
\tilde\gamma \varepsilon_{\alpha\gamma\delta} {(m_0)}_{i\gamma} A_{(i\delta)(j\beta)} + \tilde\gamma \varepsilon_{\alpha\gamma\beta} h_{i\gamma} & \text{for $i=j$} \\
\tilde\gamma  \varepsilon_{\alpha\gamma\delta} {(m_0)}_{i\gamma} A_{(i\delta)(j\beta)} & \text{for $i \neq j$}
\end{array}
\right.\text{.}
\end{equation}
(Summation is implied for repeated Greek indices but not for repeated Roman indices.)
Since we work in Cartesian coordinates, $A$ is typically of a relatively simple form. Indeed, many micromagnetic models use a Hamiltonian that is quadratic in the Cartesian coordinates, in which case $A$ does not depend on $m_0$.

We assume that $m_0$ is an equilibrium configuration, $[ m_0,h ]=0$.
As a result, \equaref{eq:linLLG} is of the form \eqref{eq:geneom}.
The matrix $\Omega$ \eqref{eq:genomega} is given by
\begin{equation}
\Omega_{(i\alpha)(j\beta)} = -{\lbrace m_{i\alpha}, m_{j\beta} \rbrace|}_{m=m_0} =
\left\lbrace
\begin{array}{l l}
\tilde\gamma \varepsilon_{\alpha\beta\gamma} {(m_0)}_{i\gamma} & \text{for $i=j$} \\
0 &\text{for $i \neq j$}
\end{array}
\right.\text{,}
\end{equation}
or equivalently,
\begin{equation}\label{eq:spinJ}
\Omega x=-\tilde\gamma [ m_0,x ]\text{.}
\end{equation}
The $2n$-dimensional image space of $\Omega$ consists of vectors $x\in \mathbb{R}^{3n}$ for which the displacement $\mathbf{x}_i \in \mathbb{R}^3$ is orthogonal at each position $i$ to the equilibrium direction $\mathbf{m}_{0i}$.
Notice also that the equilibrium effective field $\mathbf{h}_i$ must be parallel at each position $i$ to the equilibrium direction $\mathbf{m}_{0i}$.
Combining \equarefs{eq:spinM} and~\eqref{eq:spinJ}, the matrix $M\Omega$, which is symmetric (see \secref{sec:hamsysbrief}), is given by
\begin{equation}\label{eq:spinMJ}
M\Omega x = -\tilde\gamma^2 \Bigl( [m_0,A[m_0,x]] + [h,[m_0,x]] \Bigr) = \Bigl( \Omega^\trans A \Omega +\tilde\gamma[h,\cdot] \Omega  \Bigr) x\text{.}
\end{equation}
The second term, which contains $h$, originates from the fact that the Hessian matrix $A$ is calculated in Cartesian coordinates, while the symplectic leaf (a product of $n$ spheres) is curved.

\section{\label{sec:redHDGEP}Reduction to the \HDGEP{}}

In this \refsection{}, we present a method for the normal-mode problem of a general Hamiltonian system near a local minimum of the Hamiltonian. This includes the normal-mode problem of the conservative spin system as a special case.
We show that the normal-mode problem can be reduced to the \HDGEP{}, in which form it can be efficiently solved (see \secref{sec:implementation}).
Our method calculates both the positive modes and any zero modes of the system.
If zero modes are present, the method detects these and automatically determines their types (special or inertial).

The conservative spin system differs from an important subclass of Hamiltonian systems, which includes systems of coupled point masses, for which the normal-mode problem can be written as a \emph{symmetric definite generalized eigenvalue problem} (\SDGEP{}) in an obvious way (see \refappendix{}).
Such Hamiltonian systems are defined on a natural set of canonical momenta and coordinates.
In terms of these, the Hamiltonian is of the form $\mathcal H(\{p_i\},\{q_i\}) = \mathcal T(\{p_i\}) + \mathcal V(\{q_i\})$, where the potential-energy term $\mathcal V$ depends only on the coordinates $q_i$, while the kinetic-energy term $\mathcal T$ is a positive-definite quadratic function depending only on the momenta $p_i$ (typically, $\mathcal T = \sum_i p_i^2 / (2m_i)$).
The spin system is not of this special form.
All that is given is the Poisson bracket \eqref{eq:spinpoissonfull} and the Hamiltonian $\mathcal H(\{\mathbf{m}_i\})$ as a function of the magnetic moments $\mathbf{m}_i$.
Even though it is possible to construct canonical momenta and coordinates \cite{Dobrovitski2003} for this system,
an \emph{a priori} separation of kinetic energy and potential energy is not normally known.

We shall first consider the case that $\langle M\Omega \rangle$ is positive definite (no zero modes). Later in this \refsection{}, we treat the general case where $\langle M\Omega \rangle$ is positive semidefinite. This generalization is essential for spin systems such as those discussed in \ssecrefs{sec:inertial}--\ref{sec:exskyrmion}.

We remind the reader that an \HDGEP{} has the general form
\begin{equation}\label{eq:genHDGEP}
Dx = \lambda S x\text{,}
\end{equation}
where $D$ is Hermitian and $S$ is Hermitian and positive definite, which requirements guarantee that all eigenvalues $\lambda_i$ are real.
The usual Hermitian eigenvalue problem is a special case of the \HDGEP{} (set $S=I$).
If $D$ and $S$ are real matrices, so that $D$ and $S$ are symmetric, we use the abbreviation \SDGEP{}.
The eigenvectors $x_i$ of a \HDGEP{} may be chosen to satisfy $x_i^\dagger S x_j = \delta_{ij}$, a generalized orthonormality relation.
Alternatively, we may choose to normalize the eigenvectors $x_i$ in such a way that
\begin{equation}\label{eq:orthoD}
x_i^\dagger D x_j = \pm \delta_{ij}\text{,}
\end{equation}
provided that $D$ is invertible, in which case the eigenvalues $\lambda_i$ are nonzero.

Let us represent a positive normal mode \eqref{eq:posnormmode} as a single vector
\begin{equation}
w = w_1 + i w_2 \in \mathbb{C}^{2n}\text{,}
\end{equation}
where $w_1$ and $w_2$ are such that
\begin{equation*}
u_1 = \langle \Omega \rangle w_1 \hspace{1.3em}\text{ and }\hspace{1.3em} u_2 = \langle \Omega \rangle w_2\text{.}
\end{equation*}
It is easy to see that in this notation, a solution of the generalized eigenvalue problem
\begin{equation}\label{eq:eigenmodeHDGEP}
\langle M\Omega \rangle w = - i \omega \langle \Omega \rangle w
\end{equation}
with $\omega > 0$ is also a solution of \equaref{eq:posnormmode} (after conversion of the vectors in $\mathbb{R}^{2n}$ to vectors in the image space of $\Omega$): take real and imaginary parts.
If we \emph{assume} that $\langle M\Omega \rangle$ is positive definite, \equaref{eq:eigenmodeHDGEP} is a \HDGEP{} \eqref{eq:genHDGEP} with $D=-i\langle\Omega\rangle$, $S=\langle M\Omega \rangle$ and $\lambda=\omega^{-1}$, since $\Omega$ is antisymmetric and $M\Omega$ is symmetric (see \secref{sec:hamsysbrief}).
The \HDGEP{} form \eqref{eq:eigenmodeHDGEP} makes the problem suitable for efficient numerical computation.
Notice that $\lambda \neq 0$, since $\langle \Omega \rangle$ is invertible.
Notice also that each positive normal mode gives rise to two independent solutions of \equaref{eq:eigenmodeHDGEP}: if $w = w_1 + i w_2$ is an eigenvector with eigenvalue $\omega>0$, then $w^* = w_1 - i w_2$ is an eigenvector with eigenvalue $-\omega$.
By \equaref{eq:orthoD}, we may normalize the eigenvectors so that they satisfy
\begin{subequations}
\begin{align}
\label{eq:iJortho}    {(w_{1k} + i w_{2k})}^\dagger {(-i \langle \Omega \rangle)} {(w_{1l} + i w_{2l})}& = 2 \delta_{kl} \\
\label{eq:iJorthocc}  {(w_{1k} - i w_{2k})}^\dagger {(-i \langle \Omega \rangle)} {(w_{1l} + i w_{2l})}& = 0\text{,}
\end{align}
\end{subequations}
which equations together are equivalent to \equarefs{eq:orthopq} and~\eqref{eq:orthopp}.

\Equaref{eq:eigenmodeHDGEP} can be seen as a generalization of \extequarefs{27}--\extequarefpart{30} in \explcite{Zivieri2012}, which were given for the normal-mode problem of the conservative spin system, to a general Hamiltonian system.
Our formulation has the additional advantage that it does not require the use of spherical coordinates.
By itself, the method only works if $\langle M\Omega \rangle$ is positive definite.
If $\langle M\Omega \rangle$ is merely positive semidefinite, \equaref{eq:eigenmodeHDGEP} is no longer a \HDGEP{}. Zero normal modes may appear and the matrix $M$ is not even guaranteed to be diagonalizable.
These zero modes have important consequences for the dynamics of, for example, domain walls or Skyrmions in magnetic systems (see \ssecrefs{sec:inertial}--\ref{sec:exskyrmion}).
We present here a robust scheme that also works in this more general case. Thus, our method can solve the normal-mode problem of \emph{any} Hamiltonian system linearized at a local minimum of the Hamiltonian.

The main idea of our approach is that we first find the special and inertial zero normal modes and then exclude them from the problem.
The algorithm consists of the steps outlined below.
The only two `large' ($2n$-dimensional) problems in this procedure are steps~\ref{itm:findnullspace} and~\ref{itm:solveMJ}.
In step~\ref{itm:findnullspace}, we need to find the lowest eigenvalues and corresponding eigenvectors of a symmetric matrix. In step~\ref{itm:solveMJ}, we need to solve a symmetric linear system.
Both sub-problems can be efficiently solved using iterative methods. How this may be done is discussed in more detail in \secref{sec:implementation}.
The diagonalizations in steps~\ref{itm:diagnullJ} and~\ref{itm:diagnullMJ} concern small matrices and can be performed using standard routines.
\begin{enumerate}
\item \label{itm:findnullspace}
Sequentially find the eigenvectors $y_1,y_2,\ldots \in \mathbb{R}^{2n}$ of $\langle M\Omega \rangle$ that correspond to the lowest eigenvalues (see \secref{sec:implementation}).
Stop when an eigenvector appears with an eigenvalue that is larger than zero (by a certain small tolerance).
Notice that this is an ordinary (nongeneralized) symmetric eigenvalue problem, so that the fact that $\langle M\Omega \rangle$ is not necessarily positive definite is not a problem.
By positive semidefiniteness of $\langle M\Omega \rangle$, all eigenvalues are larger than or equal to zero.
Suppose that there are $s$ eigenvectors with eigenvalue zero.
Then $y_1,\ldots,y_s$ form a basis of the null space of $\langle M\Omega \rangle$.
In most cases, $s$ will be a small number.
Remember that thanks to the restriction of $M\Omega$ to $\langle M\Omega \rangle$, we have already excluded
all null vectors of $M\Omega$ that are also null vectors of $\Omega$ and thus correspond to a displacement of the system out of the symplectic leaf.
\item \label{itm:diagnullJ}
Define the $s \times s$ matrix $\varmat{\Omega}_{ij} = y_i^\trans \langle \Omega \rangle y_j$. Since $\varmat{\Omega}_{ij}$ is antisymmetric, $i \varmat{\Omega}_{ij}$ is Hermitian and can be diagonalized by a standard routine for Hermitian matrices, which guarantees that the eigenvectors are orthonormal.
Let $s_\text{d}$ be the number of independent eigenvectors $B_{i(k)}$ of $\varmat{\Omega}_{ij}$ with eigenvalue zero (up to a small tolerance). We have $\sum_{j=1}^s \varmat{\Omega}_{ij} B_{j(k)} = 0$ for $k=1,\ldots,s_\text{d}$. We may take these eigenvectors $B_{i(k)}$ to be real. The remaining nonnull eigenvectors come in $s_\text{o}$ pairs. Let $C_{i(l)} + i D_{i(l)}$ be an eigenvector of $\varmat{\Omega}_{ij}$ with eigenvalue $i\lambda_{(l)}$, where $\lambda_{(l)}>0$ and $C_{i(l)}$ and $D_{i(l)}$ are real. We have $\sum_{j=1}^s \varmat{\Omega}_{ij} ( C_{j(l)} + i D_{j(l)} ) = i\lambda_{(l)} ( C_{i(l)} + i D_{i(l)} )$ for $l=1,\ldots,s_\text{o}$. Then $C_{i(l)} - i D_{i(l)}$ is an eigenvector with eigenvalue $-i\lambda_{(l)}$. The total number of eigenvectors is $s=s_\text{d}+2s_\text{o}$.
\item
Construct the vectors $c_l = \sum_{i=1}^s C_{i(l)} y_i$ and $d_l = \sum_{i=1}^s D_{i(l)} y_i$ for $l=1,\ldots,s_\text{o}$ and $\bar{b}_k = \sum_{i=1}^s B_{i(k)} y_i$ for $k=1,\ldots,s_\text{d}$.
Notice that we have $c_l^\trans \langle \Omega \rangle d_{l'} = 0$ for $l \neq l'$ and $c_l^\trans \langle \Omega \rangle d_l > 0$.
Moreover, for all $l,l',k,k'$ we have $c_l^\trans \langle \Omega \rangle c_{l'} = d_l^\trans \langle \Omega \rangle d_{l'} = 0$, $c_l^\trans \langle \Omega \rangle \bar{b}_k = d_l^\trans \langle \Omega \rangle \bar{b}_k = 0$ and $\bar{b}_k^\trans \langle \Omega \rangle \bar{b}_{k'} = 0$.
\item \label{itm:solveMJ}
For each $k=1,\ldots,s_\text{d}$, find a vector $\bar{a}_k$ such that $\langle M\Omega \rangle \bar{a}_k = \langle \Omega \rangle \bar{b}_k$ (see \secref{sec:implementation}).
We know that such a vector exists, since by construction $\langle \Omega \rangle \bar{b}_k$ lies in the orthogonal complement of the null space of $\langle M\Omega \rangle$, a symmetric matrix, and hence in the image space of $\langle M\Omega \rangle$.
Although this vector $\bar{a}_k$ is not uniquely defined, there is a unique solution $\bar{a}_k$ that lies in the image space of $\langle M\Omega \rangle$, which is the solution that is obtained by the method given in \secref{sec:implementation}.
\item \label{itm:diagnullMJ}
Define the symmetric $s_\text{d} \times s_\text{d}$ matrix $\varmat{M\Omega}_{kk'} = \bar{a}_k^\trans \langle M\Omega \rangle \bar{a}_{k'}$ and diagonalize it using a standard routine for symmetric matrices.
Let the orthonormal eigenvectors be $G_{i(k)}$. We have $\sum_{j=1}^{s_\text{d}} \varmat{M\Omega}_{ij} G_{j(k)} = \mu_{(k)} G_{i(k)}$ with $\mu_{(k)} > 0$ for $k=1,\ldots,s_\text{d}$. The eigenvalues $\mu_{(k)}$ are positive, since $\langle M\Omega \rangle$ is positive semidefinite and the vectors $\bar{a}_k$ are independent vectors in the image space of $\langle M\Omega \rangle$.
\item
Construct the vectors $b_k = \sum_{i=1}^{s_\text{d}} G_{i(k)} \bar{b}_i$ and $\breve{a}_k = \sum_{i=1}^{s_\text{d}} G_{i(k)} \bar{a}_i$ for $k=1,\ldots,s_\text{d}$. Since $\breve{a}_k^\trans \langle \Omega \rangle b_{k'} = \breve{a}_k^\trans \langle M\Omega \rangle \breve{a}_{k'}$, we have $\breve{a}_k^\trans \langle \Omega \rangle b_{k'} = 0$ for $k \neq k'$ and $\breve{a}_k^\trans \langle \Omega \rangle b_k > 0$.
\item
Redefine $\breve{a}_k$ as $\breve{a}_k / \sqrt{\alpha_k}$, $b_k$ as $b_k / \sqrt{\alpha_k}$, $c_l$ as $c_l / \sqrt{\beta_l}$ and $d_l$ as $d_l / \sqrt{\beta_l}$,
where $\alpha_k = \breve{a}_k^\trans \langle \Omega \rangle b_k = \mu_{(k)}$ and $\beta_l = c_l^\trans \langle \Omega \rangle d_l = \lambda_{(l)}/2$.
This normalizes the modes so that $\breve{a}_k^\trans \langle \Omega \rangle b_k = 1$ for each $k$ and $c_l^{\trans} \langle \Omega \rangle d_l = 1$ for each $l$.
\item
Set $\hat{a}_k = \breve{a}_k - \sum_{l=1}^{s_\text{o}} (c_l^\trans \langle \Omega \rangle \breve{a}_k) d_l + \sum_{l=1}^{s_\text{o}} (d_l^\trans \langle \Omega \rangle \breve{a}_k) c_l$.
We have $c_l^\trans \langle \Omega \rangle \hat{a}_k = d_l^\trans \langle \Omega \rangle \hat{a}_k = 0$ for all $l,k$.
\item
Set $a_k = \hat{a}_k - \sum_{k'=1}^{s_\text{d}} \frac{1}{2}(\hat{a}_{k'}^\trans \langle \Omega \rangle \hat{a}_k) b_{k'}$. We have $a_k^\trans \langle \Omega \rangle a_{k'} = 0$ for all $k,k'$.
\item \label{itm:zeroconclusion}
The pairs $(u_1,u_2) = (\Omega a_k, \Omega b_k)$ are the inertial zero normal modes \eqref{eq:defzeronormmode}. The pairs $(u_1,u_2) = (\Omega c_l, \Omega d_l)$ are the special zero normal modes \eqref{eq:ordzeronormmode}.
All zero normal modes now satisfy the relations~\eqref{eq:orthopq} and~\equaref{eq:orthopp}.
\end{enumerate}
Let us define the zero normal modes, of which there are $s_\text{d} + s_\text{o}$, as the first modes in the list of all modes: set $w_{1i}=a_i$, $w_{2i}=b_i$ for $i=1,\ldots,s_\text{d}$ and $w_{1(s_\text{d}+i)}=c_i$, $w_{2(s_\text{d}+i)}=d_i$ for $i=1,\ldots,s_\text{o}$.
All normal modes must satisfy the relations~\eqref{eq:iJortho} and~\eqref{eq:iJorthocc}, which are equivalent to \equarefs{eq:orthopq} and~\eqref{eq:orthopp}.
Once the zero normal modes have been obtained, we may thus restrict the generalized eigenvalue problem \eqref{eq:eigenmodeHDGEP} to trial vectors $w$ that satisfy
\begin{subequations}
\begin{align}
\label{eq:orthozeropos} {(w_{1i} + i w_{2i})}^\dagger \langle \Omega \rangle w& = 0 \\
\label{eq:orthozeroneg} {(w_{1i} - i w_{2i})}^\dagger \langle \Omega \rangle w& = 0
\end{align}
\end{subequations}
for all zero normal modes $i=1,\ldots,s_\text{d} + s_\text{o}$.
These constraints can be implemented in the iterative \HDGEP{} solver in a very natural way (see \secref{sec:implementation}).
On this subspace, \equaref{eq:eigenmodeHDGEP} constitutes an \HDGEP{}, so we can efficiently find the remaining modes $i=s_\text{d} + s_\text{o} + 1,\ldots,n$.

\section{\label{sec:spindamped}Damped spin systems}

We have seen that the magnetic normal modes of a conservative spin system, which is Hamiltonian, can be obtained by solving a \HDGEP{}.
However, typical magnetic systems can be modeled more realistically using the LLG equation \cite{Gilbert2004} with a nonzero damping parameter $\eta > 0$,
\begin{equation}\label{eq:LLGfull}
\dot{\mathbf{m}}_i = -\tilde\gamma \mathbf{m}_i \times ( - \nabla_{\mathbf{m}_i}\mathcal H - \eta \dot{\mathbf{m}}_i )\text{}
\end{equation}
(compare \equaref{eq:consLLG}).
Note that some texts write the LLG equation with damping \eqref{eq:LLGfull} in a somewhat different, explicit form \cite{Landau1935,Gilbert2004}.
The damping term affects the magnetic normal modes and the eigenfrequencies $\omega$, which now acquire an imaginary part.
Our method for the magnetic normal-mode problem can be used even in this nonconservative case if we treat the damping term of the LLG equation as a perturbation.
We are justified in doing so, since $\eta$ is often small ($\eta \ll 1/\tilde \gamma m_\text{S}$, where $m_\text{S}$ is the typical magnitude $\|\mathbf{m}_i\|$ of the spins).
In this \refsection{}, we derive expressions for the damped modes in first-order perturbation theory. In particular, we obtain very simple and elegant first-order expressions~\eqref{eq:poscorrxi} and~\eqref{eq:defcorrxi} for the decay rate of the amplitude of a mode under damping.
Decay rates of modes are especially important as they determine the widths of the corresponding peaks in dynamic magnetic susceptibility functions (see \figref{fig:respfunc}), which can be measured.
Our expressions for the first-order corrections to the modes also cover those cases where special or inertial zero normal modes are present, or where the unperturbed normal modes are degenerate.

Again considering the deviation $x = m - m_0 \in \mathbb{R}^{3n}$ from a fixed equilibrium configuration $m_0$ in Cartesian coordinates (see \secref{sec:spincons}), the LLG equation with damping \eqref{eq:LLGfull} becomes, using that $\dot x = \Or({\| x \|})$,
\begin{equation}
\dot x = M x - \eta \Omega \dot x + \Or({\| x \|}^2)\text{,}
\end{equation}
with $M$ and $\Omega$ as defined in \equarefs{eq:spinM} and~\eqref{eq:spinJ}.
We can write this in explicit form as $\dot x = M' x + \Or({\| x \|}^2)$, where
\begin{equation}\label{eq:spinK}
M' = {\left(I_{3n} + \eta \Omega \right)}^{-1} M = {\left( I_{3n} + \eta^2 \Omega^\trans \Omega \right)}^{-1} \left( M - \eta \Omega M \right) \text{.}
\end{equation}
We see that to first order in $\eta$, the matrix $M'\Omega$ results from perturbation of $M\Omega$ by a term $-\eta \Omega M \Omega$.
Since the LLG equation with damping \eqref{eq:LLGfull} respects the constraint that the magnitude $\| \mathbf{m}_i \|$ of each magnetic moment be constant, we may still assume that the physically relevant vectors $x \in \mathbb{R}^{3n}$ lie in the $2n$-dimensional image space of $\Omega$.

The presence of (a not too large amount of) damping modifies the three types of normal modes as follows.
We use primes for the modes of the damped system.
\begin{enumerate}
\item A postive normal mode \eqref{eq:posnormmode} becomes a damped mode of the form
\begin{equation}\label{eq:eigenmodeext}
\left\{
\begin{array}{l}
M' u'_1 = \phantom{-}\omega' u'_2 - \xi' u'_1 \\
M' u'_2 =          - \omega' u'_1 - \xi' u'_2
\end{array}
\right.\text{.}
\end{equation}
The fundamental solutions that correspond to a damped positive mode \eqref{eq:eigenmodeext} are
\begin{equation}\label{eq:fundsoldampedpos}
\begin{array}{l}
x_1(t) = e^{-\xi' t}\left[\phantom{-}\cos(\omega' t)u'_1 + \sin(\omega' t)u'_2\right]\text{,}\\
x_2(t) = e^{-\xi' t}\left[         - \sin(\omega' t)u'_1 + \cos(\omega' t)u'_2\right]\text{.}
\end{array}
\end{equation}
\item A special zero normal mode \eqref{eq:ordzeronormmode} remains unchanged in the presence of damping.
\item An inertial zero normal mode \eqref{eq:defzeronormmode} becomes a damped mode of the form
\begin{equation}\label{eq:defmodeext}
\left\{
\begin{array}{l}
M' u'_1 = u_2 - \xi' u'_1 \\
M' u_2 = 0
\end{array}
\right.
\text{.}
\end{equation}
The corresponding fundamental solutions are
\begin{equation}\label{eq:fundsoldampeddef}
\begin{array}{l}
x_1(t) = e^{-\xi' t}u'_1 + [(1-e^{-\xi' t})/\xi'] u_2\text{,}\\
x_2(t) = u_2\text{.}
\end{array}
\end{equation}
Notice that the $u_2$ part of an inertial normal mode remains unchanged in the presence of damping.
\end{enumerate}
\Equaref{eq:eigenmodeext} is equivalent to \equaref{eq:eigenmodeHDGEP} if we replace $M$ in \equaref{eq:eigenmodeHDGEP} with $M'$ and $\omega$ with $\omega' - i \xi'$; it is in this sense that the frequency of a damped positive mode acquires an imaginary part.
Notice that with damping, \equaref{eq:eigenmodeHDGEP} is no longer a \HDGEP{}. As a result, the damped modes do not necessarily satisfy the relations~\eqref{eq:orthopq} and~\eqref{eq:orthopp}.

In first-order perturbation theory, we write a damped positive mode \eqref{eq:eigenmodeext} as
\begin{subequations}
\begin{align*}
 \omega'_k& = \omega_k + \Or(\eta^2) \\
 \xi'_k& = \eta\xi_k^{(1)} + \Or(\eta^2) \\
 u'_{1k}& = u_{1k} + \eta u_{1k}^{(1)} + \Or(\eta^2) \\
 u'_{2k}& = u_{2k} + \eta u_{2k}^{(1)} + \Or(\eta^2)\text{.}
\end{align*}
\end{subequations}
where $k$ is the mode index and $u_{1k}$, $u_{2k}$ and $\omega_k$ is the unperturbed normal mode and frequency.
We assume that the vectors $w_1,w_2$ of all unperturbed normal modes \eqref{eq:defw} satisfy the relations~\eqref{eq:orthopq} and~\eqref{eq:orthopp}.
Moreover, we assume that if any of the unperturbed normal modes are degenerate, they satisfy certain additional conditions (stated below).
Using these assumptions and the definitions~\eqref{eq:posnormmode}, \eqref{eq:ordzeronormmode} and~\eqref{eq:defzeronormmode}, it can be derived, by a rather lengthy calculation, that the first-order corrections to a positive mode $k$ are given by
\begin{equation}\label{eq:poscorrxi}
\xi_k^{(1)} = \frac{1}{2}\omega_k \left( u_{1k}^{\trans} u_{1k} + u_{2k}^{\trans} u_{2k} \right)
            = \frac{1}{2}\omega_k \left( {\| u_{1k} \|}^2 + {\| u_{2k} \|}^2 \right)
\end{equation}
\begin{subequations}
\begin{align}
\label{eq:poscorru}
\begin{split}
u_{1k}^{(1)} =
  & \sum_{l \, (\omega_l =  \omega_k)} \frac{1}{4}\left[ -(u_{1l}^\trans u_{2k} + u_{2l}^\trans u_{1k})u_{1l} + (u_{1l}^\trans u_{1k} - u_{2l}^\trans u_{2k})u_{2l} \right] \\
  & +        \sum_{l \, (\omega_l\neq\omega_k)} \frac{1}{2} \omega_k \Bigl[
          \Bigl( \phantom{-} \frac{u_{1l}^\trans u_{2k} - u_{2l}^\trans u_{1k}}{\omega_k-\omega_l} - \frac{u_{1l}^\trans u_{2k} + u_{2l}^\trans u_{1k}}{\omega_k+\omega_l} \Bigr)u_{1l} \\
  & \phantom{\sum_{l \, (\omega_l\neq\omega_k)} \frac{1}{2} \omega_k \Bigl[}
        + \Bigl( \phantom{-} \frac{u_{1l}^\trans u_{1k} + u_{2l}^\trans u_{2k}}{\omega_k-\omega_l} + \frac{u_{1l}^\trans u_{1k} - u_{2l}^\trans u_{2k}}{\omega_k+\omega_l} \Bigr)u_{2l} \Bigr] \\
  & + \sum_{l\text{ ord.}} \left[ -(u_{2l}^\trans u_{1k})u_{1l} + (u_{1l}^\trans u_{1k})u_{2l} \right] \\
  & + \sum_{l\text{ def.}} \left[ -(u_{2l}^\trans u_{1k})u_{1l} + (u_{1l}^\trans u_{1k} + \omega_k^{-1} u_{2l}^\trans u_{2k})u_{2l} \right]
\end{split} \\
\begin{split}
\label{eq:poscorrv} u_{2k}^{(1)} =
  & \sum_{l \, (\omega_l =  \omega_k)} \frac{1}{4}\left[ (u_{1l}^\trans u_{1k} - u_{2l}^\trans u_{2k})u_{1l} + (u_{1l}^\trans u_{2k} + u_{2l}^\trans u_{1k})u_{2l} \right] \\
  & +        \sum_{l \, (\omega_l\neq\omega_k)} \frac{1}{2} \omega_k \Bigl[
          \Bigl( - \frac{u_{1l}^\trans u_{1k} + u_{2l}^\trans u_{2k}}{\omega_k-\omega_l} + \frac{u_{1l}^\trans u_{1k} - u_{2l}^\trans u_{2k}}{\omega_k+\omega_l} \Bigr)u_{1l} \\
  & \phantom{\sum_{l \, (\omega_l\neq\omega_k)} \frac{1}{2} \omega_k \Bigl[}
        + \Bigl(\phantom{-} \frac{u_{1l}^\trans u_{2k} - u_{2l}^\trans u_{1k}}{\omega_k-\omega_l} + \frac{u_{1l}^\trans u_{2k} + u_{2l}^\trans u_{1k}}{\omega_k+\omega_l} \Bigr)u_{2l} \Bigr] \\
  & + \sum_{l\text{ ord.}} \left[ -(u_{2l}^\trans u_{2k})u_{1l} + (u_{1l}^\trans u_{2k})u_{2l} \right] \\
  & + \sum_{l\text{ def.}} \left[ -(u_{2l}^\trans u_{2k})u_{1l} + (u_{1l}^\trans u_{2k} - \omega_k^{-1} u_{2l}^\trans u_{1k})u_{2l} \right]\text{,}
\end{split}
\end{align}
\end{subequations}
where the first sum in \equaref{eq:poscorru} or~\eqref{eq:poscorrv} is over any modes $l$ that are degenerate with the positive normal mode $k$, plus $k$ itself; the second sum is over all other positive normal modes; the third sum is over the special zero normal modes; and the fourth sum is over the inertial zero normal modes.
For the damped inertial zero mode \eqref{eq:defmodeext}, we have
\begin{subequations}
\begin{align*}
 \xi'_k& = \eta\xi_k^{(1)} + \Or(\eta^2) \\
 u'_{1k}& = u_{1k} + \eta u_{1k}^{(1)} + \Or(\eta^2)\text{.}
\end{align*}
\end{subequations}
The first-order corrections are given by
\begin{equation}\label{eq:defcorrxi}
\xi_k^{(1)} = u_{2k}^\trans u_{2k} = {\| u_{2k} \|}^2
\end{equation}
\begin{equation}\label{eq:defcorru}
u_{1k}^{(1)} = \sum_{l\text{ pos.}} -\omega_l^{-1} \left[ (u_{1l}^\trans u_{2k})u_{1l} + (u_{2l}^\trans u_{2k})u_{2l} \right]
 + \sum_{l\text{ def.}} -(u_{1l}^\trans u_{2k})u_{1l}
\text{,}
\end{equation}
where the first sum in \equaref{eq:defcorru} is over all positive normal modes and the second sum is over all inertial zero normal modes.
We see that in both cases~\eqref{eq:poscorrxi} and~\eqref{eq:defcorrxi}, $\xi^{(1)}$ is guaranteed to be positive: for a positive damping parameter $\eta$, amplitudes of modes decrease in time.
Notice that the frequency $\omega'$ of a damped positive mode is constant to first order in $\eta$; however, there will be a second-order correction (normally negative).

If all magnetic moments in the equilibrium configuration $m_0$ have the same magnitude $\|\mathbf{m}_{0i}\| = m_\text{S}$, we have $\langle \Omega^\trans \Omega \rangle = \tilde{\gamma}^2 m_\text{S}^2 I_{2n}$,
and \equaref{eq:spinK} becomes
\begin{equation}
M'\Omega = \frac{1}{1+{(\eta \tilde\gamma m_\text{S})}^2} \left( M\Omega - \eta \Omega M\Omega \right)\text{.}
\end{equation}
We can then often further reduce the residual error in the damped positive modes, which is of second order in $\eta$, simply by dividing the $\omega'$ and $\xi'$ as obtained to first order by $1+{(\eta \tilde\gamma m_\text{S})}^2$. For a damped inertial zero mode, divide the value $\xi'$ by $1+{(\eta \tilde\gamma m_\text{S})}^2$ and multiply the vector $u'_1$ by the same factor.
These corrections do not eliminate the error of second order completely, but are very easy to implement.

If there are several distinct positive normal modes with the same frequency $\omega$, or if the dimension $s$ of the null space of $\langle M\Omega \rangle$ is larger than one, the normal-mode problem is degenerate. The damping perturbation may lift this degeneracy.
For the correctness of the expressions for the first-order corrections it is essential to choose the degenerate unperturbed normal modes in such a way that the perturbation does not mix them.
We amend the procedure of \secref{sec:redHDGEP} as follows.
Given any symmetric and positive-definite matrix $A$, we may choose the null-space vectors $y_1,\ldots,y_s$ in step~\ref{itm:findnullspace} of \secref{sec:redHDGEP} in such a way that they satisfy $y_i^\trans A y_j = \delta_{ij}$.
For the spin system with damping, we must use $A = \langle\Omega^\trans \Omega \rangle$.
The rest of the algorithm then automatically ensures that the vectors $b_k$, $c_l$ and $d_l$ (see step \ref{itm:zeroconclusion}) of the zero normal modes satisfy
\begin{subequations}
\begin{align}
                   b_k^\trans A b_{k'} = 0                        &\hspace{1.3em}\text{(for $k \neq k'$)}\\
                   c_l^\trans A c_{l'} = d_l^\trans A d_{l'} = 0  &\hspace{1.3em}\text{(for $l \neq l'$)} \\
                   b_k^\trans A c_l = b_k^\trans A d_l = 0        &\hspace{1.3em}\text{(for all $k,l$)} \\
\label{eq:liftcAd} c_l^\trans A d_{l'} = 0                        &\hspace{1.3em}\text{(for all $l,l'$)}\text{,}
\end{align}
\end{subequations}
where $k,k'$ index the inertial zero normal modes and $l,l'$ index the special zero normal modes.
For example, \equaref{eq:liftcAd} is equivalent to the condition that $u_{1l}^\trans u_{2l'}=0$ for all pairs of special zero normal modes $l,l'$.
As for the positive normal modes, if we have a block of $r$ degenerate modes at frequency $\omega>0$, we can, without breaking the conditions~\eqref{eq:orthopq} and~\eqref{eq:orthopp}, choose them in such a way that the Hermitian $r \times r$ matrix $\varmat{A}_{ij} = (w_{1i} + iw_{2i})^\dagger A (w_{1j} + iw_{2j})$ is diagonal. Here $i,j$ index those modes that are part of the degenerate block.
Again, we must use $A = \langle\Omega^\trans \Omega \rangle$.
As a result, the components $w_{1i}$ and $w_{2i}$ satisfy
\begin{subequations}
\begin{align}
                     w_{1i}^\trans A w_{1j} + w_{2i}^\trans A w_{2j} = 0 &\hspace{1.3em}\text{(for $i \neq j$)} \\
\label{eq:liftuvmin} w_{1i}^\trans A w_{2j} - w_{2i}^\trans A w_{1j} = 0 &\hspace{1.3em}\text{(for all $i,j$).}
\end{align}
\end{subequations}
For example, \equaref{eq:liftuvmin} is equivalent to the condition that $u_{1i}^\trans u_{2j} - u_{2i}^\trans u_{1j} = 0$ for all pairs of positive normal modes $i,j$ that are part of the degenerate block.

\section{\label{sec:implementation}Implementation}

The procedure for finding the magnetic normal modes can be summarized as follows.
\begin{enumerate}[A.]
\item \label{itm:procminconf}
Find a configuration $m=m_0$ that is a local minimum of the Hamiltonian $\mathcal H$, under the constraint that $\| \mathbf{m}_i \| = \text{constant}$ for each position $i$.
\item \label{itm:proczero}
If necessary, follow the procedure in \secref{sec:redHDGEP} to detect and compute any zero normal modes.
\item \label{itm:procpos}
Solve the \HDGEP{} of \equaref{eq:eigenmodeHDGEP} to find the (low-energy) positive normal modes.
\item \label{itm:procdamp}
If a damping parameter $\eta > 0$ is used, correct the normal modes using the expressions in \secref{sec:spindamped}.
\end{enumerate}
All important steps can be efficiently implemented using iterative methods for large Hermitian problems.
For concreteness, we shall discuss the iterative methods based on conjugate gradients in a bit more detail.
Alternative approaches, such as matrix-free versions of the \Lanczos{} eigenvalue algorithm \cite{Knyazev2000}, have similar properties.

Let us first remark that our scheme can also be used to find the magnetic normal modes near a local energy minimum of some continuum model. One discretizes the system using, for example, the finite-difference method or a geometric finite-element method \cite{Scholz2003}, which give effective systems that are mathematically equivalent to a finite system \cite{dAquino2005}. It is important to use a mesh that is smooth enough, to avoid effects such as an artificial Peierls pinning of domain walls \cite{Novoselov2003,Herz1977}. (This effect decreases exponentially in the inverse lattice constant \cite{Herz1977}, so there is no fundamental problem.)

In its simplest form, the conjugate-gradient method \cite{Shewchuk1994} is an iterative method for solving systems of linear equations,
\begin{equation}\label{eq:genlinsys}
Ax = b\text{,}
\end{equation}
where $A$ is a symmetric or Hermitian $N \times N$ matrix and $x$ and $b$ are vectors in $\mathbb{R}^N$ or $\mathbb{C}^N$.
$A$ and $b$ are given; $x$ is asked.
\Equaref{eq:genlinsys} is considered solved when the magnitude $\| r \|$ of the \emph{residual vector}
\begin{equation}\label{eq:resivec}
r = b - Ax
\end{equation}
is less than a certain (very small) tolerance.
In each iteration $i=1,2,\ldots$, the trial solution $x_i$ is updated,
\begin{equation}
\left\lbrace
\arraycolsep=1.4pt
\begin{array}{l l}
x_0 & = 0 \\
x_{i+1} & = x_i + \alpha_i p_i\text{,}
\end{array}
\right.
\end{equation}
where
\begin{equation}
\left\lbrace
\arraycolsep=1.4pt
\begin{array}{l l l}
p_0 & = r_0 & = b \\
p_i & = r_i + \beta_i p_{i-1} & = (b - Ax_i) + \beta_i p_{i-1}\text{.}
\end{array}
\right.
\end{equation}
A more detailed discussion of the algorithm, with expressions for the coefficients $\alpha_i$ and $\beta_i$, can be found in most textbooks on numerical methods \cite{Shewchuk1994}. What is relevant here is the following.
\begin{inparaenum}[\itshape a\upshape)]
\item We do not need to store the $N^2$ elements of $A$. All we need is a routine that can evaluate $Ax$ for any given $x$ (the \emph{action} $x \mapsto Ax$ of $A$). The conjugate-gradient algorithms use this routine as a `black box'.
\item Every trial solution $x_i$ is a linear combination of $b,Ab,A^2b,\ldots,A^{i-1}b$; the conjugate-gradient method is a \emph{\Krylov{}-subspace method}.
\end{inparaenum}

A variant of the conjugate-gradient method can be used to solve nonlinear optimization problems \cite{Shewchuk1994}, where a local minimum of a multivariate function $f(x)$ is asked.
Here the gradient $\nabla f$ plays the role of the residual vector \eqref{eq:resivec}.
This method is also suitable for minimization problems under constraints $g_1(x)=\ldots=g_k(x)=0$.
In that case, one should project the residual vector $r$ onto the tangent space:
\begin{equation}
r = \nabla f - \sum_{i=1}^k \lambda_i\nabla g_i \text{,}
\end{equation}
where
\begin{equation}
\lambda_i = \frac{(\nabla f) \cdot (\nabla g_i)}{{\| \nabla g_i \|}^2}
\end{equation}
is a Lagrange multiplier.

The conjugate-gradient eigenvalue algorithm \cite{Bradbury1966} can be seen as a special case of constrained nonlinear optimization.
If we minimize the function
\begin{equation}
f(x) = x^\dagger D x
\end{equation}
under the constraint
\begin{equation}
g_1(x) = x^\dagger S x = 1\hspace{1.2em}\text{ (normalization),}
\end{equation}
where $D$ and $S$ are Hermitian matrices, we obtain the lowest eigenvalue $\lambda_1$ and the corresponding eigenvector $x_1$ of the \HDGEP{} $Dx=\lambda S x$.
(The \SDGEP{} case, where $D$, $S$ and $x$ are real, is entirely analogous.)
$S$ must be positive definite to guarantee that a minimum exists.
Once we have the first eigenvector $x_1$, we can obtain the next eigenvector by repeating the minimization under an additional constraint:
\begin{equation}
g_2(x) = x_1^\dagger S x = 0\hspace{1.2em}\text{ (orthogonality).}
\end{equation}
For $\lambda \neq 0$, this is equivalent to the constraint
\begin{equation}\label{eq:conjgradorthoconsalt}
g_2'(x) = \lambda_1 x_1^\dagger S x = x_1^\dagger D x = 0\text{.}
\end{equation}
Once we have found the second eigenvector, we move on to the third, and so on, applying constraints of the form \eqref{eq:conjgradorthoconsalt} for all previously obtained eigenvectors.
We continue until we have found as many eigenvectors $x_1,x_2,\ldots$ with eigenvalues $\lambda_1 < \lambda_2 < \ldots$ as we need.

The fact that we do not need to explicitly store the matrices in memory is a crucial advantage.
For simplicity, let us first consider a one-dimensional $n$-spin chain with only exchange and uniaxial anisotropy energy,
\begin{equation}
\mathcal H = E_\text{ex} + E_\text{ani} = \sum_{i=1}^{n-1} -2J \mathbf{m}_i \cdot \mathbf{m}_{i+1} -\sum_{i=1}^n K m_{iz}^2\text{.}
\end{equation}
The Hessian matrix $A$ (see \secref{sec:spincons}) is given by
\begin{equation}\label{eq:Ashortrange}
A_{(i\alpha)(j\beta)} = \left\lbrace
\begin{array}{l l}
-2K & \text{if $i=j$ and $\alpha=\beta=z$} \\
-2J & \text{if $i=j-1,j+1$ and $\alpha=\beta$} \\
0 & \text{otherwise}
\end{array}
\right.\text{;}
\end{equation}
equivalently, it may be defined by its action $x \mapsto Ax$,
\begin{equation}
(Ax)_{i\alpha} = \left\lbrace
\begin{array}{ll}
-2J (x_{(i-1)\alpha} + x_{(i+1)\alpha}) & \text{if $\alpha=x,y$} \\
-2J (x_{(i-1)\alpha} + x_{(i+1)\alpha}) - 2K x_{i\alpha} & \text{if $\alpha=z$}
\end{array}
\right.\text{,}
\end{equation}
where we take $x_{i\alpha}=0$ for $i=0$ and $i=n+1$.
We see that the evaluation of the action of $A$ on an arbitrary vector $x$ takes only $\Or(N)$ time, while any manipulation with or decomposition of the $3n \times 3n$ matrix $A$ obviously takes at least $\Or(N^2)$ time if it is explicitly stored in memory in full.
That is why \Krylov{}-subspace methods are a popular choice for linear equations or eigenvalue problems of sparse matrices \cite{Saad2003}.
If long-range interactions are taken into consideration, the matrix $A$ is dense. Nevertheless, the action of $A$ can still be evaluated in much less than $\Or(N^2)$ time, as follows.
For nearly all physical systems, $A$ can be separated into a short-ranged part $A_\text{s}$ such as \equaref{eq:Ashortrange}, which is sparse, and a long-ranged part $A_\text{l}$, which is invariant under spatial translations (it is a convolution) \cite{Labbe1999}.
To perform the action on a given vector $x$, we separately evaluate the contributions $A_\text{s}x$ and $A_\text{l}x$ and then add them up to obtain $Ax = A_\text{s}x + A_\text{l}x$.
In typical magnetic systems, the relevant long-range interaction is the dipolar interaction. We can evaluate $A_\text{l}x$ by performing the convolution in the Fourier representation of $x$, where it becomes a simple elementwise multiplication. The two Fourier transformations that are necessary take $\Or (N\log N)$ time \cite{Frigo2005}. A similar mixed real-space--reciprocal-space approach is taken in most plane-wave electronic-structure codes \cite{Payne1992}.
Even if the system is not perfectly translationally invariant, for instance because it has some nonrectangular finite geometry, we can efficiently evaluate $A_\text{l}x$ by reducing the dipolar problem to the Poisson problem \cite{Fidler2001} and solving it using multigrid methods \cite{Saad2003}. The complexity analysis is similar.
It is thus possible to implement a routine that can evaluate $Ax$, and hence $M\Omega x$ \eqref{eq:spinMJ}, for any given $x$ in $\Or(N\log N)$ rather than $\Or(N^2)$ time.

In the remainder of this \refsection{}, we discuss the specific implementation of each of the four stages listed above.

\paragraph*{Stage~\ref{itm:procminconf}.}

A minimum-energy configuration $m_0$ can be found using, for example, the nonlinear conjugate-gradient optimization method, which is implemented in existing micromagnetics codes. Note that many magnetic systems have multiple local energy minima.
In this article, we regard one particular $m_0$ as given.

\paragraph*{Stage~\ref{itm:proczero} -- step~\ref{itm:findnullspace}.}

In step~\ref{itm:findnullspace} of \secref{sec:redHDGEP}, we need to calculate the null vectors $y_1,\ldots,y_s$ of $\langle M\Omega \rangle$.
This is in fact a symmetric eigenvalue problem. It might be solved as a particular case of the conjugate-gradient \SDGEP{} algorithm (set $D=\langle M\Omega \rangle$ and $S=I$).
The sequential nature of this method means that we can efficiently obtain the lowest few eigenvectors. We stop once we find the first positive eigenvalue. The eigenvectors with eigenvalue zero constitute a basis of the null space of $\langle M\Omega \rangle$.

In our definition of the restricted matrix $\langle M\Omega \rangle$, we formally require construction of a basis of the image space of $\Omega$. In practice, we do not normally need to construct the basis explicitly. We may simply set $D=M\Omega$, provided our initial guess $x_0$ is in the image space of $\Omega$ (that is, we set $x_0 = \Omega y_0$, where $y_0$ is a random vector). Since $x_0$, $M\Omega x_0$, $\Omega x_0$, ${(M\Omega)}^2 x_0$, etc.\ all lie in the image space of $\Omega$, the minimization will automatically be restricted to trial solutions in this space.
We remark that for numerical stability, it may be necessary occasionally to project the trial vector $x_i$ back onto the image space of $\Omega$.

\paragraph*{Stage~\ref{itm:proczero} -- step~\ref{itm:solveMJ}.}

In step~\ref{itm:solveMJ} of \secref{sec:redHDGEP}, we need to solve the linear system $\langle M\Omega \rangle x = g$, where $g = \langle \Omega \rangle b_k$.
This problem may seem ill posed, since $\langle M\Omega \rangle$ is not invertible (even with the angular brackets).
However, we know that a solution exists ($g$ lies in the image space of $\langle M\Omega \rangle$). Since the solution-vector iterates are always linear combinations of $g, \langle M\Omega \rangle g, {(\langle M\Omega \rangle)}^2 g, \ldots$, we in effect restrict our search to trial solutions $x$ in the image space of $\langle M\Omega \rangle$. In this linear subspace, the solution $x$ is unique.

In practice, $g$ will not lie in the image space of $\langle M\Omega \rangle$ numerically exactly, but only up to a small tolerance, so that the solver may fail once the magnitude of the residual vector becomes on the order of this tolerance.
We may remedy this as follows.
Project $g$ onto the orthogonal complement of $y_1,...,y_s$, and do the same for $\langle M\Omega \rangle x$ in each iteration.
Effectively, we now find a solution of $P \langle M\Omega \rangle P = P h$, where $P$ (symmetric) performs the projection.

For the sake of completeness, we remark that again, we may use $M\Omega$ instead of $\langle M\Omega \rangle$, as the image space of $M\Omega$ is contained in the image space of $\Omega$.

\paragraph*{Stage~\ref{itm:procpos}.}

The problem \eqref{eq:eigenmodeHDGEP} can be solved using the conjugate-gradient \HDGEP{} scheme, where in \equaref{eq:genHDGEP} we set
\begin{equation}
\arraycolsep=1.4pt
\begin{array}{l l}
D &= -i\langle\Omega\rangle\text{,}\\
S &= \langle M\Omega \rangle\text{,}\\
\lambda &= \omega^{-1}
\text{.}
\end{array}
\end{equation}
Notice that we only need the (action of the) matrices $\Omega$ and $M\Omega$, which have simple forms~\eqref{eq:spinJ} and~\eqref{eq:spinMJ}.
Again, we do not need to implement the restrictions $\langle \cdot \rangle$ explicitly, provided that our initial guess is in the image space of $\Omega$.
For each positive normal mode \eqref{eq:posnormmode}, there are two solutions of \equaref{eq:genHDGEP}: one with $\lambda = \omega^{-1}$ and one with $\lambda = -\omega^{-1}$. We obviously need to find only one of the two. If we find a negative-$\lambda$ solution $x$, we must take the complex conjugate of $x$ to obtain the positive-$\lambda$ solution.
Notice that the eigenvalue $\lambda$ that the \HDGEP{} algorithm finds is the reciprocal of the angular frequency $\omega$.
The \HDGEP{} algorithm normalizes the solutions $x$ so that $x_i^\dagger S x_j = x_i^\dagger \langle M\Omega \rangle x_j = \delta_{ij}$.
To obtain the correct normalization \eqref{eq:iJortho}, we must divide each (positive-$\lambda$) solution $x_i$ by $\sqrt{\lambda_i/2}$; we have
\begin{equation}
w_{1i} + iw_{2i} = \sqrt{2 / \lambda_i} \, x_i\text{,}
\end{equation}
where $w_{1i}$ and $w_{2i}$ are the real vectors defined in \equaref{eq:defw}.

The eigenvalues $\lambda$ at the extremes of the spectrum are $\lambda=-\omega_0^{-1}$ and $\lambda=\omega_0^{-1}$, where $\omega_0$ is the angular frequency of the lowest-frequency positive normal mode.
\HDGEP{} algorithms such as the conjugate-gradient scheme find the solutions of \equaref{eq:genHDGEP} with either the lowest or the highest eigenvalues $\lambda$. We see that it does not matter if we let the algorithm minimize $\lambda$ (as we do above) or maximize $\lambda$: in either case, we obtain the \emph{lowest-frequency} normal modes first.
If we minimize the eigenvalue $\lambda$, we find the negative-$\lambda$ solutions and we must apply to the trial solution $x$ a constraint ${(w_{1k} - iw_{2k})}^\dagger {(-i \langle \Omega \rangle)} x=0$ for each previously obtained positive normal mode $k$ (see \equaref{eq:conjgradorthoconsalt}). If we choose to maximize the eigenvalue $\lambda$, we must apply a constraint ${(w_{1k} + iw_{2k})}^\dagger {(-i \langle \Omega \rangle)} x=0$ for each previously obtained positive normal mode $k$.
If any zero normal modes were found in stage~\ref{itm:proczero}, we need to eliminate those from the problem to ensure that $S = \langle M\Omega \rangle$ is positive definite on the space of trial solutions. The constraints~\eqref{eq:orthozeropos} and~\eqref{eq:orthozeroneg} that accomplish this are of exactly the same form as the constraints for previously obtained positive normal modes.

The simple conjugate-gradient \HDGEP{} scheme outlined above may be improved in several ways.
It is well known that matrix-free eigenvalue methods require good preconditioning to be efficient \cite{Knyazev2000,Saad2003,Feng1996,Payne1992}.
Indeed, we find that preconditioning as described below greatly improves performance, especially if the exchange constant between adjacent sites is large as compared to the anisotropy constant. This is the case in most atomistic simulations and in continuum systems discretized with a reasonably high spatial resolution.
(Only for relatively modest systems, say $n \sim 1000$, preconditioning is unnecessary; methods that use explicit matrix decompositions \cite{Demmel2000} are likely to be more efficient.)
How a preconditioner can be incorporated into the conjugate-gradient \HDGEP{} scheme is described in many texts \cite{Knyazev2000,Saad2003,Feng1996,Payne1992}.
In addition, efficiency may be improved by using a \emph{simultaneous} conjugate-gradient scheme \cite{Knyazev2000,Payne1992}, especially if some of the eigenvalues are closely spaced.

We use a preconditioner that is based on an inversion of the spin-wave dispersion relation \eqref{eq:spindispersion} in reciprocal space, similar to the preconditioners used to solve the Schr\"{o}dinger equation in electronic-structure calculations \cite{Payne1992}. In other words, the preconditioner approximates the spectrum of the system with the spin-wave spectrum of a homogeneous system and uses this to speed up convergence of the trial solution.
Note that since a typical spin-wave dispersion relation has no zeros (see \figref{fig:dispperfect}), the preconditioner acts in real space as a convolution with some kernel that decays exponentially, with a characteristic decay distance on the order of the domain-wall width. Thus, we could in principle even implement the preconditioner in $\Or(N)$ rather than $\Or(N\log N)$ time.
If the explicit restrictions $\langle \cdot \rangle$ of $M\Omega$ and $\Omega$ are not used, it is of course important to ensure that the preconditioned reciprocal vector is projected back onto the image space of $\Omega$ in order to ensure that the trial solution $x$ does not move out of the image space of $\Omega$.
Preconditioning can also greatly speed up convergence for steps~\ref{itm:findnullspace} and~\ref{itm:solveMJ} of stage~\ref{itm:proczero}.

\paragraph*{Stage~\ref{itm:procdamp}.}

In principle, the full set of unperturbed magnetic normal modes needs to be available to calculate the correction due to damping for any given mode. This could be a problem, since we usually know only the normal modes near the bottom of the spectrum.
This forces us to truncate the sums in \equarefs{eq:poscorru}, \eqref{eq:poscorrv} and~\eqref{eq:defcorru}.
We verify in \ssecref{sec:accuracy} for a realistic system that this approximation is justified.
In practice, the high-wavenumber spin-wave modes are increasingly oscillatory and have an overlap with the lower, smoother modes that decreases exponentially in wavenumber.

Notice that the damped modes do not, in general, satisfy the relations~\eqref{eq:orthopq} and~\eqref{eq:orthopp}. To carry out a mode analysis of some configuration near $m_0$, first obtain the coefficients of the unperturbed modes using \equaref{eq:decompx} and then use \equarefs{eq:poscorru}, \eqref{eq:poscorrv} and~\eqref{eq:defcorru} to convert these into the coefficients of the damped modes.

\section{\label{sec:examples}Examples}

\begin{figure*}
  \includegraphics[width=\figwidthfull]{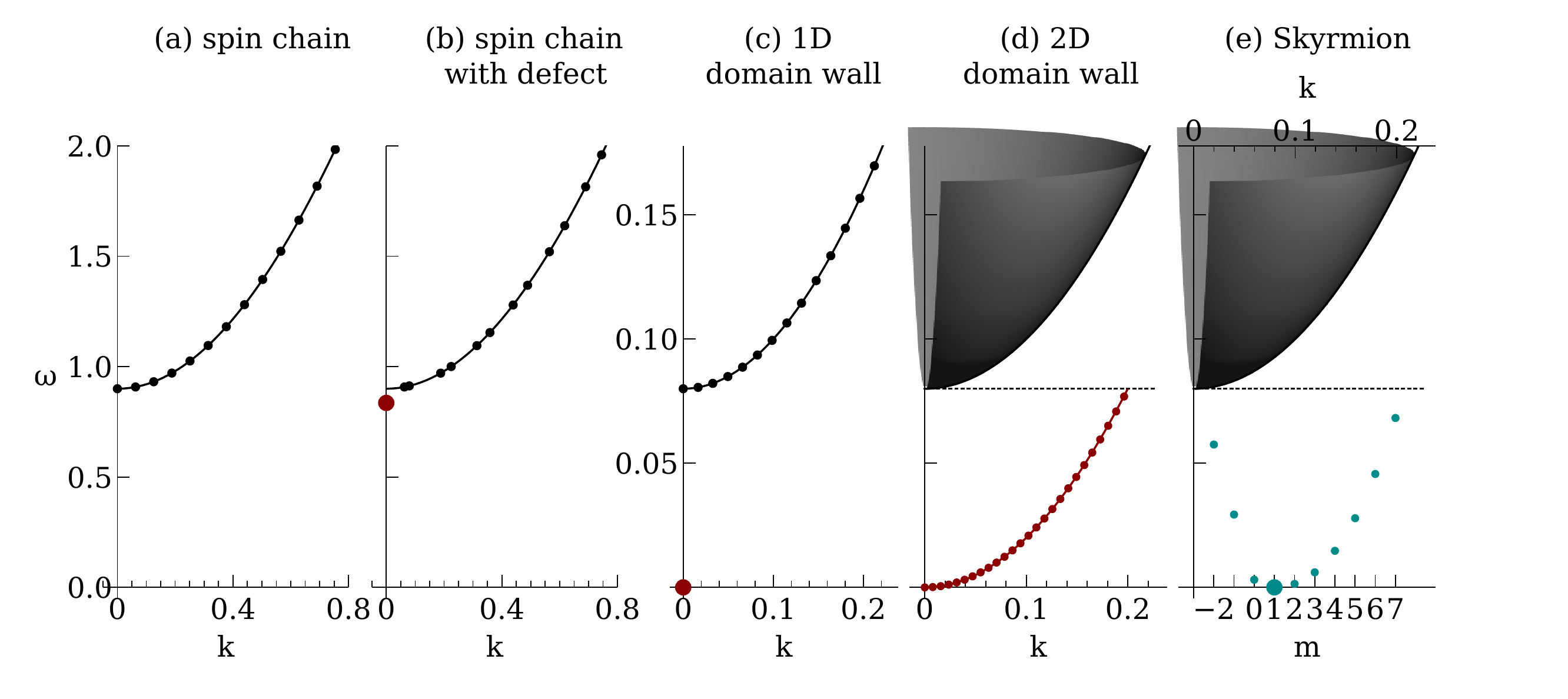}
  \caption{\label{fig:dispersion}
    Color) Comparison of the spectra of all systems considered. The presence of defects leads to localized modes with frequencies below the spin-wave continuum. The 1D spin chain is described in \ssecref{sec:experfectchain}, the 1D and 2D domain walls in \ssecrefs{sec:1Ddw} and~\ref{sec:2Ddw} and the Skyrmion in \ssecref{sec:exskyrmion}. The bottom of the spin-wave continuum is at $\omega=2\tilde\gamma m_\text{S} K$ (for uniaxial anisotropy). In the plots the wavevector $k$ is given in units of $a^{-1}$ and the angular frequency $\omega$ in units of $\tilde\gamma m_\text{S} J$.
    All continuous branches of modes are discretized (become quasicontinuous) because of the finite dimensions of the systems.
  }
\end{figure*}
In this \refsection{}, we study some key examples that are illustrative of the general properties of magnetic normal modes and make evident the fundamental distinction between inertial and special zero normal modes. We also discuss how to calculate effective masses for the inertial zero normal modes.
\Figref{fig:dispersion} provides an overview of the spectra of all systems we consider here.

We begin by studying the one-dimensional (1D) spin chain, possibly with a defect, in \ssecref{sec:experfectchain}.
We specifically look at the effect the effect of damping (\ssecref{sec:respfunc}), and we demonstrate how the expressions in \secref{sec:spindamped} can be used to calculate dynamical magnetic susceptibilities.
In \ssecref{sec:inertial}, we discuss how the fundamentally different types of dynamics of magnetic structures can be related to the two types of zero modes (special and inertial). In particular, we show how to calculate effective masses.
We focus on the properties of zero modes in spin systems with a 1D or 2D domain wall or a Skyrmion (\ssecrefs{sec:1Ddw}--\ref{sec:exskyrmion}), and we investigate a general relation between zero modes and the dispersion relations of extended systems (\ssecref{sec:2Ddw}). \Ssecref{sec:accuracy} evaluates the accuracy of our perturbative treatment of damping.

The 2D systems are of a size ($40\thinspace000$ spins) for which we begin to appreciate the scalability of the iterative \HDGEP{} methods (see \secref{sec:implementation}).
With our code, we are able to find the 20 or 30 lowest modes of these systems in a matter of minutes on just a single CPU core. (We remark that the calculation time could be reduced further by parallelization. Matrix-free iterative methods such as the conjugate-gradient \HDGEP{} scheme, especially the simultaneous versions, are known for being highly parallelizable \cite{Knyazev2000, Saad2003}.)
While for the sake of simplicity the examples only take short-range interactions into account, they could be extended to include magnetostatic (dipolar) and other interactions. This may be done in an efficient manner without any fundamental change to the method (see \secref{sec:implementation}).
Inclusion of magnetostatic interactions in rectangular systems of a similar size would not lead to much longer calculation times, since for the purpose of preconditioning our present code already performs a full FFT of the trial solution in each iteration.

\subsection{\label{sec:experfectchain}Spin waves in 1D spin chains}

\begin{figure}
  \includegraphics[width=\figwidthhalf]{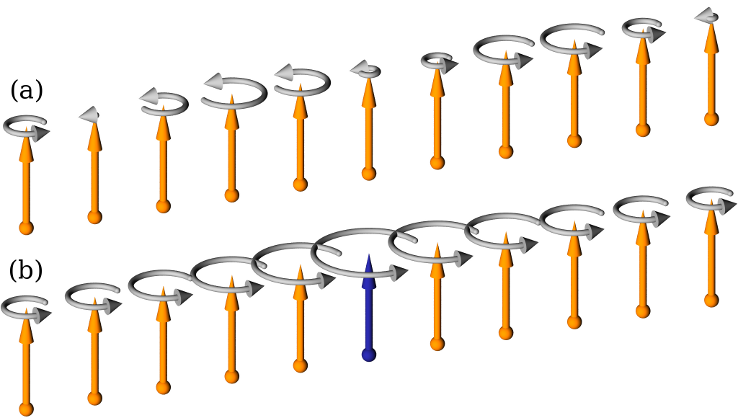}
  \caption{\label{fig:spinwaves}Color) Normal modes of a 1D ferromagnetic spin chain with \Neumann{} boundary conditions. Only a part of the chain is shown. The big straight arrows indicate the equilibrium orientations of the spins. The circular arrows indicate the path traced by the spins if the normal mode is excited. While the normal modes are calculated in the linear (small-amplitude) approximation, we show them with a large amplitude for clarity. \figsub{a}~A spin-wave mode with $k = 0.71a^{-1}$ 
  in a perfect spin chain. \figsub{b}~The lowest mode in a spin chain with a defect, located at the site shown in blue, where the anisotropy is reduced from $K=0.45J$ to $K=0.09J$.}
\end{figure}

We first consider a finite, $n$-atom spin chain without defects. We set $\| \mathbf{m}_i \| = m_\text{S}$ for all spins. The Hamiltonian
$\mathcal H = E_\text{ex} + E_\text{ani}$
consists of nearest-neighbor exchange coupling
\begin{equation}\label{eq:Eex}
E_\text{ex} = \sum_{i=1}^{n-1} -2J \mathbf{m}_i \cdot \mathbf{m}_{i+1}
\end{equation}
with an exchange constant $J>0$ (ferromagnetic) and uniaxial anisotropy
\begin{equation}\label{eq:Eaniuni}
E_\text{ani} = \sum_{i=1}^n -K {(\mathbf{m}_i \cdot \hat{ \mathbf{z}})}^2
\end{equation}
with $K>0$ (easy-axis type).
We number the spins as $i=1,\ldots,n$. There is no external magnetic field.
We linearize around the uniform, collinear equilibrium configuration $\mathbf{m}_{0i} = m_\text{S} \hat{\mathbf{z}}$, shown in \figref{fig:spinwaves}, which is one of the two ground-state configurations ($\mathbf{m}_{0i} = -m_\text{S} \hat{\mathbf{z}}$ is the other).
Our truncation of the exchange couplings \eqref{eq:Eex} at the ends of the chain results in \Neumann{} boundary conditions for the spin waves.

\begin{figure}
  \includegraphics[width=\figwidthhalf]{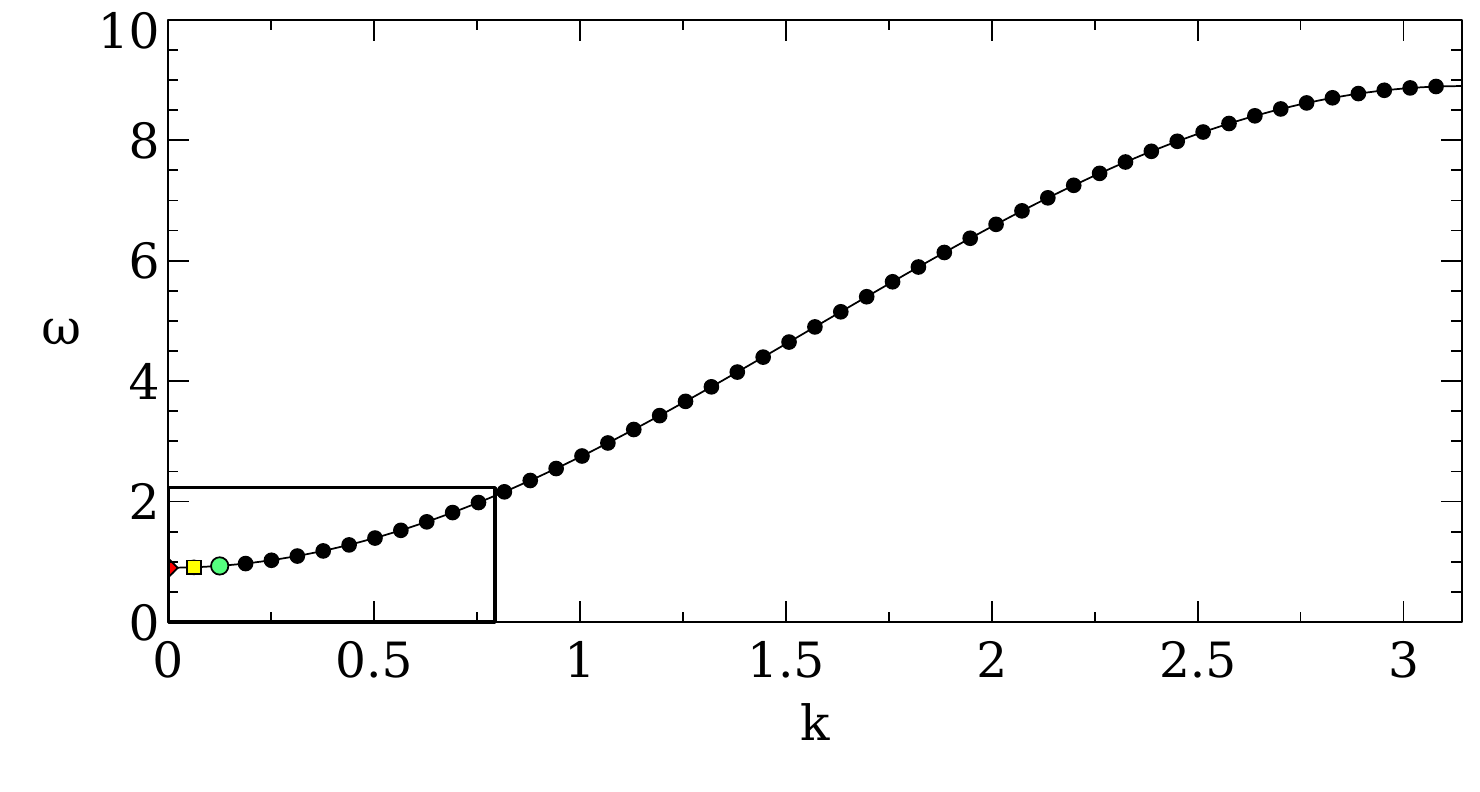}
  \caption{\label{fig:dispperfect}
    Color) Dispersion $\omega(k)$  of the perfect 1D spin chain. The wavenumber $k$ is given in units of $a^{-1}$. The angular frequency $\omega$ is given in units of $\tilde\gamma m_\text{S} J$.
    The solid line is the analytical dispersion relation \eqref{eq:spindispersion} and the dots show the spectrum of a $50$-spin chain.
  The Hamiltonian consists of exchange \eqref{eq:Eex} and uniaxial anisotropy \eqref{eq:Eaniuni} with  $K=0.45J$.
   The area in the rectangle is expanded in \figrefsub{fig:dispersion}{a}.
   The colored dots correspond to \figrefsub{fig:waveforms}{a}.}
\end{figure}

\begin{figure}
  \includegraphics[width=\figwidthhalf]{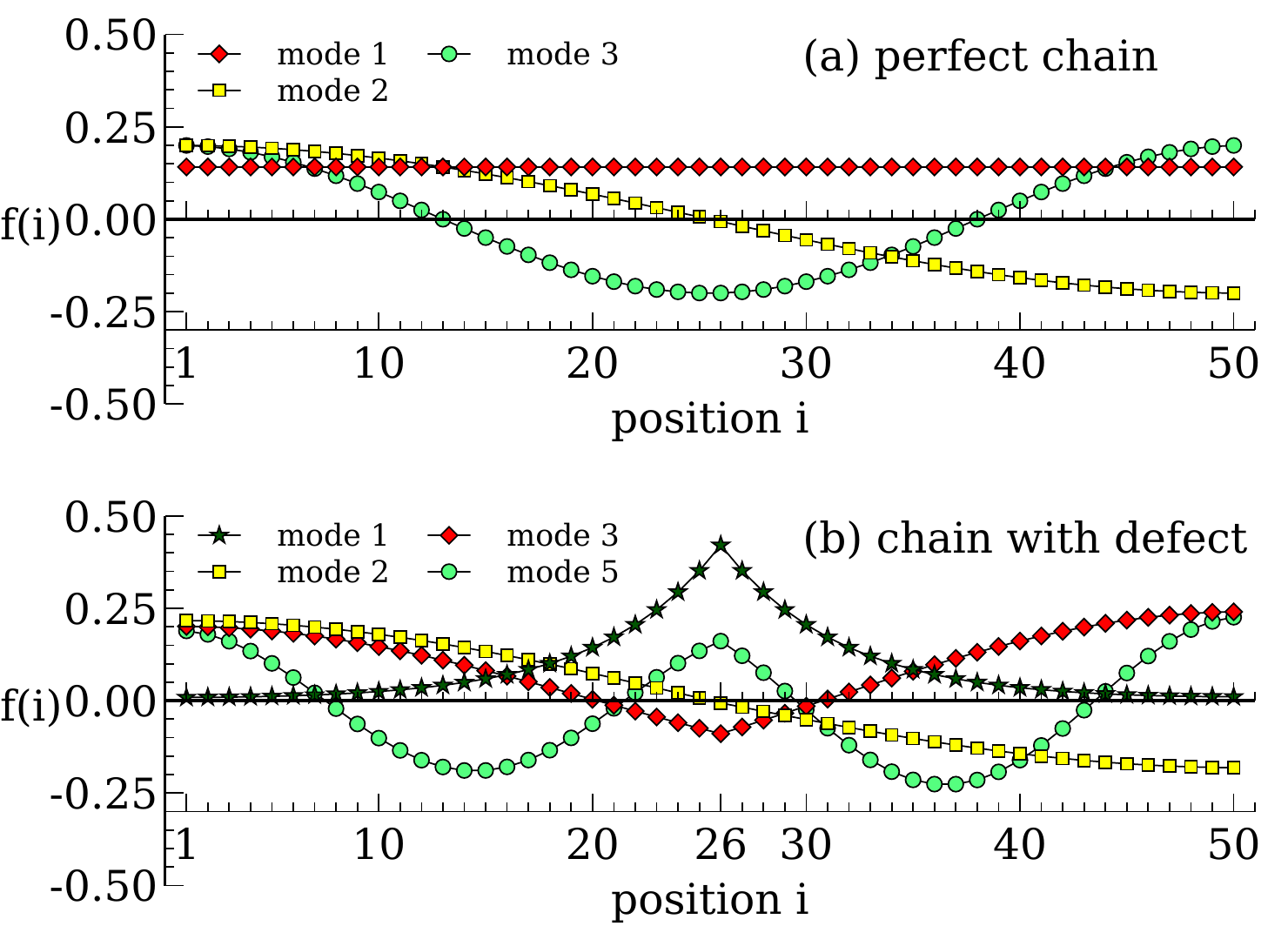}
  \caption{\label{fig:waveforms}
  Color) Amplitude profiles $f(i)$ \eqref{eq:simplewaveform} for some low-energy normal modes of a 1D $50$-spin chain, \figsub{a}~without and \figsub{b}~with a defect.
  The spin chain with defect is different from the perfect spin chain only at a single site $i=26$, where $K=0.09J$ instead of $0.45J$.
  Notice that the defect gives rise to a localized mode (see also \figrefssub{fig:dispersion}{b} and~\figrefsubpart{fig:spinwaves}{b}).
  }
\end{figure}

The magnetic normal modes of a 1D spin chain are well known, but we reproduce them here for comparison (see \figrefssub{fig:dispersion}{a}, \figrefsubpart{fig:spinwaves}{a}, \ref{fig:dispperfect} and~\figrefsubpart{fig:waveforms}{a}).
By taking a general linear combination of the fundamental solutions \eqref{eq:fundsolconspos}, we see that the dynamics of any positive normal mode \eqref{eq:posnormmode} is given by
\begin{equation}\label{eq:spinmodedyn}
\mathbf{x}_i(t) = A \cos(\omega t + \phi) \mathbf{u}_{1i} + A \sin(\omega t + \phi) \mathbf{u}_{2i} + \Or(A^2)\text{,}
\end{equation}
where $A$ is the amplitude and $\phi$ is the phase of the mode.
The variable $\mathbf{x}_i = \mathbf{m}_i - \mathbf{m}_{0i}$ is the deviation of the magnetic moment at site $i$ from its equilibrium position.
For the 1D collinear spin chain with Neumann-type boundary conditions, we have spin-wave modes \eqref{eq:spinmodedyn} with
\begin{equation}\label{eq:simplewaveform}
\mathbf{u}_{1i} = f(i)\hat{\mathbf x}\hspace{0.7em}\text{and}\hspace{0.7em}\mathbf{u}_{2i} = f(i)\hat{\mathbf y}\text{,}
\end{equation}
where
\begin{equation}
f(i) = \cos\left[ak_l \left(i-\frac{1}{2}\right)\right]
\end{equation}
(standing waves).
The dispersion relation is given by
\begin{equation}\label{eq:spindispersion}
\omega(k) = 2\tilde\gamma m_\text{S} [K + 2 J(1-\cos {ak})]\text{,}
\end{equation}
where $k$ is the wavenumber and $a$ is the spacing between lattice sites. The bottom of the spin wave continuum is thus at $\omega=2\tilde\gamma m_\text{S} K$.
The wavenumber of the mode with index $l=1,\ldots,n$ is given by $k_l = \pi (l-1) / a n$.
Our code finds the right frequencies $\omega(k_l)$ (see \figref{fig:dispperfect}) and the right form of the spin waves (see \figrefsub{fig:waveforms}{a}).

We now consider the effect of a defect, modeled by reducing the anisotropy constant $K$ at a single site.
The normal modes are still of the form \eqref{eq:simplewaveform}, but have different profiles $f(i)$ (see \figrefsub{fig:waveforms}{b}).
The lowest mode is localized at the defect site and decays exponentially away from it (evanescent spin wave; see also \figrefsub{fig:spinwaves}{b}); its frequency is just below the spin wave continuum (see \figrefsub{fig:dispersion}{b}).
The other $n-1$ modes are spin-wave modes.
They are perturbed with respect to the normal modes of the perfect spin chain.
Since in the example of \figrefsub{fig:waveforms}{b} we place the defect almost in the middle ($i=26$) of a chain of $n=50$ spins, the odd-numbered spin-wave modes have a `kink' at the defect site while the even-numbered spin-wave modes are almost identical to those of the perfect spin chain.

\subsection{\label{sec:respfunc}Dynamical magnetic susceptibility}

\begin{figure}
  \includegraphics[width=\figwidthhalf]{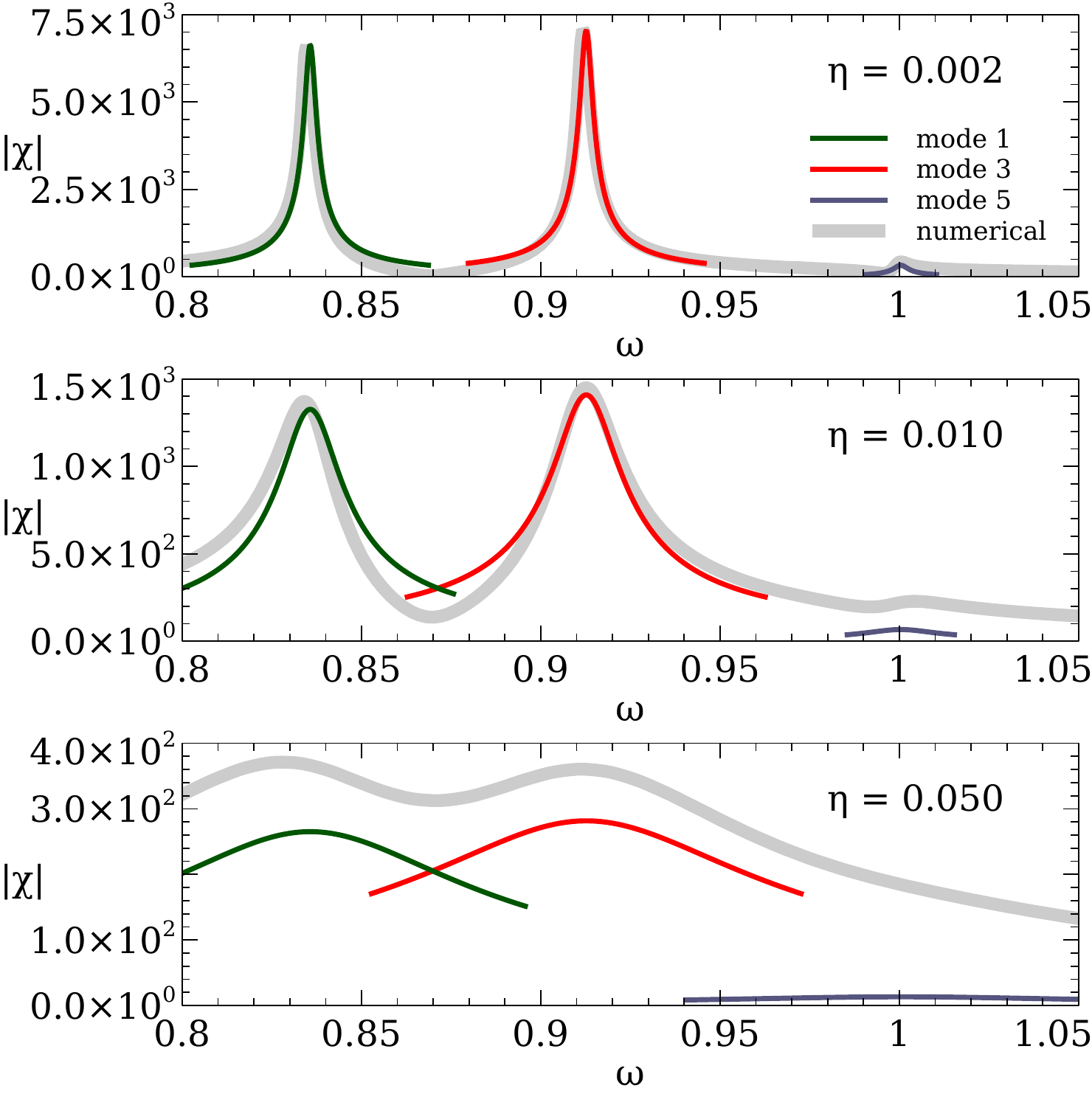}
  \caption{\label{fig:respfunc}
   Color) Absolute value of the magnetic susceptibility function $\chi(\omega)$ of the 1D 50-spin chain with a defect ($K=0.09J$ at site $i=26$; $K=0.45J$ everywhere else), for three different values of the damping parameter $\eta$.
   The driving frequency $\omega$ is given in units of $\tilde\gamma m_\text{S} J$, $\chi$ in $J^{-1}$ and $\eta$ in ${(\tilde\gamma m_\text{S})}^{-1}$.
   The absolute value of the contribution of each mode is plotted alongside the overall magnetic susceptibility function as obtained in spin-dynamics simulations.
   The width of the peak that corresponds to a mode $l$ is proportional to its decay rate $\xi'_l$ \eqref{eq:eigenmodeext}.
   We find an excellent agreement between the peak widths calculated by means of \equaref{eq:poscorrxi} and the results from spin-dynamics simulations.
   The discrepancy in between peaks is due to the fact that the contributions of all modes must be added up with their complex phases and may interfere with each other. 
   In this particular example, only the modes 1, 3 and 5 contribute significantly to the dynamical susceptibility. Other modes generate a magnetization that is negligible when integrated over the whole length of the chain (see, for example, mode 2 in \figrefsub{fig:waveforms}{b}).}
\end{figure}
Response functions, such as dynamical magnetic susceptibilities, allow comparison of calculated spectra to experimental observables (see, for example, \explcite{Giesen2007}).
Using the expressions in \secref{sec:spindamped}, our scheme allows one to calculate dynamical susceptibility functions in a way that is usually much less computationally expensive than with spin-dynamics simulations.
Here we consider the response of the magnetization in the $\hat{\mathbf{x}}$-direction (perpendicular to the equilibrium magnetization, which is in the $\hat{\mathbf{z}}$-direction) to an oscillating external magnetic field, also in the $\hat{\mathbf{x}}$-direction.
The dynamical susceptibility $\chi$ is the ratio between the complex amplitudes of the applied field and the resulting magnetization.

Each positive mode \eqref{eq:eigenmodeext} contributes to $\chi$ a term $\propto 1/(\xi'_l + i(\omega - \omega'_l))$ \cite{Akhiezer1968}, where $\omega'_l$ and $\xi'_l$ are the frequency and decay rate of mode $l$ and $\omega$ is the driving frequency. This means that the width of a peak in the dynamical magnetic susceptibility function is directly proportional to the decay rate $\xi'_l$. The same mode gives a similar contribution near $\omega = -\omega'_l$.

In \figref{fig:respfunc}, we plot the absolute value of the dynamical susceptibility $\chi(\omega)$ of the $50$-spin system with defect (see \ssecref{sec:experfectchain} and \figrefsub{fig:waveforms}{b}), for several values of the damping parameter $\eta$.
We have obtained $\chi(\omega)$ from spin-dynamics simulations, where we integrate the LLG equation \eqref{eq:LLGfull} using the implicit-midpoint timestepping scheme \cite{dAquino2005,Mentink2010} (no stochastic term).
We compare this to the contributions from the individual modes calculated using our scheme.
We see that the contribution of each mode to $\chi(\omega)$ is very well approximated by calculating $\xi'_l$ by means of \equaref{eq:poscorrxi}.

\subsection{\label{sec:inertial}Inertial versus noninertial behavior of topological defects}

The fundamental distinction between inertial and special zero normal modes described in \secref{sec:hamsysbrief} is further clarified by examining the effect of an external potential on the dynamics of a topological defect. The general considerations we present here are applied to specific systems in \ssecrefs{sec:1Ddw}--\ref{sec:exskyrmion}.


Zero modes typically appear as a consequence of a broken continuous symmetry of the system.
For example, the energy of a system with a domain wall (see \ssecrefs{sec:1Ddw} and~\ref{sec:2Ddw}) or a Skyrmion (see \ssecref{sec:exskyrmion}) in a homogeneous material is invariant under translation of the topological defect.
Since no microscopic energy scale is associated with changes of the system that respect the symmetry, weak external perturbations of the Hamiltonian that couple to such changes can have a significant effect over time.
By studying the response of the system to such external forces, we establish its effective (that is, low-energy or long-time) dynamics.
For example, an effective force on a topological defect may result from an external magnetic field or from dipolar interactions within the system.

We consider the dynamics of just a single degree of freedom, corresponding to a zero normal mode $(u_1,u_2)$.
The deviation $x = m - m_0$ of the system from its equilibrium configuration is given by (see \equaref{eq:decomppq})
\begin{equation}\label{eq:xinpq}
x = p_q u_1 + q u_2 + \Or(p_q^2 + q^2) \text{.}
\end{equation}
We write $p_q$ instead of just $p$ to emphasize that this variable is canonically conjugate to $q$.
Let us for concreteness assume that the vector $u_2$ generates an infinitesimal translation of a topological defect. Thus we have, for a certain constant $\alpha$,
\begin{equation}
s = \alpha q\text{,}
\end{equation}
where $s$ is the position of the center of the defect, in units of length.
Using \equaref{eq:xinpq}, it is straightforward to obtain the coefficient $\alpha$ from the calculated normal mode.
The variable canonically conjugate to $s$ is
\begin{equation}
p_s = \alpha^{-1} p_q\text{.}
\end{equation}

Let us first consider the case that the zero normal mode is \emph{inertial}.
The unperturbed Hamiltonian is then given, to second order, by (see \equaref{eq:Hquadratic})
\begin{equation}\label{eq:parthamdef}
\mathcal{H}^\text{iner} = \frac{1}{2} p_q^2 = \frac{1}{2} \alpha^{2} p_s^2 = p_s^2 / (2m_\text{eff}) \text{,}
\end{equation}
where we have
\begin{equation}\label{eq:effmassdef}
m_\text{eff} = \alpha^{-2}\text{,}
\end{equation}
the \emph{effective mass} of the degree of freedom.
Suppose that the Hamiltonian \eqref{eq:parthamdef} is perturbed by an external potential $V(s)$ which depends only on the position of the defect, so that we have $\mathcal{H}=\mathcal{H}^\text{iner} + V(s)$. We get
\begin{equation}\label{eq:eomNewton}
\ddot{s} = \frac{d}{dt} \frac{\partial \mathcal{H}}{\partial p_s} = \frac{1}{m_\text{eff}} \dot{p}_s = -\frac{1}{m_\text{eff}} \frac{\partial \mathcal{H}}{\partial s} = -\frac{1}{m_\text{eff}} \frac{dV}{ds}\text{,}
\end{equation}
which is Newton's equation of motion.

For a \emph{special} zero normal mode, the picture is different.
The unperturbed Hamiltonian is then given, to second order, by (see \equaref{eq:Hquadratic})
\begin{equation}\label{eq:parthamord}
\mathcal{H}^\text{spec} = 0 \text{,}
\end{equation}
which implies, in a sense, an infinite effective mass.
Since for the special zero mode no energy term is associated with $p_s$,
an effective force in the $s$-direction does not, by itself, cause an acceleration in the $s$-direction.
It does generate a motion in the canonically conjugate variable; however, here the first, not second, time derivative is proportional to the force.
Let us consider a case where $p_q$ and $q$ correspond to orthogonal displacements of a 2D magnetic defect, such as a Skyrmion (see \ssecref{sec:exskyrmion}). We have, for certain constants $\alpha$ and $\beta$,
\begin{equation}\label{eq:defsxsy}
\left\lbrace
\begin{array}{l l}
s_x = \alpha q\\
s_y = \beta p_q
\end{array}
\right.\text{,}
\end{equation}
where $s_x$ and $s_y$ respectively represent the $x$- and $y$-coordinate of the position of the defect. Again, we can straightforwardly obtain $\alpha$ and $\beta$ from the calculated normal mode using \equaref{eq:xinpq}.
If the Hamiltonian \eqref{eq:parthamord} is perturbed by an external potential $V(s_x,s_y)$, we get
\begin{equation}\label{eq:eomThiele}
\left\lbrace
\begin{array}{l}
\dot{s}_x = \phantom{-}\alpha\beta (\partial V / \partial s_y) \\
\dot{s}_y =          - \alpha\beta (\partial V / \partial s_x)
\end{array}
\right.\text{.}
\end{equation}
Notice that the velocity (not acceleration!) in the $s_y$-direction is proportional to the force in the positive $s_x$-direction, while the velocity in the $s_x$ direction is proportional to the force in the negative $s_y$-direction with the same constant of proportionality.
We see that we can interpret effective dynamical behavior described by Thiele's equation of motion \cite{Thiele1973A,Makhfudz2012} as a direct consequence of the existence of a special zero mode.

\subsection{\label{sec:1Ddw}1D domain wall}

\begin{figure}
  \includegraphics[width=\figwidthhalf]{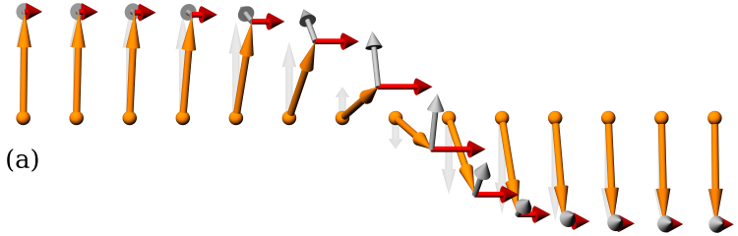}

  \includegraphics[width=\figwidthhalf]{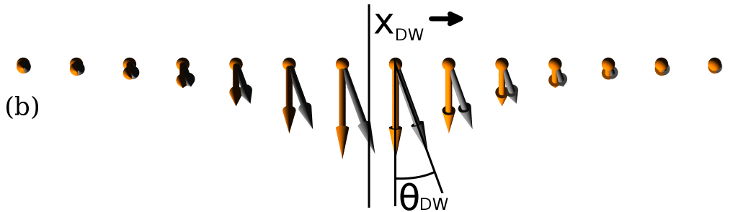}

  \includegraphics[width=\figwidthhalf]{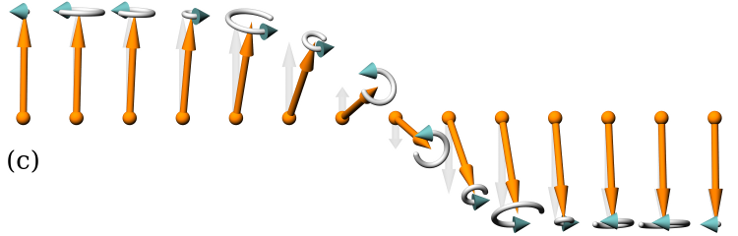}
    \caption{\label{fig:dw1Ddef}
    Color) Spin chain with domain wall. The domain wall separates two domains, magnetized in either the positive or the negative $\hat{\mathbf{z}}$-direction. Only one in every three spins is shown in the picture; the spin chain should be considered as effectively continuous. We set $K_1=0.04J$, giving the domain wall a characteristic width \cite{Skomski2008} of $\delta \propto \sqrt{J/K}a = 5.0a$.
    The big arrows show the equilibrium configuration $m_0$.
    The vectors $u_1$ and $u_2$ of the zero mode of the domain wall are indicated in~\figsub{a} with red ($u_1$) and gray ($u_2$) arrows. The actual magnitudes of $u_1$ and $u_2$ depend on the type of mode (special or inertial) and the effective mass.
    \figsub{b}~Top view of the domain wall. The position $x_\text{DW}$ of the domain wall and the angle $\theta_\text{DW}$ are indicated.
    \figsub{c}~Spin-wave mode with $k=0.37a^{-1}$ for $K_2=0.004J$. 
    }
\end{figure}
Even if the Hamiltonian as such is translationally invariant (the material properties are homogeneous), translational symmetry may be broken by the equilibrium configuration $m_0$, for instance if $m_0$ contains a domain wall. We consider a 1D spin chain with a domain wall like the one in \figref{fig:dw1Ddef}. We ensure that the equilibrium width of the domain wall is large enough to make the system effectively continuous (Peierls pinning \cite{Novoselov2003,Herz1977} is negligible).
The 1D domain wall is the simplest case where the two types of zero modes arise. As in the previous examples, the Hamiltonian consists of exchange and anisotropy terms, which are taken the same for all spins in the system. We will consider, however, two types of anisotropy that yield one or the other type of zero mode.
We shall see that the inertial dynamics of many domain walls \cite{Doring1948,Tatara2008} can be interpreted as a consequence of the existence of an inertial zero mode.

For a 1D domain wall, we find below the spin-wave continuum only a single zero mode (see \figrefsub{fig:dispersion}{c}).
If the Hamiltonian is the form considered up to now, with exchange and uniaxial anisotropy, this mode is a \emph{special} zero mode. In \figrefsub{fig:dw1Ddef}{a} we show the two components $u_1$ and $u_2$ of the zero mode. The component $u_2$ generates an infinitesimal increase  of the position $x_\text{DW}$ of the domain wall whereas $u_1$ generates an infinitesimal increase of the angle $\theta_\text{DW}$ (see \figrefsub{fig:dw1Ddef}{b}). An angle $\theta_\text{DW}=0$ or $\theta_\text{DW}=\pi$ corresponds to a Bloch domain wall, whereas $\theta_\text{DW}=\pm \pi/2$ corresponds to a N\'eel domain wall \cite{Tatara2008}. The coordinate $x_\text{DW}$ is canonically conjugate \cite{Tatara2008} to
\begin{equation}
p_\text{DW}=\frac{2 m_\text{S}}{a\tilde{\gamma} }\theta_\text{DW}\text{.}
\end{equation}
\begin{figure}
  \includegraphics[width=\figwidthhalf]{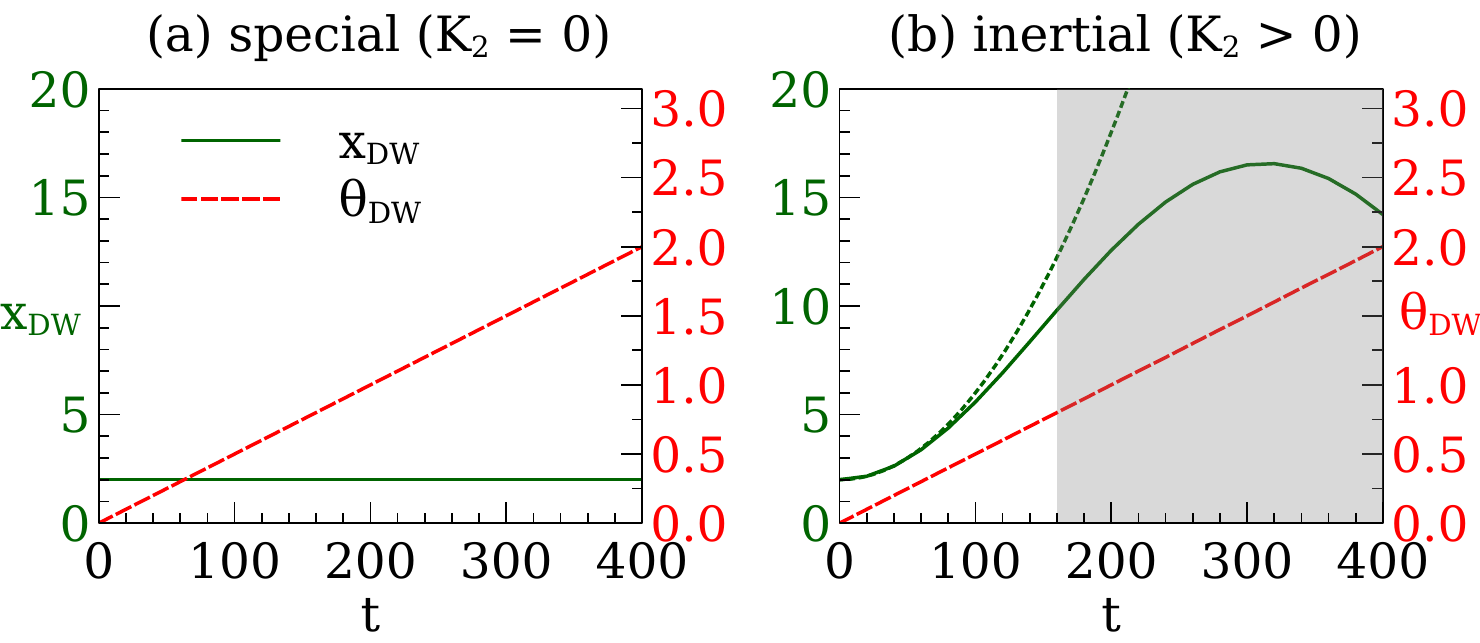}
  \caption{\label{fig:dwdynamics}
    Color) Time evolution of $x_\text{DW}$ and $\theta_\text{DW}$ in the presence of an external magnetic field $\mathbf{h}=0.005m_\text{S}J \hat{\mathbf{z}}$, for $K_1=0.04J$.
    Due to the external field, the domain wall experiences an effective external potential \eqref{eq:effDWpot}.
    The position $x_\text{DW}$ is given in units of $a$, angle $\theta_\text{DW}$ in radians and time $t$ in units of $\tau = {(\tilde\gamma m_\text{S} J)}^{-1}$.
    The plots have been obtained in spin-dynamics simulations.
    \figsub{a}~Behavior of a domain wall with a special zero mode for $K_2=0$.
    \figsub{b}~Behavior of a domain wall with an inertial zero mode for $K_2=0.016J$.
    The dotted line is a quadratic fit to the behavior of $x_\text{DW}$, which satisfies Newton's law \eqref{eq:eomNewton} in the limit of small $\theta_\text{DW}$. The shaded area indicates where deviations occur (see text).}
\end{figure}
If we apply an external magnetic field in the $\hat{\mathbf{z}}$-direction, which adds to the Hamiltonian a contribution (Zeeman energy) of the form
\begin{equation}
E_\text{Zeeman} = \sum_{i=1}^n -\mathbf{h} \cdot \mathbf{m}_i = \sum_{i=1}^n - h_z (\mathbf{m}_i \cdot \hat{\mathbf{z}}) \text{,}
\end{equation}
the domain wall experiences an effective force that acts on the $x_\text{DW}$ coordinate. In fact, a displacement of the domain wall by one site (distance $a$) leads to one more spin aligned along the field and one fewer spin antialigned. This results in an effective external potential
\begin{equation}\label{eq:effDWpot}
V(x_\text{DW})=-2 h_z m_\text{S} x_\text{DW} / a\text{.}
\end{equation}
Nevertheless, the domain wall position $x_\text{DW}$ remains constant, as shown in \figrefsub{fig:dwdynamics}{a}. The conjugated $\theta_\text{DW}$ increases linearly (the spins near the center of the domain wall rotate around the $\hat{\mathbf{z}}$ axis).
This is in line with the general dynamical behavior predicted for systems with a special zero mode (see \secref{sec:inertial}).

Motion of the domain wall in an external magnetic field along $\hat{\mathbf{z}}$ occurs if we add to the Hamiltonian a term that breaks the symmetry under rotation of the magnetic moments around $\hat{\mathbf{z}}$. In many magnetic systems, magnetostatic interactions favor Bloch domain walls, where the magnetization is in the plane of the domain wall. We model this effect by introducing a second term to the anisotropy energy \eqref{eq:Eaniuni}. We use \cite{Wieser2010,Thiele1973B}
\begin{equation}\label{eq:Eaniext}
E_\text{ani} = \sum_i \bigl[ -K_1 {(\mathbf{m}_i \cdot \hat{ \mathbf{z}})}^2 + K_2{(\mathbf{m}_i \cdot \hat{ \mathbf{x}})}^2 \bigr]
\end{equation}
with $K_1,K_2>0$.
In this case, we find an \emph{inertial} zero mode, with the components $u_1$ and $u_2$ again as in \figrefsub{fig:dw1Ddef}{a} but with a different dynamics. Even in the absence of an external field, a small deviation of $\theta_\text{DW}$ from its equilibrium value $\theta_\text{DW}=0$ now causes a linear motion of the domain wall, $\dot{x}_\text{DW}=\text{constant}$. In the presence of an external magnetic field in the $\hat{\mathbf{z}}$-direction, which creates a constant effective force $-\partial V / \partial x_\text{DW} = 2 h_z m_\text{S} / a$ \eqref{eq:effDWpot}, we find that $x_\text{DW}$ initially increases quadratically in time (see \figrefsub{fig:dwdynamics}{b}), in perfect agreement with the general dynamical behavior \eqref{eq:eomNewton} predicted for inertial zero modes.

\begin{figure}
  \includegraphics[width=\figwidthhalf]{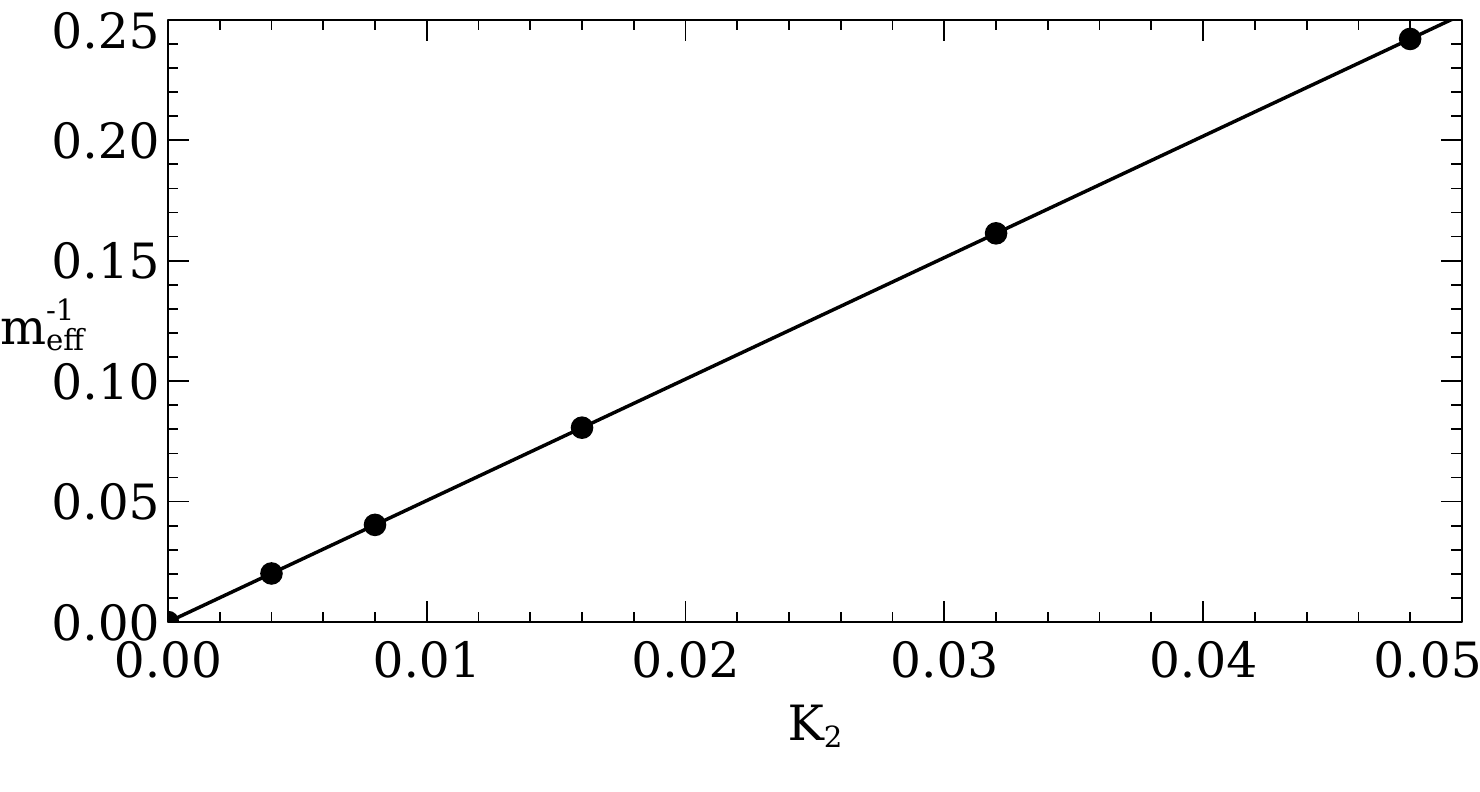}
  \caption{\label{fig:effmass}
    Inverse D\"{o}ring effective mass \cite{Doring1948} of a domain wall as a function of $K_2$, for $K_1=0.04 J$.
    We determine the effective masses from the calculated zero modes using \equaref{eq:effmassdef}.
    $K_2$ is given in units of $J$, $m^{-1}_\text{eff}$ in units of $a^2 \tilde{\gamma}^2 J$.}
\end{figure}
In \figref{fig:effmass} we show how the presence of nonuniaxial anisotropy leads to a finite effective mass, transforming a special zero mode ($K_2=0$) into an inertial zero mode ($K_2>0$). The notion of the effective mass of a domain wall was first introduced by D\"{o}ring \cite{Doring1948}.
The deviations from quadratic behavior calculated at large times (shaded area in \figrefsub{fig:dwdynamics}{b}) are beyond the linearized approach.
In principle, the effective mass, defined as the inverse of the second derivative of the Hamiltonian $\mathcal H$ with respect to the momentum $p_\text{DW}$ conjugate to $x_\text{DW}$, depends on $\theta_\text{DW}$.
Eventually, in a conservative system the domain wall starts reverting to its original position when $\theta_\text{DW}$ reaches $\pi/2$.
This type of motion of the domain wall, which occurs when damping is absent or small as compared to the effective force, is responsible for the phenomenon called Walker breakdown \cite{Schryer1974}.

\begin{figure}
  \includegraphics[width=\figwidthhalf]{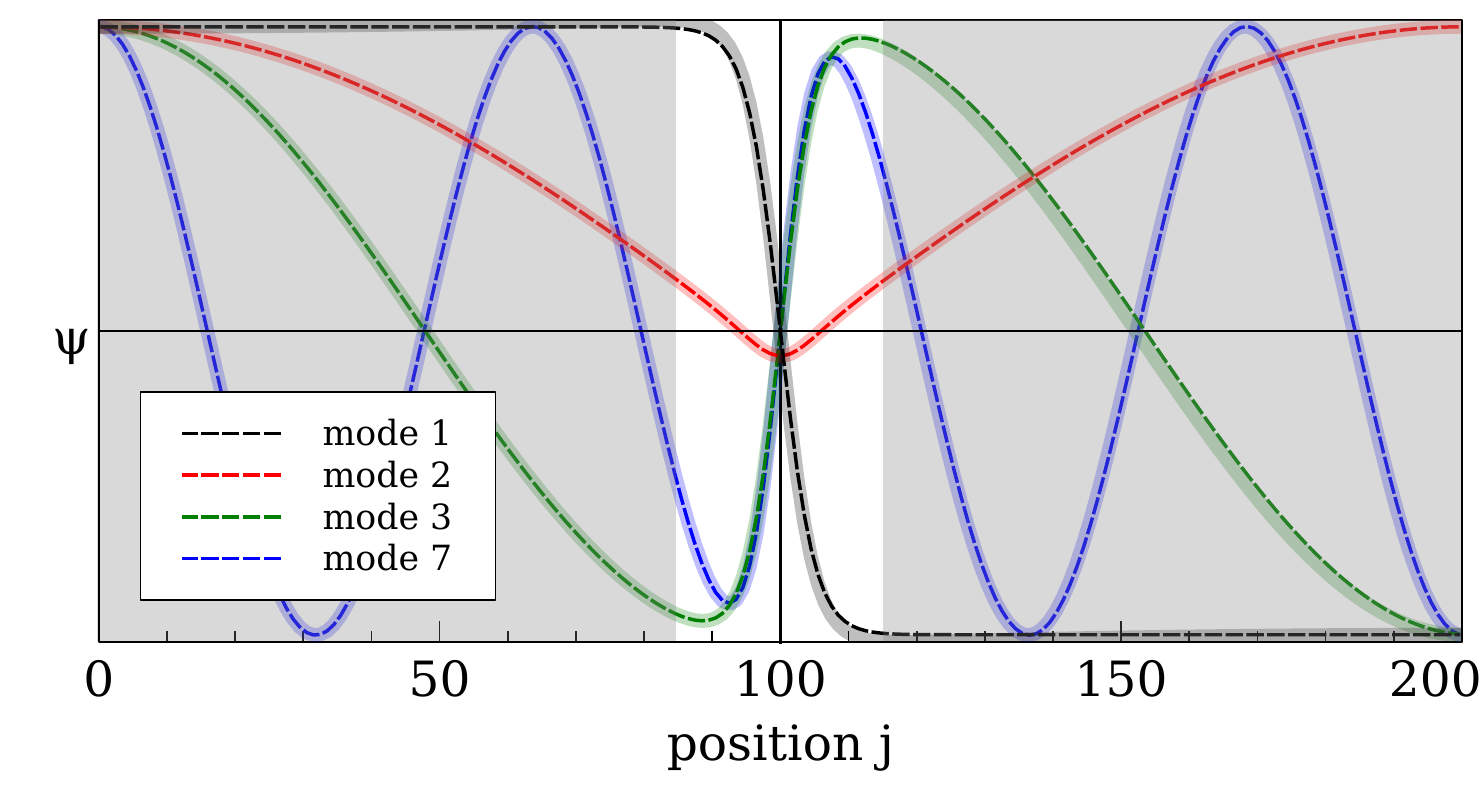}
  \caption{\label{fig:DW1Dwaveforms}
    Color) Spin-wave modes in a 1D uniaxial ($K_1=0.04 J$, $K_2=0$) 200-spin chain with a domain wall at the center.
    We compare the numerical results (thick solid lines) to the analytical form \cite{Helman1991} for the continuum model (thin dashed lines).
    Away from the domain wall, in the shaded area, the spin-wave modes resemble those of a perfect chain and can be characterized by a wavevector $k$.
    In addition to the spin-wave modes, the system has a localized special zero normal mode (not shown here; see \figrefsub{fig:dw1Ddef}{a}).}
\end{figure}
In addition to the zero mode, we have a spin-wave continuum (see \figrefsub{fig:dispersion}{c}).
In general, it is hard to find analytical solutions of the magnetic normal-mode problem for systems such as these, where the magnetic moments in the equilibrium configuration are not collinear. However, in this simple case, an analytical solution for the spin-wave modes has been found \cite{Helman1991}, which we can use to verify the numerical results.
In \figref{fig:DW1Dwaveforms}, we compare the calculated spin-wave modes successfully to this analytical solution.
It is convenient to express the analytical solution in the coordinate system \cite{Helman1991}
\begin{equation}\label{eq:dwcoordsys}
\left\lbrace \begin{array}{l}
  m_{jx} = {(\cosh \zeta_j)}^{-1} \cos \phi_j \\
  m_{jy} = {(\cosh \zeta_j)}^{-1} \sin \phi_j \\
  m_{jz} = -\tanh \zeta_j
\end{array}\right. \text{.}
\end{equation}
In this system the equilibrium configuration $m_0$ of the domain wall is given by the linear functions $\zeta_j = (aj-x_\text{DM})/\delta$ and $\phi_j = \text{constant}$, where $j$ is the index of the spin, $x_\text{DM}$ is the position of the center of the domain wall and $\delta = \sqrt{J/K}a$ is the characteristic domain-wall length.
We convert the Cartesian deviations from the equilibrium orientations, $\mathbf{x}_j = \mathbf{m}_j - \mathbf{m}_{0j}$, into values $d\zeta_j,d\phi_j$ in the coordinate system \eqref{eq:dwcoordsys}. For any given mode $l$, both functions $d\zeta_j,d\phi_j$ and both parts $u_1,u_2$ of the normal mode \eqref{eq:posnormmode} all have a common shape $f^{(l)}_j$, though the amplitudes may be different. We plot $\psi^{(l)}_j = f^{(l)}_j / \cosh(\zeta_j)$.
The fundamental solutions are given by $\psi_j = [ -ik + \tanh \zeta_j ]e^{ik\zeta_j}$ \cite{Helman1991}, where $k\in\mathbb{R}$ is the wavenumber of the spin wave away from the domain wall, in units of $\delta^{-1}$.
In our finite system, the spin-wave spectrum is discretized.
We calculate the right $k$-values for the analytical solutions from the numerically obtained values of $\omega$ via \equaref{eq:spindispersion}. A linear combination of the solutions for $k$ and $-k$ is taken in such a way that a real solution is obtained with a vanishing derivative at the boundaries of the chain.

\subsection{\label{sec:2Ddw}2D domain wall}

If a domain wall is extended to two dimensions, the zero mode of the 1D domain wall turns into a continuum of low-frequency modes \cite{Winter1961,Thiele1973B}.
\begin{figure}
  \includegraphics[width=\figwidthhalf]{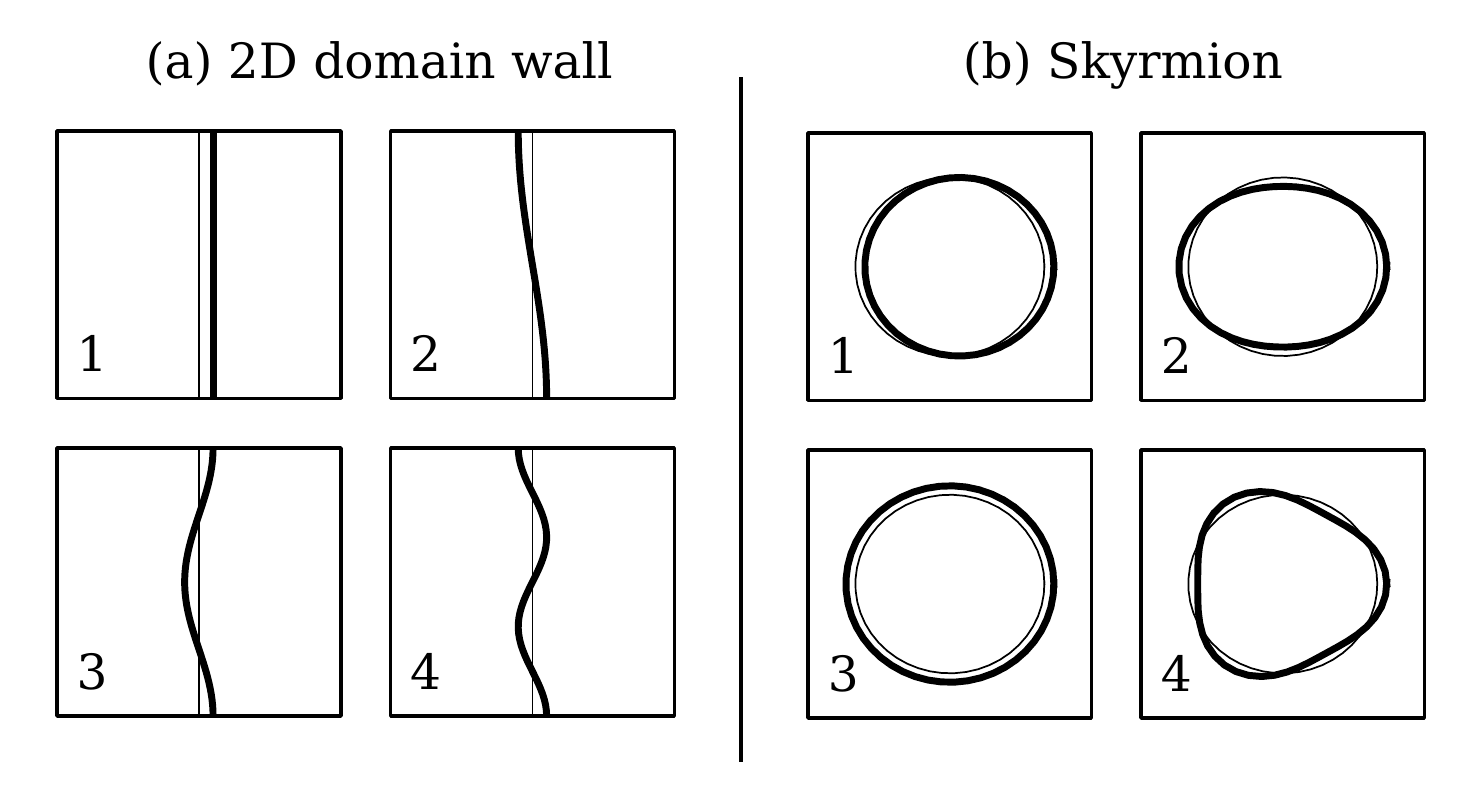}
  \caption{\label{fig:schematic}
    Sketches of the lowest modes of
    \figsub{a}~the 2D domain wall and
    \figsub{b}~the Skyrmion.
    The modes are shown in order of increasing frequency.
    In both cases, the lowest mode is a zero mode that corresponds to an infinitesimal translation of the defect.
    For our choice of parameters (see text), the four lowest modes of the Skyrmion are
    1)~$m=1$,
    2)~$m=2$,
    3)~$m=0$, and
    4)~$m=3$
    (see \figrefsub{fig:dispersion}{e}).}
\end{figure}
These modes correspond to bending of the domain wall, as sketched in \figrefsub{fig:schematic}{a}; in other words, they represent small spatial variations of the position $x_\text{DW}$ of the domain wall.
The domain-wall modes, which form a one-dimensional continuum with a vanishing frequency in the low-$k$ limit, exist alongside the two-dimensional continuum of spin-wave modes (see \figrefsub{fig:dispersion}{d}). A domain-wall mode can only exist if its frequency is below the bottom of the spin-wave continuum, which puts a maximum on its wavenumber.
The dispersion relation of the domain-wall modes in a system with arbitrary (possibly nonuniaxial) anisotropy was derived in \explcite{Thiele1973B}.
Here we show, using very general arguments, that the qualitative features of this dispersion relation follow immediately from the type of zero mode present in the system.

The domain-wall modes are a good example of physically interesting low-energy excitations of large systems, which can be found very efficiently using our method.
\begin{figure*}
  \includegraphics[width=\figwidthfull]{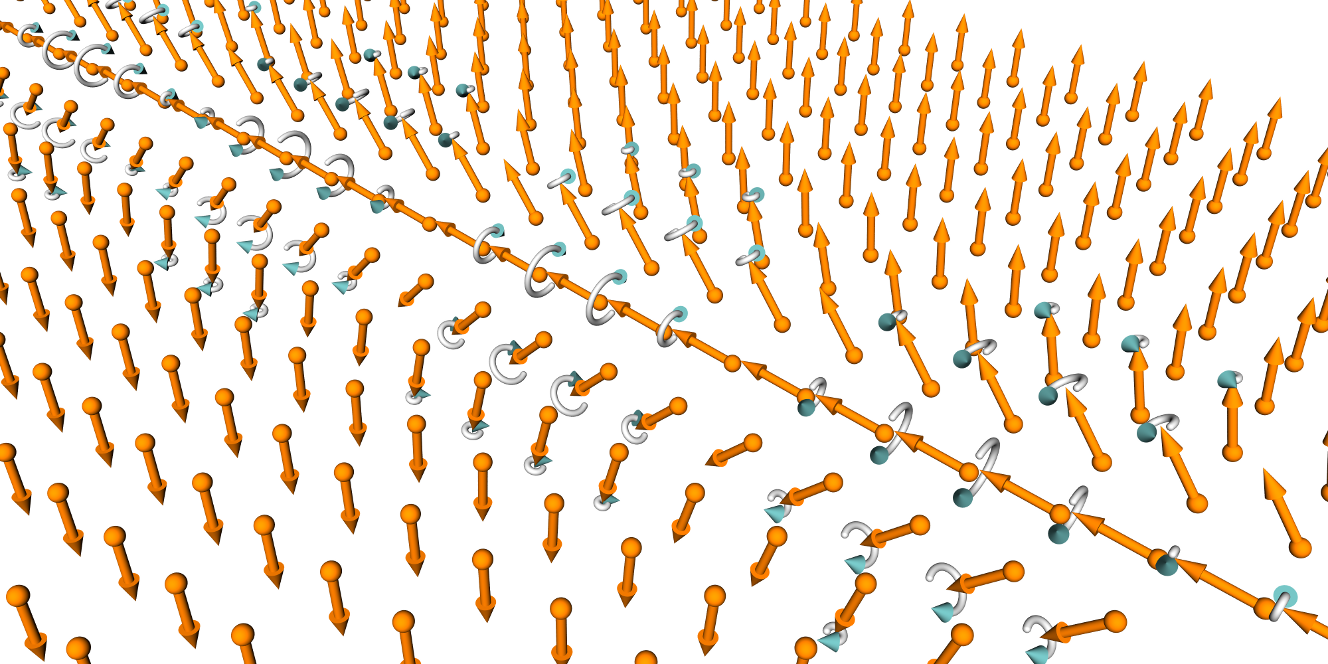}
    \caption{
    Color) A domain-wall mode. Only one spin is shown for each block of $5 \times 5$ spins; the system ($100\times400$ spins) may be considered as effectively continuous. Here we show the 16th domain-wall mode, with wavenumber $k=15\pi/(400a)$.
    Notice that the motion of the spins is in phase, since the boundary conditions used result in standing waves.
    When the deviation of the spins at the center of the domain wall from their equilibrium orientations is vertical, the domain wall is bent in a way similar to what is shown in \figrefsub{fig:schematic}{a}. When it is horizontal, the domain wall is not bent; at this point, the energy of the mode is stored as a spatial variation of $\theta_\text{DW}$ rather than of $x_\text{DW}$.
    }\label{fig:locdwmode}
\end{figure*}
The domain-wall mode in \figref{fig:locdwmode} was calculated in a system of $100 \times 400$ spins (square lattice) with exchange and uniaxial anisotropy ($K = 0.04J$).
As in \ssecref{sec:experfectchain}, our truncation of the expression for the exchange energy results in Neumann boundary conditions.
We find that the lowest 26 modes (including the zero mode) of this system are domain-wall modes (see \figrefsub{fig:dispersion}{d}).

The distinction between special and inertial zero modes has important consequences for the dispersion relations that correspond to them, as we show in the following.
For the case with uniaxial anisotropy, shown in \figrefsub{fig:dispersion}{d}, we see that the zero mode of the domain wall, which is a \emph{special} zero mode, turns into a continuum with \emph{quadratic} dispersion.
\begin{figure}
  \includegraphics[width=\figwidthhalf]{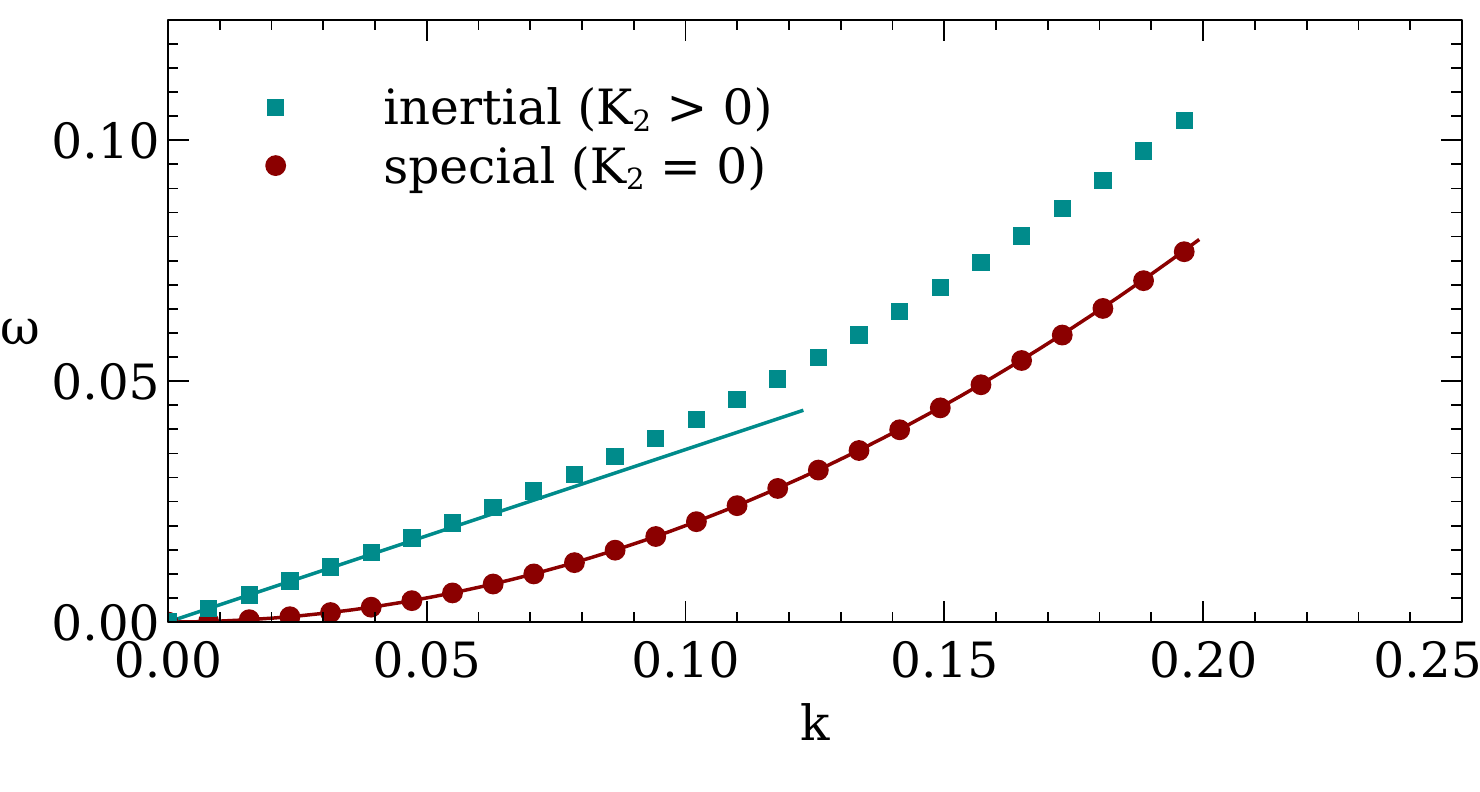}
  \caption{\label{fig:dwdispersions}
    Color) Dispersion curves of the domain-wall modes \cite{Thiele1973B} of a 2D system with uniaxial anisotropy ($K_1=0.04J$, $K_2=0$; see \figrefsub{fig:dispersion}{d}) or nonuniaxial anisotropy ($K_1=0.04J$, $K_2=0.032J$). 
    The wavenumber $k$ is in units of $a^{-1}$ and the angular frequency $\omega$ in units of $ \tilde\gamma m_\text{S} J$.
    In the uniaxial case, where the domain wall has a special zero mode, the dispersion is quadratic.
    In the nonuniaxial case, where the domain wall has an inertial zero mode, the dispersion is linear in the limit of small $k$.
    The modes were calculated in a system of $100\times400$ spins.}
\end{figure}
In \figref{fig:dwdispersions}, we compare this case to a similar system with nonuniaxial anisotropy.
For $K_2>0$, where the domain wall has an \emph{inertial} zero mode, the dispersion relation $\omega(k)$ is \emph{linear} in $k$ in the limit of low wavenumber $k$.
This suggests that long-wavelength waves in a system with an inertial zero mode propagate with a finite group velocity. Indeed, a finite group velocity is also observed for acoustic waves in crystals, which agrees with the fact that zero modes of systems of coupled point masses are always inertial (see \refappendix{}).

It is easy to understand the link between the type of zero mode and the low-$k$ behavior of the dispersion relation.
Suppose we have a system with a zero mode, such as the 1D spin chain with a domain wall.
We describe the relevant dynamics of this system with just two variables, the canonical momentum $p$ and coordinate $q$ \eqref{eq:xinpq} corresponding to the zero mode. In the case of the domain wall, $p$ and $q$ are proportional to $x_\text{DW}$ and $\theta_\text{DW}$ respectively (see \ssecref{sec:1Ddw}).
We now extend the system to a higher dimension.
The variables $p$ and $q$ become functions of position: we have $p(\mathbf{r})$ and $q(\mathbf{r})$.
(In the case of the 2D domain wall, $\mathbf{r} \in \mathbb{R}^1$ represents a position along the length of the domain wall.)
It is reasonable to assume that for functions $p(\mathbf{r})$ and $q(\mathbf{r})$ that vary very smoothly in $\mathbf{r}$ and for short-range interactions, the Hamiltonian of the system can be approximated by the functional
\begin{equation}
\mathcal H = \int \left[ f(p,q) + \frac{1}{2}\rho{\| \nabla q \|}^2 + \frac{1}{2}\sigma{\| \nabla p \|}^2 \right] \,d\mathbf{r}
\end{equation}
for certain constants $\rho,\sigma>0$.
In the limit of small $p$ and $q$ we have that $f(p,q) = 0$ for a special zero mode, $f(p,q) = \frac{1}{2}p^2$ for an inertial zero mode and $f(p,q) = \frac{1}{2}\omega(p^2 + q^2)$ for a positive mode (see \equaref{eq:Hquadratic}).
After Fourier transformation in $\mathbf r$, spatial variations with different wavevectors $\mathbf{k}$ decouple and we get
\begin{equation}
\mathcal{H}_\mathbf{k}^\text{spec} = \phantom{\frac{1}{2} p_\mathbf{k}^2 + {}}\frac{1}{2}\rho k^2 p_\mathbf{k}^2 + \frac{1}{2}\sigma k^2 q_\mathbf{k}^2
\end{equation}
for the system with a special zero mode and
\begin{equation}
\mathcal{H}_\mathbf{k}^\text{iner} =          \frac{1}{2} p_\mathbf{k}^2 + {} \frac{1}{2}\rho k^2 p_\mathbf{k}^2 + \frac{1}{2}\sigma k^2 q_\mathbf{k}^2
\end{equation}
for the system with an inertial zero mode, where we define $k = \|\mathbf k\|$.
It follows immediately from Hamilton's equations that for $\mathcal{H}_\mathbf{k}^\text{spec}$, we get
\begin{equation}
\left\lbrace
\begin{array}{l}
\dot{p}_{\mathbf k} =          - \partial \mathcal H / \partial q_{\mathbf k} =          - \sigma  k^2 q_{\mathbf k} \\
\dot{q}_{\mathbf k} = \phantom{-}\partial \mathcal H / \partial p_{\mathbf k} = \phantom{-}\rho k^2 p_{\mathbf k}
\end{array}
\right.
\text{,}
\end{equation}
while for $\mathcal{H}_\mathbf{k}^\text{iner}$, we get
\begin{equation}
\left\lbrace
\begin{array}{l}
\dot{p}_{\mathbf k} = \phantom{p_{\mathbf k}} - \sigma  k^2 q_{\mathbf k} \\
\dot{q}_{\mathbf k} =          p_{\mathbf k}  + \rho k^2 p_{\mathbf k}
\end{array}
\right.
\text{.}
\end{equation}
The momenta $p_{\mathbf k}$ can be eliminated from both systems of equations, yielding equations of motion of the form
\begin{equation}
\ddot{q}_{\mathbf k} = - {\omega(\mathbf k)}^2 q_{\mathbf k}\text{.}
\end{equation}
For the special zero mode we get a quadratic dispersion relation
\begin{equation}
\omega(\mathbf k) = \sqrt{\rho\sigma} {\| \mathbf k \|}^2 \text{,}
\end{equation}
whereas for  the inertial zero mode we get a linear dispersion relation
\begin{equation}
\omega(\mathbf k) = {\left[(1+\rho{\| \mathbf k \|}^2)\sigma\right]}^{1/2} \| \mathbf k \| = \sqrt{\sigma} \| \mathbf k \| + \Or({\| \mathbf k \|}^2)\text{.}
\end{equation}

\subsection{\label{sec:exskyrmion}Skyrmion}

\begin{figure*}
  \includegraphics[width=0.8\figwidthfull]{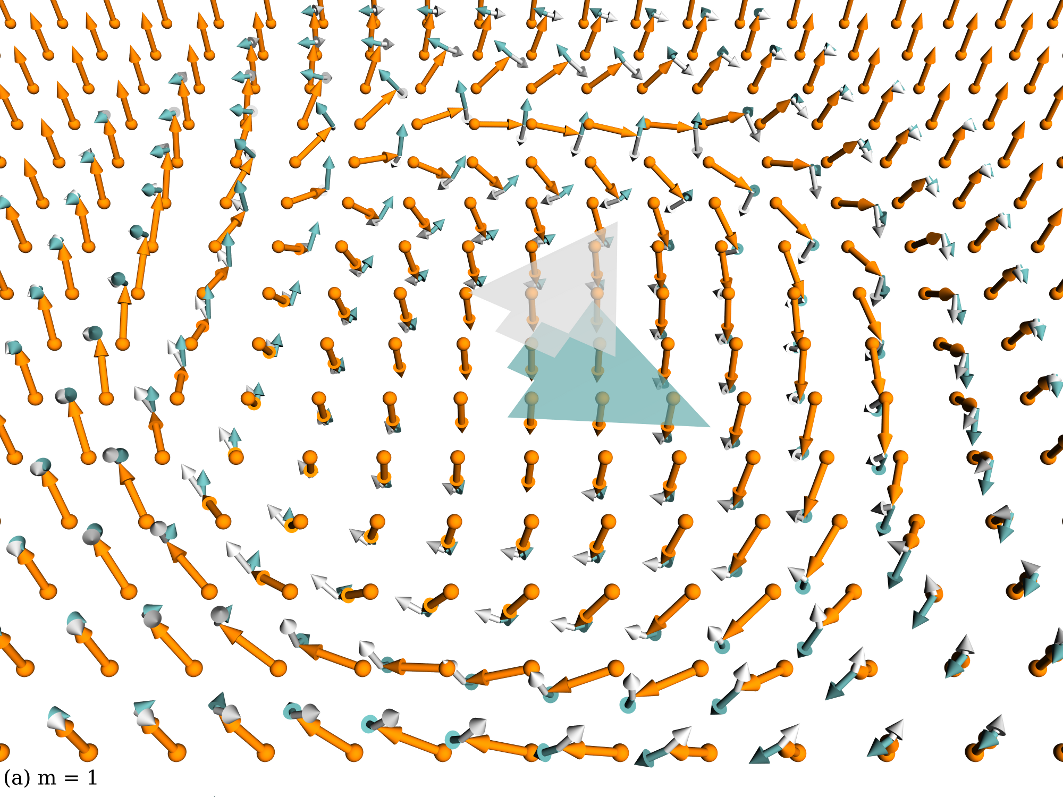}

  \vspace{1em}

  \includegraphics[width=0.8\figwidthfull]{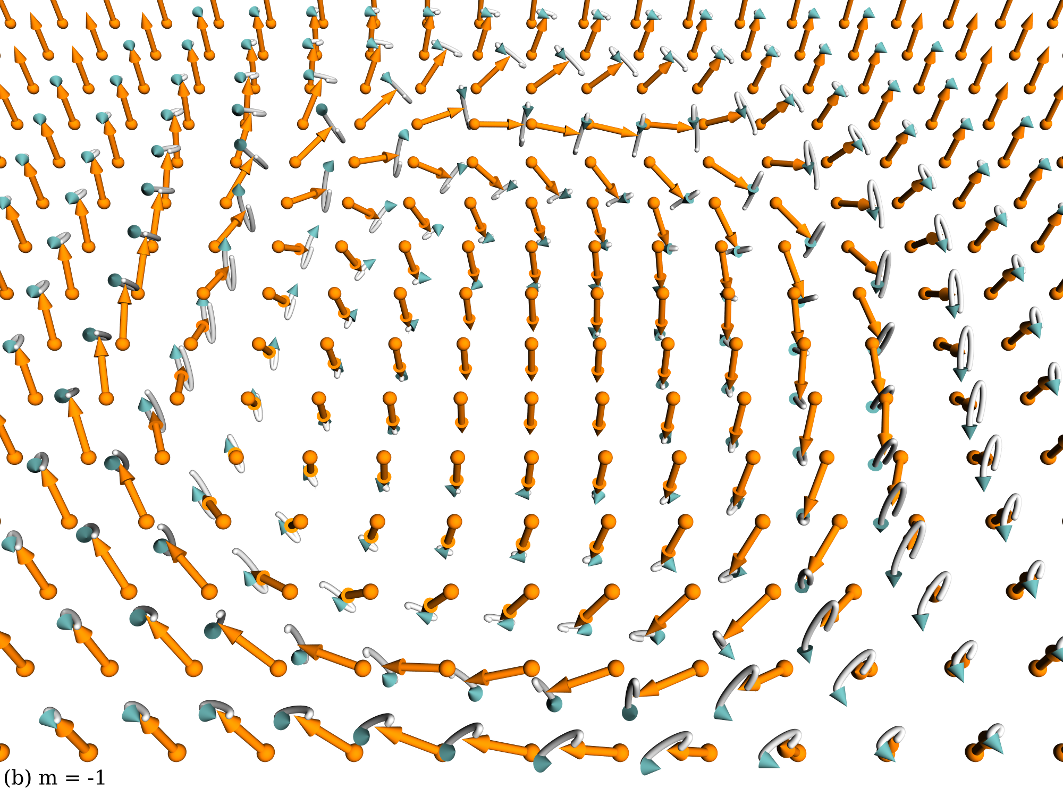}

  \caption{
    Color) A Skyrmion shown with \figsub{a}~the $m=1$ special zero mode and \figsub{b}~the $m=-1$ positive mode.
    Only one spin is shown for each block of $5 \times 5$ spins; the system may be considered as effectively continuous.
    For our choice of parameters (see text), the $m=-1$ mode is the 7th-lowest mode of the spectrum ($\omega = 0.0293\tilde\gamma m_\text{S} J$).
  }
  \label{fig:skyrmionXY}
\end{figure*}

Magnetic bubbles or Skyrmions can be seen as circular domain walls (see \figref{fig:skyrmionXY}).
The dynamics of a Skyrmion in an effective potential can be understood very well in terms of its normal modes \cite{Makhfudz2012}. We shall see that our algorithm for normal-mode analysis provides a direct way to calculate the effective mass $\mathcal M$ and gyrocoupling constant $\mathcal G$ of any Skyrmion, regardless of the details of the interactions present in the system.
This is another example of how a normal-mode analysis that includes the zero modes gives the effective equation of motion of some magnetic structure almost immediately.

Skyrmion structures can be stabilized by magnetostatic interactions \cite{Dell1986} or by the \DM{} (DM) interaction \cite{Muhlbauer2009}.
In the latter case, which we shall consider here, the equilibrium radius is fixed by the material parameters.
In the example shown in \figref{fig:skyrmionXY}, we consider a system of $200 \times 200$ spins (square lattice) with only exchange, uniaxial anisotropy ($K=0.04J$) and DM interactions (no external field).
We write the DM interaction as \cite{Mochizuki2012}
\begin{equation}\label{eq:EDM}
E_\text{DM} = \sum_{\mathbf{x}} -D \mathbf{m}_\mathbf{x} \cdot \left( \sum_{\hat{\mathbf{r}}} \hat{\mathbf{r}} \times \mathbf{m}_{\mathbf{x} + a\hat{\mathbf{r}}} \right)\text{,}
\end{equation}
where $\mathbf{m}_\mathbf{x}$ is the magnetic moment at the lattice site $\mathbf{x}$,
$\hat{\mathbf{r}}$ sums over the directions of the nearest neighbors in the square lattice ($\hat{\mathbf{r}}=\hat{\mathbf{x}},\hat{\mathbf{y}},-\hat{\mathbf{x}},-\hat{\mathbf{y}}$), $a$ is the lattice constant and $D$ is the interaction strength. Here, we set $D = 0.125J$. We use periodic boundary conditions.
We construct a Skyrmion configuration and relax it.
For the given parameters, we get an equilibrium radius of $\sim 27a$.
We find 10 normal modes below the spin-wave continuum (edge modes \cite{Makhfudz2012}), as shown in \figrefsub{fig:dispersion}{e}.
We may interpret these modes as domain-wall modes traveling around the Skyrmion (see \figrefsub{fig:schematic}{b}).
The edge modes can be identified by the number of periods $m$ seen when going around the Skyrmion once.
We use a positive or negative value of $m$ to indicate the direction in which the wave travels around the Skyrmion \cite{Makhfudz2012}.
As pointed out in \explcite{Makhfudz2012}, the edge-mode spectrum is chiral: it is not symmetric for positive and negative $m$ (see \figrefsub{fig:dispersion}{e}).

The system has translational symmetry in two spatial directions. This symmetry gives rise to a special zero mode ($m=1$). The vector $u_1$ of this mode generates an infinitesimal translation in the $\hat{\mathbf{y}}$-direction and the vector $u_2$ generates an infinitesimal translation in the $\hat{\mathbf{x}}$-direction (see \figrefsub{fig:skyrmionXY}{a}).
The existence of this special zero mode suggests that the dynamical behavior of the position of the Skyrmion in an external potential is described by the noninertial equation of motion \eqref{eq:eomThiele}, which is equivalent to Thiele's equation \cite{Thiele1973A} without damping.
It has recently been observed that the actual behavior of the Skyrmion position is more accurately described by an equation which has an additional inertial term \cite{Moutafis2009,Makhfudz2012}, a result which seems to contradict our statement that the Skyrmion possesses a special zero mode and has noninertial character.
It is therefore important to make a detailed comparison with the interesting findings of \explcite{Makhfudz2012}, as we do in the following.
An analysis of the normal modes indicates that the inertial term results in this case from the positive mode $m=-1$ \cite{Makhfudz2012}. Despite its finite frequency, this mode gives rise to a displacement of the Skyrmion similar to that of the $m=1$ zero mode, albeit with a concomitant change of the Skyrmion configuration (see \figrefsub{fig:skyrmionXY}{b}).
A derivation of the equation of motion of the Skyrmion was given in the Lagrangian formalism in \explcite{Makhfudz2012}.
Here we reproduce this result in the Hamiltonian formalism and show how the parameters $\mathcal G$ and $\mathcal M$ of the equation of motion can be obtained immediately from the normal-mode calculation.

The crucial observation, which also underlies the derivation in \explcite{Makhfudz2012}, is that if we define the position of the Skyrmion as
\begin{equation}\label{eq:posXY}
X = \frac{\int \left( m_z(\mathbf{r}) - m_S \right)x\,d\mathbf{r}}{\int \left( m_z(\mathbf{r}) - m_S \right)\,d\mathbf{r}} \hspace{2em}
Y = \frac{\int \left( m_z(\mathbf{r}) - m_S \right)y\,d\mathbf{r}}{\int \left( m_z(\mathbf{r}) - m_S \right)\,d\mathbf{r}}\text{,}
\end{equation}
as was done in \explcite{Moutafis2009}, then not only the $m=1$ zero mode (see \figrefsub{fig:skyrmionXY}{a}) but also the $m=-1$ mode (see \figrefsub{fig:skyrmionXY}{b}), which is a positive mode, generates a change in position.
In fact, we find from our normal-mode calculation that
\begin{equation}\label{eq:skyrmionXYpq}
\left\lbrace
\arraycolsep=1.4pt
\begin{array}{l l}
\hspace{3.6pt}X - X_0& = \alpha p_+ + \alpha q_- \\
\hspace{3.6pt}Y - Y_0& = \alpha p_- + \alpha q_+
\end{array}\right.\text{,}
\end{equation}
where $p_+,q_+$ are the canonical momentum and coordinate \eqref{eq:decomppq} corresponding to the special zero mode $m=1$, $p_-,q_-$ correspond to the positive mode $m=-1$, $(X_0,Y_0)$ is the position of the Skyrmion in the equilibrium configuration and $\alpha$ is a constant that can be obtained easily from the calculated normal modes. In our calculation, we find $\alpha=0.282\,a \tilde{\gamma}^{1/2} m_\text{S}^{-1/2}$. Since the system is rotationally invariant, the normal modes output by the computer code may be oriented in any direction but we can always rotate them to satisfy \equaref{eq:skyrmionXYpq}.
Since we consider only the modes $m=\pm 1$ that couple to position, the unperturbed Hamiltonian is given, to second order, by (see \equaref{eq:Hquadratic})
\begin{equation}\label{eq:hamskyrunp}
\mathcal H = \frac{1}{2} \omega ( p_-^2 + q_-^2 ) \text{,}
\end{equation}
where $\omega$ is the angular frequency of the positive mode $m=-1$ (in our example, $\omega = 0.0293\tilde\gamma m_\text{S} J$). There is no energy term associated with the special zero mode $m=1$.
If we introduce an external potential that depends only on the position $(X,Y)$, \equaref{eq:hamskyrunp} becomes
\begin{equation}\label{eq:hamskyr}
\mathcal H = \frac{1}{2} \omega ( p_-^2 + q_-^2 ) + V(X,Y) \text{.}
\end{equation}
Using Hamilton's equations, \equaref{eq:hamskyr} gives
\begin{equation}
\begin{array}{l}
\dot{p}_+ =          - \partial \mathcal H / \partial q_+ = -\alpha \partial V / \partial Y \\
\dot{q}_+ = \phantom{-}\partial \mathcal H / \partial p_+ = \phantom{-}\alpha \partial V / \partial X \\
\dot{p}_- =          - \partial \mathcal H / \partial q_- =          - \omega q_- - \alpha \partial V / \partial X \\
\dot{q}_- = \phantom{-}\partial \mathcal H / \partial p_- = \phantom{-}\omega p_- + \alpha \partial V / \partial Y
\end{array}\text{,}
\end{equation}
from which it follows that
\begin{equation}
\begin{array}{l}
\dot{X} = \alpha \dot{p}_+ + \alpha \dot{q}_- = \phantom{-}\alpha\omega p_-\\
\dot{Y} = \alpha \dot{p}_- + \alpha \dot{q}_+ =          - \alpha\omega q_-
\end{array}\text{.}
\end{equation}
Again taking the time derivative and applying Hamilton's equations, this becomes
\begin{equation}\label{eq:skyrmioneom}
\left\lbrace
\begin{array}{l}
\ddot{X} = \phantom{-}\omega \dot{Y} - \alpha^2\omega \partial V / \partial X\\
\ddot{Y} =          - \omega \dot{X} - \alpha^2\omega \partial V / \partial Y
\end{array}\right.\text{.}
\end{equation}
These equations of motion are equivalent to \extequaref{3} in \explcite{Makhfudz2012} if we set
\begin{align}
\label{eq:skyrM} \mathcal M&= \phantom{-}1/(\alpha^2\omega)\text{,} \\
\label{eq:skyrG} \mathcal G&=          - 1/ \alpha^2\text{.}
\end{align}
The equations of motion \eqref{eq:skyrmioneom} consist of a `gyrocoupling' term, which is also present in Thiele's equation, and an additional inertial term, which gives a contribution to the acceleration proportional to the force.
For the parameters used in our example, we find $\mathcal M=4.29\times10^2\,a^{-2} \tilde{\gamma}^{-2} J^{-1}$ and $\mathcal G = -12.6\,a^{-2}\tilde{\gamma}^{-1}m_\text{S} \approx -4\pi a^{-2}\tilde{\gamma}^{-1}m_\text{S}$. For $\mathcal G$, an analytical expression was given in \explcites{Makhfudz2012} and~\explcitepart{Thiele1973A}, with which our calculated value is in excellent agreement.
From \equarefs{eq:skyrM} and~\eqref{eq:skyrG} we also recover $\omega = -\mathcal G / \mathcal M$, which is indeed the frequency of the $m=-1$ mode found in \extequaref{4} in \explcite{Makhfudz2012} in the absence of an external potential (set $\mathcal K=0$ in that equation).

Notice that the above derivation does not contradict the general statement made in \ssecref{sec:inertial} that a system with a special zero mode should have noninertial dynamics \eqref{eq:eomThiele}. In \equaref{eq:defsxsy}, we defined the position $(s_x,s_y)$ in terms of a perfect translation of the magnetic structure. The positive mode $m=-1$, however, simultaneously induces a change in the configuration of the Skyrmion and is not a perfect translation. In fact, the $m=-1$ mode causes the spins in the center of the circular domain wall to deviate from their Bloch-type equilibrium orientation, which is tangential to the domain wall. This mode therefore does not represent a change in $(s_x,s_y)$, while it does represent a change in the Skyrmion position $(X,Y)$ in the sense of \equaref{eq:posXY}. If we define the position according to \equaref{eq:posXY}, one obtains the partially inertial behavior derived above.
In many practical situations $(X,Y)$ is the right definition of position, since the effective potential couples to the location of the bubble domain and is mostly insensitive to the domain wall.
However, on timescales much longer than $\omega$ the cyclic effect of the positive mode on the position averages out, and $(s_x,s_y)$ is again the best representation of the position of the Skyrmion.

\subsection{\label{sec:accuracy}Accuracy of the corrections to the modes due to damping}

If we introduce damping ($\eta>0$), this has an effect not only on the amplitudes of the modes, which now decay in time, but also on the mode vectors $u_1,u_2$ (see \secref{sec:spindamped}).
Since for large systems we can usually only calculate a number of the lowest modes of the system, which are of the greatest interest, we are forced to truncate the perturbative expressions~\eqref{eq:poscorru}, \eqref{eq:poscorrv} and~\eqref{eq:defcorru} for these corrections to those modes that are available.
In principle, this approximation is uncontrolled.
However, we may argue that modes with very different frequencies also have very different characteristic wavelengths and hence have a very small overlap, so that the contribution of high-frequency modes to the damping correction of the low-frequency modes that we are interested in is likely to be negligible.
Here, we test the accuracy of the damping correction by comparing the actual time evolution of a Skyrmion system to the linearized solutions \eqref{eq:fundsoldampedpos} obtained from normal-mode analysis.
This also serves as a test of the expressions~\eqref{eq:poscorrxi}, \eqref{eq:poscorru} and~\eqref{eq:poscorrv}.

\begin{figure}
  \includegraphics[width=\figwidthhalf]{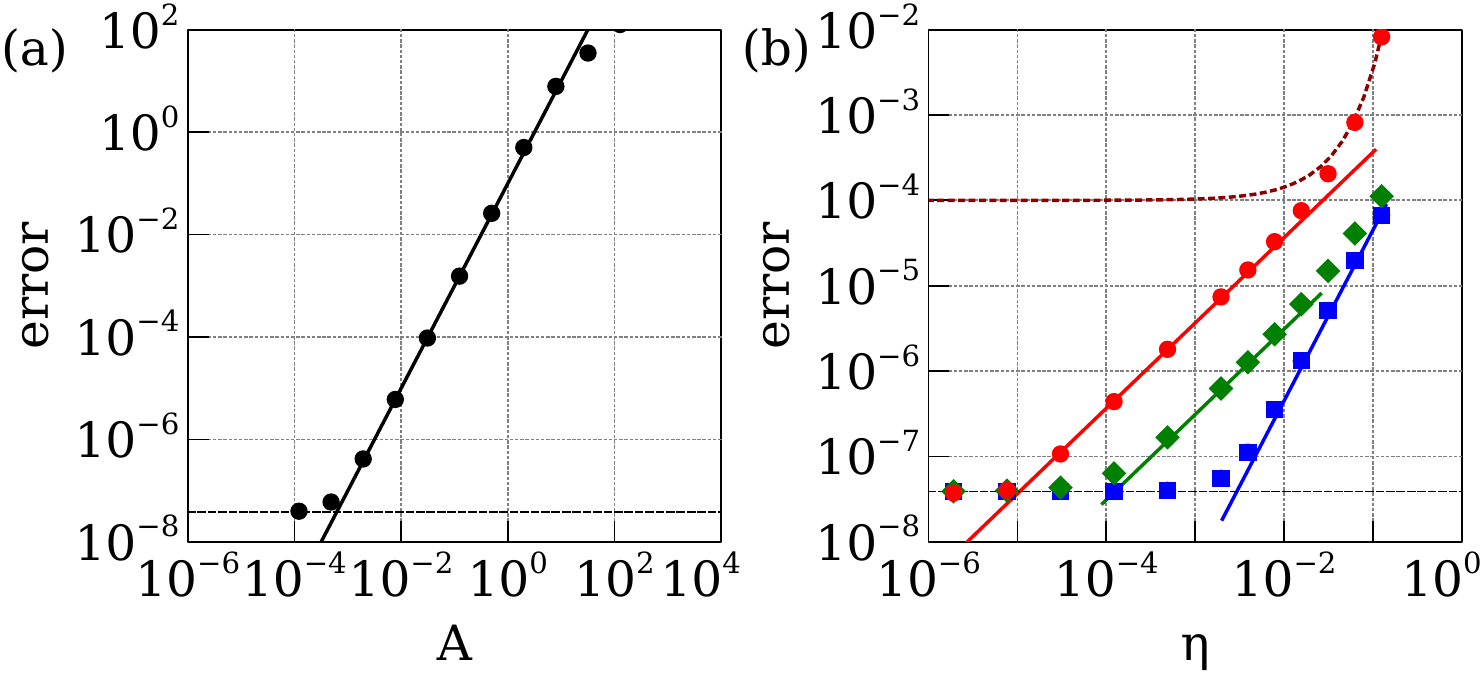}
  \caption{\label{fig:accuracy}
    Color) Accuracy of the linearized solution of an initial-value problem for the LLG equation (see text).
    \figsub{a}~Error as a function of the amplitude $A$ for zero damping.
    \figsub{b}~Error as a function of the damping parameter $\eta$ for a very small, fixed initial amplitude $A=1\times10^{-4} \, m_\text{S}^{1/2} \tilde\gamma^{-1/2}$. We consider three different levels of correction for damping in the linearized solution.
    Red:~damping is not taken into account at all.
    Green:~the decay rate $\xi'$ from first-order perturbation theory is taken into account, but the zero-damping modes $(u_1,u_2)$ are used.
    Blue:~both the modes $(u'_1,u'_2)$ and the decay rate $\xi'$ are corrected to first order of perturbation theory.
    For a fair comparison between different values of $\eta$, we have scaled the errors by the decay factor between the initial and final amplitudes, as indicated by the upper horizontal curve.
    The lower horizontal line indicates the accuracy of the numerical timestepping solution.}
\end{figure}
We consider the time evolution of an initial configuration $m = m_0 + A u_{1k}$, where mode $k$ is given an initial amplitude $A$. The details of the simulated system are specified below.
In \figref{fig:accuracy}, we plot the difference between the results of a numerical time integration of the LLG equation \eqref{eq:LLGfull} and the linearized solution \eqref{eq:fundsolconspos} or \eqref{eq:fundsoldampedpos}.
Since the error in the numerical solution can be made very small, we may use this difference to evaluate the accuracy of the normal modes.
The error stems from two sources. First, the linearization of the LLG equation necessary for normal-mode analysis results in an error of second order in the amplitude $A$. Second, the fact that the modes of a damped system are calculated in perturbation theory gives an additional error that depends on the damping parameter $\eta$.

\Figrefsub{fig:accuracy}{a} shows the error as a function of the amplitude $A$ for a conservative spin system ($\eta = 0$). We find a quadratic dependence, as expected \eqref{eq:geneom}.
\Figrefsub{fig:accuracy}{b} shows the error as a function of $\eta$, for an amplitude $A$ that is chosen so small that the error from nonlinearity is smaller than the error in the numerical solution. The error that we see in the plot is thus due to the error in the perturbative solution of the damped modes.
We see that if we do not take damping into account at all \eqref{eq:fundsolconspos}, the error in the solution is of first order in $\eta$, as expected.
If we take damping into account by using the calculated decay rate $\xi'$ \eqref{eq:fundsoldampedpos}, but without correcting the modes $(u_1,u_2)$, the error is much smaller, but it is still of first order in $\eta$.
If we also correct the modes $(u'_1,u'_2)$, so that we use the full first-order perturbation theory, we get an error of second order in $\eta$.
Notice that we get this order of accuracy even though we use only the contributions of the lowest modes to the perturbative correction.
We conclude that, at least in this case, any first-order error due to this truncation is so small as to be insignificant.

The results in \figref{fig:accuracy} are obtained in a system of $86 \times 86$ spins (square lattice) with uniaxial anisotropy ($K=0.12J$) and the DM interaction \eqref{eq:EDM} ($D=0.20J$). We use an equilibrium configuration $m_0$ containing a single Skyrmion. After relaxation, the Skyrmion is similar to the one in \figref{fig:skyrmionXY} but has a different radius (owing to the different interaction parameters used).
We construct a configuration $m = m_0 + A u_{1k}$ in which the second mode (
$\omega = 3.56\times10^{-2}\,\tilde\gamma m_\text{S} J$,
$\xi^{(1)}=1.18\times10^{-1}\,\tilde\gamma m_\text{S} J$) is given a finite initial amplitude $A$. The magnitudes of the magnetic moments are normalized to $m_\text{S}$.
We then numerically integrate the LLG equation \eqref{eq:LLGfull} starting from this initial configuration, using the implicit-midpoint timestepping scheme \cite{dAquino2005,Mentink2010} with $\Delta t = 0.1\tau$ where $\tau = {(\tilde\gamma m_\text{S} J)}^{-1}$.
The simulated time is $300\tau$.

The above results suggest that it is useful to take into account damping in a calculation of the normal modes. Using the expressions in \secref{sec:spindamped}, this can be done relatively easily and at a low computational cost. It is already very useful to take into account the decay rate $\xi'$ calculated to first order. An even better accuracy can be achieved by also using the first-order corrections to the modes $(u'_1,u'_2)$.
We find that in practice, we get an error of second order in $\eta$ in the time evolution of a low-frequency mode even when only a relatively small set of other low-frequency modes were used to calculate the correction.
The results also suggest that it is unnecessary to go beyond first-order perturbation theory for damping unless $\eta$ is unusually large.

\section{\label{sec:summary}Concluding remarks}

Using explicitly the symplectic structure of the dynamical spin system, we have developed a method  that allows us to solve the magnetic normal-mode problem in a very general situation, with the only assumption that the equilibrium magnetic structure corresponds to a local minimum of the Hamiltonian.
The examples we have considered (1D and 2D domain walls, Skyrmions) clearly demonstrate that zero modes are an essential part of this normal-mode analysis and can be very useful for understanding dynamics.

Systems with zero modes were difficult to treat within the framework of previous approaches for the magnetic normal-mode problem.
To our knowledge, all previous approaches assume that the dynamical matrix of the spin system is diagonalizable. This is not the case when inertial zero modes are present, which may occur even for the 1D domain wall.
Our approach allows one to calculate, in an efficient and scalable manner, all magnetic normal modes, including the spin-wave modes and those modes that are related, for example, to the motion of localized or extended defects (domain walls, Skyrmions,~\ldots). For the latter case, we give a clear and computationally efficient procedure to calculate the parameters that determine the motion under external forces, such as effective masses.
Last but not least, we have developed an efficient perturbation scheme to take into account dissipation effects and calculate dynamical magnetic susceptibilities.

We believe that this approach can be useful in many further problems of spin dynamics, especially those dealing with the motion of Skyrmions and other defects in the magnetic configuration under an external field, their collision (momentum transfer), pinning, dissipation, and so on.


%



\appendix*
\section{The normal-mode problem of Hamiltonian systems}

In this \refappendix{}, we investigate the general properties of linear and nonlinear Hamiltonian systems and their normal modes. An advantage of our general approach is that it \emph{explains why} it is possible to reduce the magnetic normal-mode problem to the \HDGEP{}: this is a natural consequence of the symplectic structure of the conservative spin system. Moreover, it means that the same method may be reused for other kinds of Hamiltonian systems.

The best-known example of a linear Hamiltonian system is a set of point masses coupled by harmonic springs. For this system the normal-mode problem can be reduced to the symmetric definite generalized eigenvalue problem (\SDGEP{}) in an obvious way (see \sappref{app:harmpointmass}). However, the general normal-mode problem of linear Hamiltonian systems is much richer \cite{Arnold1989}. We consider here the normal-mode problem of a linear Hamiltonian system with a postive semidefinite Hamiltonian, as results from the linearization of a general Hamiltonian system near an energy minimum.

This \refappendix{} is organized as follows.
In \sappref{app:linhamsys}, we reproduce the definition of a linear Hamiltonian system and consider the types of normal modes that it may have. We also explain how the Hamiltonian structure can be used to efficiently perform mode analysis on a given state vector once the normal modes of the system have been calculated.
For comparison, we discuss the well-known special case of a system of coupled point masses in \sappref{app:harmpointmass}.
In \sappref{app:genhamsys}, we generalize the results to a nonlinear Hamiltonian system, possibly defined on a Poisson manifold rather than a symplectic manifold. We show that the linearization of a general Hamiltonian system near a local energy minimum results in a linear Hamiltonian system with a positive-semidefinite matrix $\langle M\Omega \rangle$ (defined in \secref{sec:hamsysbrief}).

\subsection{\label{app:linhamsys}Linear Hamiltonian systems}

Let us first reproduce the definition of a linear Hamiltonian system on the vector space $\mathbb{R}^{2n}$.
Fix some arbitrary basis set $\mathbf{e}_1,\ldots,\mathbf{e}_{2n}$, and let $x^1,\ldots,x^{2n}$ represent the coefficients of a vector $\mathbf{x}$ in this basis.
Let the matrix $\Omega$ be antisymmetric ($\Omega^{ij}=-\Omega^{ji}$) and invertible.
(We will relax the latter condition in \sappref{app:genhamsys}.)
Then the \emph{symplectic form}
\begin{equation}\label{eq:sympform}
\omega(\mathbf{e}_i,\mathbf{e}_j) = (\Omega^{-1})_{ij}
\end{equation}
defines a symplectic structure on $\mathbb{R}^{2n}$.
Since symplectic forms are bilinear, \equaref{eq:sympform} fixes the value of the form for any pair of vectors.
The symplectic structure induces a \emph{Poisson bracket}
\begin{equation}\label{eq:linpoissonbracket}
\lbrace x^i, x^j \rbrace = -\Omega^{ij}
\end{equation}
between the variables $x_i,x_j$; more generally, for any two functions $f,g$,
\begin{equation}
\lbrace f, g \rbrace = - \frac{\partial f}{\partial x^i} \Omega^{ij} \frac{\partial g}{\partial x^j}\text{.}
\end{equation}
The condition that $\Omega$ be invertible ensures that the symplectic form on $\mathbb{R}^{2n}$ is \emph{nondegenerate}.
In the special case that the $x^i$ with $1 \le i \le n$ represent canonical momenta and the $x^i$ with $n+1 \le i \le 2n$ represent the corresponding canonical coordinates, $\Omega$ takes the familiar form
\begin{equation}\label{eq:Jsimple}
\Omega = \left[
\begin{array}{cc}
0 & I_n \\
-I_n & 0
\end{array}\right]\text{,}
\end{equation}
where $I_n$ is the identity matrix,
and in particular, $\Omega$ is orthogonal ($\bar{\Omega} = \Omega$, where we define $\bar{\Omega} = \Omega^{-\trans}$); but we shall not make this assumption in this paper.

It is always possible in principle to construct a system of canonical momenta and coordinates of a symplectic vector space.
Let us write our momenta and coordinates as linear combinations
\begin{align}
p_{(k)}& =          -  w_{2(k)i} x^i\text{,} \\
q_{(k)}& = \phantom{-} w_{1(k)i} x^i
\end{align}
of the variables $x^i$ for certain vectors $w_{1(k)}$ and $w_{2(k)}$.
By definition, we must have $\lbrace q_{(k)},p_{(l)} \rbrace = \delta_{kl}$ and $\lbrace p_{(k)},p_{(l)} \rbrace = \lbrace q_{(k)},q_{(l)} \rbrace = 0$ for all $k,l$.
Using \equaref{eq:linpoissonbracket}, this can be rewritten as
\begin{subequations}
\begin{align}
\label{eq:orthopqfull}                                     w_{1(k)i} \Omega^{ij} w_{2(l)j}& = \delta_{kl} \\
\label{eq:orthoppfull}   w_{1(k)i} \Omega^{ij} w_{1(l)j} = w_{2(k)i} \Omega^{ij} w_{2(l)j}& = 0
\end{align}
\end{subequations}
(see \equarefs{eq:orthopq} and~\eqref{eq:orthopp}).
As a result, we may decompose an arbitrary state vector $x$ as
\begin{subequations}
\begin{align}
\label{eq:decompxfull}  x^i& = \sum_{k=1}^n \left[ -(w_{2(k)h} x^h) \, \Omega^{ij} w_{1(k)j} + (w_{1(k)h} x^h) \, \Omega^{ij} w_{2(k)j} \right] \\
\label{eq:decomppqfull}    & = \sum_{k=1}^n \left( p_{(k)} \, \Omega^{ij} w_{1(k)j} + q_{(k)} \, \Omega^{ij} w_{2(k)j} \right)
\end{align}
\end{subequations}
(see \equarefs{eq:decompx} and~\eqref{eq:decomppq}).
The vectors $\mathbf{e}_{p(k)} = \Omega^{ij} w_{1(k)j} \mathbf{e}_i$ and $\mathbf{e}_{q(k)} = \Omega^{ij} w_{2(k)j} \mathbf{e}_i$ form a \emph{symplectic basis} of the symplectic vector space.

Let us return to the original system of variables $x^1,\ldots,x^{2n}$ (not necessarily canonical) of our symplectic vector space.
We define a (time-invariant) Hamiltonian function
\begin{equation}\label{eq:Hgeneral}
\mathcal{H} = \frac{1}{2} x^i H_{ij} x^j\text{,}
\end{equation}
where $H$ is symmetric.
Using the generalized form of Hamilton's equations and the properties of Poisson brackets, we now derive the equation of motion
\begin{equation}\label{eq:linhamsys}
\dot{x}^i = \lbrace x^i, \mathcal{H} \rbrace = -\Omega^{ij} H_{jk} x^k\text{,}
\end{equation}
where the dot denotes the time derivative.
We may rewrite \equaref{eq:linhamsys} as
\begin{equation}\label{eq:lineom}
\dot{x}^i = {M^i}_j x^j\text{,}
\end{equation}
with
\begin{equation}\label{eq:lindefM}
M = -\Omega H\text{.}
\end{equation}
We see that for a linear Hamiltonian system, $M\Omega = \Omega^\trans H \Omega$ is symmetric.
Conversely, if a given matrix $M$ is such that $M\Omega$ is symmetric (or equivalently, if $\bar{\Omega}M + M^\trans \bar{\Omega} = 0$), it is is called a \emph{Hamiltonian matrix} \cite{Burgoyne1972}. The dynamical system \eqref{eq:lineom} is then a linear Hamiltonian system on the symplectic vector space defined by $\Omega$.
In \sappref{app:genhamsys}, we generalize the result that $M\Omega$ is symmetric to Hamiltonian systems that are nonlinear or for which $\Omega$ is not necessarily invertible.

The matrix $M$ describes the dynamical behavior \eqref{eq:lineom} of the linear Hamiltonian system.
This matrix is not necessarily diagonalizable \cite{Arnold1989}; its Jordan normal form may contain Jordan blocks of high order.
Moreover, the eigenvalues of these blocks, which often but not always appear in pairs or quadruples, may be zero, real, imaginary or complex.
Linear Hamiltonian systems may thus display a wide variety of inequivalent types of motion.
An exhaustive list of possibilities is given in \explcite{Arnold1989}.
In this paper, we restrict ourselves to systems where $M\Omega$ is positive semidefinite.
Even though this condition considerably limits the forms the normal modes may take, we shall see that three inequivalent types still need to be distinguished.

It can be shown that any linear Hamiltonian system admits a special symplectic basis in which the Hamiltonian takes its \emph{normal form} \cite{Williamson1936,Arnold1989}.
In terms of the momenta $p_{(k)}$ and coordinates $q_{(k)}$ that correspond to this special symplectic basis, the Hamiltonian is a direct sum of simple terms, each of which belongs to one of the families listed in \explcite{Arnold1989}.
Note that many of those types of irreducible terms depend on not just one but two or more pairs of canonical momenta and coordinates.
Here we consider Hamiltonians that are positive semidefinite, for which the possibilities are more limited.
Indeed, we have verified that for all but three types, the irreducible term cannot be positive semidefinite by finding a counterexample where the term takes a negative value.
The only three exceptions, which are positive semidefinite, are
\begin{subequations}
\begin{align}
\label{eq:Hpartpos} \mathcal{H}^\text{pos}_k(p_{(k)},q_{(k)})& = \frac{1}{2} \omega_k (p_{(k)}^2 + q_{(k)}^2) \\
\label{eq:Hpartord} \mathcal{H}^\text{spec}_k(p_{(k)},q_{(k)})& = 0 \\
\label{eq:Hpartdef} \mathcal{H}^\text{iner}_k(p_{(k)},q_{(k)})& = \frac{1}{2} p_{(k)}^2\text{,}
\end{align}
\end{subequations}
where in \equaref{eq:Hpartpos}, $\omega_k > 0$; a term of this type is in fact positive definite.
We introduce the names \emph{positive}, \emph{special zero} and \emph{inertial zero} respectively for the three types of terms that may appear in the normal form of a positive-semidefinite  Hamiltonian.

By Hamilton's equations, $\dot{p}_{(k)} = -\partial \mathcal H / \partial q_{(k)}$ and $\dot{q}_{(k)} = \partial \mathcal H / \partial p_{(k)}$, the three types of terms correspond to the following types of motion:
\begin{subequations}
\begin{align}
\label{eq:motpos}
\text{positive:}\hspace{1.2em}&
\left\lbrace
\begin{array}{l}
 \dot{p}_{(k)} =          -  \omega_k q_{(k)} \hspace{1em} \\
 \dot{q}_{(k)} = \phantom{-} \omega_k p_{(k)}
\end{array}
\right. \\
\label{eq:motord}
\text{special zero:}\hspace{1.2em}&
\left\lbrace
\begin{array}{l}
 \dot{p}_{(k)} = 0 \\
 \dot{q}_{(k)} = 0
\end{array}
\right. \\
\label{eq:motdef}
\text{inertial zero:}\hspace{1.2em}&
\left\lbrace
\begin{array}{l}
 \dot{p}_{(k)} = 0 \\
 \dot{q}_{(k)} = p_{(k)}\text{.}
\end{array}
\right.
\end{align}
\end{subequations}
It follows immediately from \equarefs{eq:decomppqfull} and~\eqref{eq:lineom} that \equaref{eq:motpos} corresponds to a positive normal mode \eqref{eq:posnormmode} with $\omega=\omega_k$, that \equaref{eq:motord} corresponds to a special zero normal mode \eqref{eq:ordzeronormmode} and that \equaref{eq:motdef} corresponds to a inertial zero normal mode \eqref{eq:defzeronormmode}, as defined in \secref{sec:hamsysbrief}, if we set
$(u_1^i,u_2^i) = (\Omega^{ij} w_{1(k)j}, \Omega^{ij} w_{2(k)j})$.
An important practical consequence of the fact that the normal modes of a Hamiltonian system form a symplectic basis is that we have a direct expression \eqref{eq:decompxfull} for the decomposition of an arbitrary state vector into a linear combination of the normal modes.

While the special zero normal mode \eqref{eq:ordzeronormmode} can be interpreted as the $\omega \rightarrow 0$ limit of the positive normal mode \eqref{eq:posnormmode}, the inertial zero normal mode \eqref{eq:defzeronormmode} is fundamentally different. One might interpret it as the $\omega \rightarrow 0$ limit of
\begin{equation}\label{eq:posdefnormmode}
\left\lbrace
\begin{array}{l}
M \tilde u_1 = \phantom{-\omega^2} \tilde u_2 \\
M \tilde u_2 = -\omega^2 \tilde u_1\text{,}
\end{array}
\right.
\end{equation}
which for $\omega>0$ is equivalent to \equaref{eq:posnormmode} if one sets $\tilde u_1 = u_1 / \sqrt \omega$ and $\tilde u_2 = \sqrt{\omega} u_2$.

Notice that even if the original dynamical variables $x^i$ represent canonical momenta and coordinates (which is not necessary), the special canonical momenta $p_{(k)}$ and canonical coordinates $q_{(k)}$ of the normal form are still, in principle, linear combinations of \emph{all} of the $x^i$. There is thus no guarantee that $p_{(k)}$ is a linear combination of the original momenta, or that $q_{(k)}$ is a linear combination of the original coordinates, unless the system is of the special form discussed in \sappref{app:harmpointmass}.

\subsection{\label{app:harmpointmass}Harmonically coupled point masses}

The variety in the types of dynamics that linear Hamiltonian systems display (see \sappref{app:linhamsys} and \explcite{Arnold1989}) may seem surprising.
Such variety is not seen in the archetypal example of a linear Hamiltonian system, a collection of point masses coupled by harmonic springs, for which it is obvious how the normal-mode problem can be cast in the form of a \SDGEP{}.
We shall see that this type of system is considerably simplified by the special structure of its Hamiltonian, which is not present in all linear Hamiltonian systems.
We discuss the system of coupled oscillators here to show how it is special and to explain why the most common method for solving the normal-mode problem cannot be used in the more general case discussed in \secref{sec:hamsysbrief} and \sappref{app:linhamsys}.

The Hamiltonian of a system of harmonically coupled point masses is given by $\mathcal H = \sum_{i,j} \frac{1}{2} p_i {(S^{-1})}_{ij} p_j + \sum_{i,j} \frac{1}{2} q_i D_{ij} q_j $, where $D$ is the \emph{force-constant matrix} and $S$ is the \emph{mass matrix}. The matrix $S$ is positive definite; both matrices are symmetric. In the simplest case, we have $S = m I_n$, where $m$ is the mass of a single particle. The variables $p_i$ and $q_i$ represent the momentum and the displacement of particle $i=1,\ldots,n$. (In multidimensional systems, we let $i$ represent the spatial direction as well as the particle index; this does not affect the mathematical structure.)
If we write the state of the system as a single vector
\begin{equation}
x = \left[\begin{array}{c} p \\ q \end{array}\right] \in \mathbb{R}^{2n}\text{,}
\end{equation}
the matrix $\Omega$ takes its standard form \eqref{eq:Jsimple}, since the variables $p_i$ and $q_i$ form a canonical system.
The Hamiltonian takes the form \eqref{eq:Hgeneral} if we set
\begin{equation}
H = \left[
\begin{array}{cc}
S^{-1} & 0 \\
0 & D
\end{array}\right]\text{.}
\end{equation}
Notice that $H$ is block diagonal: the Hamiltonian does not contain any terms that couple coordinates to momenta.
The equation of motion is given by
\begin{equation}\label{eq:springeom}
\left[\begin{array}{c} \dot p \\ \dot q \end{array}\right]
= M \left[\begin{array}{c} p \\ q \end{array}\right]
= \left[\begin{array}{cc} 0 & -D \\ S^{-1} & 0 \end{array}\right]  \left[\begin{array}{c} p \\ q \end{array}\right]\text{,}
\end{equation}
where we have used \equaref{eq:lindefM}. The structure of \equaref{eq:springeom} is such that we can derive equations of motion for the momenta and for the coordinates separately. For the coordinates, we have
\begin{equation}
\ddot q = S^{-1} \dot p = -S^{-1}Dq\text{.}
\end{equation}
The fundamental solutions of this equation may be found by calculating the eigenvectors $q^*$, which satisfy $S^{-1}Dq^* = \lambda q^*$. This equation is equivalent to the \SDGEP{}
\begin{equation}
Dq^* = \lambda S q^*\text{.}
\end{equation}
If we assume that the Hamiltonian is positive semidefinite, so that the the classification of \sappref{app:linhamsys} is applicable, then $D$ must also be positive semidefinite. We have that $\lambda \ge 0$, and the vector pair
\begin{equation}
(\tilde u_1, \tilde u_2) = \left(
\left[\begin{array}{c} Sq^* \\ 0 \end{array}\right],
\left[\begin{array}{c} 0 \\ q^* \end{array}\right]
\right)
\end{equation}
satisfies \equaref{eq:posdefnormmode} with $\omega = \sqrt \lambda$. If $\omega>0$, this is a positive normal mode \eqref{eq:posnormmode}; if $\omega=0$, it is a inertial zero normal mode \eqref{eq:defzeronormmode}. Notice that special zero normal modes \eqref{eq:ordzeronormmode} do not occur in a system of coupled point masses.

We see that the normal-mode problem of a system of coupled point masses can be reduced to the \SDGEP{}, as is well known. The same is true for the normal-mode problems of the wave equation or in elasticity theory, which have a similar mathematical structure (and are in a sense continuum analogues of systems of harmonically coupled masses).
However, the same reduction cannot be applied to arbitrary linear Hamiltonian systems.
What makes the system of coupled point masses special is that
\begin{inparaenum}[\itshape a\upshape)]
\item \label{itm:reqnatsys} there is a natural system of canonical variables (the momenta and displacements of the individual masses);
\item \label{itm:reqsepham} in this canonical system, the Hamiltonian is the sum of a kinetic-energy term, which depends only on the momenta, and a potential-energy term, which depends only on the coordinates; and
\item \label{itm:reqposdef} the kinetic-energy term is positive definite.
\end{inparaenum}
As for the spin system, while it is not hard to construct a system of canonical momenta and coordinates (condition~\ref{itm:reqnatsys}; see \explcite{Dobrovitski2003}), in this system the Hamiltonian generally does not separate into a kinetic-energy and a potential-energy part (condition~\ref{itm:reqsepham}), especially if the equilibrium configuration is not collinear.
One might remark that if the Hamiltonian is positive semidefinite a system of momenta and coordinates that satisfies condition~\ref{itm:reqsepham} must exist: such a system is a by-product of the solution of the normal-mode problem (see \sappref{app:linhamsys}). The issue, of course, is that we do not know this system when we start. Moreover, the kinetic-energy term is not guaranteed to be positive definite (condition~\ref{itm:reqposdef}) unless the Hamiltonian is positive definite.
\Secref{sec:redHDGEP} presents a way in which the normal-mode problem of any linear Hamiltonian system can be reduced to the \HDGEP{}, provided that its Hamiltonian is positive semidefinite.

\subsection{\label{app:genhamsys}General Hamiltonian systems}

In this \refsubappendix{}, we generalize the approach of \sappref{app:linhamsys} in two ways.
First, we allow the Hamiltonian system to be nonlinear.
Second, we no longer require that the matrix $\Omega$ defining the Poisson bracket at $x=0$ is invertible.
In the language of symplectic geometry, the latter generalization means that the Hamiltonian system may be defined on a Poisson manifold rather than a symplectic manifold.
While any symplectic manifold is also a Poisson manifold, the converse is not true. The spin system in Cartesian coordinates (see \secref{sec:spincons}) is an important example.
We shall show that even under these relaxed conditions, linearization of the equation of motion of a general Hamiltonian system near an equilibrium point $x^i = 0$ results in a linear Hamiltonian system.
In particular, we shall show that the matrix $M\Omega$ (see \secref{sec:hamsysbrief}) is symmetric. Moreover, we show that $\langle M\Omega \rangle$ can be interpreted as the Hessian matrix at the equilibrium point of the restriction of the Hamiltonian function to the symplectic leaf that contains $x=0$. This implies that $\langle M\Omega \rangle$ is indeed guaranteed to be positive semidefinite, as we require, provided that we linearize at a constrained local minimum of the Hamiltonian.

We fix a nonsingular local system of variables $x^1,\ldots,x^{m}$ in such a way that $x^i = 0$ is an equilibrium point.
In this system of variables, we expand the Hamiltonian $\mathcal H$ to second order in $x$ as
\begin{equation}\label{eq:Hexpanded}
\mathcal H(x) = H_0 - h_i x^i + \frac{1}{2} x^i A_{ij} x^j + \Or({\| x \|}^3)\text{,}
\end{equation}
for a constant scalar $H_0 = \mathcal{H}(0)$, vector $h_i = {- \partial \mathcal H / \partial x^i |}_{x=0} $, and symmetric matrix $A_{ij} = {\partial^2 \mathcal H / (\partial x^i \partial x^j)|}_{x=0}$.
We expand the Poisson bracket to first order as
\begin{equation}\label{eq:Pexpanded}
\lbrace x^i,x^j\rbrace = -\Omega^{ij} + {K^{ij}}_k x^k + \Or({\| x \|}^2)\text{.}
\end{equation}
The properties of the Poisson bracket (antisymmetry, Jacobi identity) give the following conditions on the coefficients of this expansion:
$\Omega^{ij}$ must be antisymmetric ($\Omega^{ij} = -\Omega^{ji}$); ${K^{ij}}_k$ must be antisymmetric in the first two indices (${K^{ij}}_k = -{K^{ji}}_k$); and we must have \cite{Liu2009}
\begin{equation}\label{eq:genhamjacobi}
{K^{ij}}_l \Omega^{lk} + {K^{jk}}_l \Omega^{li} + {K^{ki}}_l \Omega^{lj} = 0\text{.}
\end{equation}
The last condition follows from the Jacobi identity,
\begin{equation}\label{eq:jacobipoisson}
\{x_i,\{x_j,x_k\}\} + \{x_j,\{x_k,x_i\}\} + \{x_k,\{x_i,x_j\}\} = 0\text{,}
\end{equation}
which holds for any Poisson bracket $\{\cdot,\cdot\}$.
From \equaref{eq:Pexpanded}, we get
\begin{equation}\label{eq:jacobipart}
\begin{split}
\{x_i,\{x_j,x_k\}\}& = -\Omega^{jk} \{x^i,1\} + {K^{jk}}_l \{x^i,x^l\} + \{x^i, \Or({\| x \|}^2) \} \\
                   & = {K^{jk}}_l \Omega^{li} + \Or(\| x \|)\text{.}
\end{split}
\end{equation}
Since this expression holds at any point $x$, we obtain \equaref{eq:genhamjacobi} by collecting the constant parts of the three cyclic permutations of it that appear in \equaref{eq:jacobipoisson}.

Using \equarefs{eq:Hexpanded} and~\eqref{eq:Pexpanded} and the general properties of Poisson brackets, we derive the equation of motion to first order from the generalized Hamilton equations,
\begin{equation}\label{eq:genlinEoM}
\dot{x}^i = \lbrace x^i,\mathcal H\rbrace = \Omega^{ij}h_j + {M^i}_j x^j + \Or({\| x \|}^2)\text{,}
\end{equation}
where
\begin{equation}\label{eq:gendynmat}
{M^i}_j = -\Omega^{ik}A_{kj} - {K^{ik}}_j h_k\text{.}
\end{equation}
\Equaref{eq:gendynmat} may be considered as the equivalent of \equaref{eq:spinM} for a general Hamiltonian system.
Since $\dot{x}^i = 0$ at $x^i=0$, we must have $\Omega^{ij} h_j = 0$. From this fact and \equaref{eq:genhamjacobi}, we can derive that $M\Omega$ is symmetric, as follows.
We may write ${(M\Omega)}^{ij} = F^{ij} + G^{ij}$, where $F^{ij}$ is given by
\begin{equation}
F^{ij} = - \Omega^{ik} A_{kl} \Omega^{lj} = \Omega^{ki} A_{kl} \Omega^{lj}\text{,}
\end{equation}
and $G^{ij}$ is given by
\begin{equation}
G^{ij} = -{K^{ik}}_l h_k \Omega^{lj} = {K^{ki}}_l h_k \Omega^{lj}\text{.}
\end{equation}
$F^{ij}$ is obviously symmetric ($A_{ij}$ is symmetric).
We can see that $G^{ij}$ is symmetric by rewriting it as
\begin{equation}
\begin{split}
G^{ij}& = \frac{1}{2} \left( {K^{ki}}_l \Omega^{lj} - {K^{jk}}_l \Omega^{li} - {K^{ij}}_l \Omega^{lk} \right) h_k \\
      & = \frac{1}{2} {K^{ki}}_l \Omega^{lj} h_k + \frac{1}{2}{K^{kj}}_l \Omega^{li} h_k - \frac{1}{2}{K^{ij}}_l \Omega^{lk} h_k\text{,}
\end{split}
\end{equation}
where we have used \equaref{eq:genhamjacobi}.
If $x=0$ is an equilibrium position, \equaref{eq:genlinEoM} implies $\Omega^{ij} h_j=0$ and the last term vanishes.
The other two terms together are explicitly symmetric under $i \leftrightarrow j$.

Except for the fact that $\Omega$ is not necessarily invertible, we could conclude from the symmetry of $M\Omega$ that the linearization $\dot{x}^i = {M^i}_j x^j$ of a general Hamiltonian system near an equilibrium point is a linear Hamiltonian system in the sense of \sappref{app:linhamsys}.
To be explicit, the matrix $\Omega$ of this linear Hamiltonian system is defined, according to \equaref{eq:Pexpanded}, by 
\begin{equation}
\Omega^{ij} = - {\lbrace x^i,x^j\rbrace|}_{x=0} = {\lbrace x^j,x^i\rbrace|}_{x=0}\text{,}
\end{equation}
which is the value of the Poisson bracket between $x^j$ and $x^i$ at $x=0$, while
the symmetric matrix $M\Omega$ is given by
\begin{equation}\label{eq:MJexplicit}
{(M\Omega)}^{ij} = -\Omega^{ik}A_{kl} \Omega^{lj} - {K^{ik}}_l h_k \Omega^{lj}\text{.}
\end{equation}
Since $\Omega$ is antisymmetric, its rank is always even. We write $\rank(\Omega) = 2n$.
If $m>2n$ ($\Omega$ is not invertible), we can make $\Omega$ invertible by interpreting the matrices $\Omega$ and $M\Omega$ as \emph{restricted} to the $2n$-dimensional image space of $\Omega$. In the notation of \secref{sec:hamsysbrief}, we get $\langle \Omega \rangle$ and $\langle M\Omega \rangle$. We may do this because the image space of $M\Omega$ is contained in the image space of $\Omega$.
Thus, the matrices $\langle M\Omega \rangle$ and $\langle \Omega \rangle$ together define a proper linear Hamiltonian system.

Our method for the normal-mode problem requires that $\langle M\Omega \rangle$ be positive semidefinite (see \secref{sec:redHDGEP}).
We can show that it is if $x=0$ is a (constrained) local minimum of the Hamiltonian $\mathcal H$. For simplicity, we first consider the case $m=2n$ ($\Omega$ is invertible). If $\Omega$ is invertible, we have $h=0$, so that $M\Omega = -\Omega A \Omega = \Omega^\trans A \Omega$.
Evidently, $M\Omega$ is positive (semi)definite if and only if $A$, the Hessian matrix of $\mathcal H$, is positive (semi)definite.
Consequently, if $x=0$ is a local minimum of $\mathcal H$, then $M\Omega$ is positive semidefinite.

For $m>2n$, the dynamical matrix \eqref{eq:gendynmat} is no longer determined only by the Hessian matrix $A$ of $\mathcal H$; there is an additional $h$-dependent term, which is essential.
We shall see that the matrix $\langle M\Omega \rangle$ can be interpreted as the Hessian matrix of the restriction of the Hamiltonian function $\mathcal H$ to a certain $2n$-dimensional submanifold containing $x=0$.
For example, while the Hamiltonian $\mathcal H = -\mathbf m \cdot \hat{\mathbf{z}}$ has no local minimum on $\mathbb{R}^3$, is has a constrained minimum at $\mathbf m = \hat{\mathbf{z}}$ on the set $S^2_{c=1} = \{ \mathbf{m} \in \mathbb{R}^3 : \| \mathbf{m} \|=1 \}$.
For positive semidefiniteness of $\langle M\Omega \rangle$ we do not require that $x=0$ be an actual local minimum of $\mathcal H$; it is sufficient that $x=0$ be a \emph{constrained} local minimum on this submanifold.
To define the relevant submanifold in a general setting, it is necessary to use some elements from the theory of symplectic structures and Poisson structures \cite{Weinstein1983}.

In a symplectic manifold, any point (that is, any state of the system) may be reached from any other point by following the trajectory generated by a suitably chosen Hamiltonian function $\mathcal H$, or a finite sequence of such trajectories.
In a Poisson manifold, this is not necessarily the case. However, by the splitting theorem on Poisson manifolds \cite{Weinstein1983}, a Poisson manifold can be divided into \emph{equivalence classes} of points for which this is possible.
These equivalence classes are symplectic submanifolds of the Poisson manifold and are called \emph{symplectic leaves}.
Two points of a Poisson manifold are in the same symplectic leaf if one can get from one point to the other through a finite sequence of trajectories induced by Hamiltonian functions.
For example, consider a conservative spin system (see \secref{sec:spincons}) with a single spin $\mathbf m \in \mathbb{R}^3$, which is governed by the equation of motion $\dot{\mathbf{m}} = \mathbf{m} \times \nabla \mathcal H$. Since this equation conserves $\|\mathbf{m}\|$, a spin in position $\mathbf{m} = \hat{\mathbf{z}}$ will never end up in position $\mathbf{m} = \frac{1}{2}\hat{\mathbf{z}}$, regardless of the choice of $\mathcal H$. However, it may at some point in time reach $\mathbf{m} = \hat{\mathbf{y}}$, for instance if the Hamiltonian is given by $\mathcal H = \mathbf m \cdot \hat{\mathbf{x}}$.
Thus, the Poisson manifold of the conservative single-spin system (that is, $\mathbb{R}^3$ equipped with the spin Poisson bracket; see \secref{sec:spincons}) splits into symplectic leaves of the form $S^2_c = \{ \mathbf{m} \in \mathbb{R}^3 : \| \mathbf{m} \|=c \}$ for $c \ge 0$.

It can be shown that the $2n$-dimensional symplectic leaf containing the equilibrium point $x=0$ can locally be parametrized by a vector $v_i$, which we require to lie in the image space of $\Omega^{ij}$, as
\begin{equation}\label{eq:paramx}
x^i = -\Omega^{ij} v_j - \frac{1}{2} {K^{ij}}_k \Omega^{kl} v_j v_l + \Or({\| v \|}^3)
\end{equation}
if we assume that the Poisson bracket of the Poisson manifold is of the form \eqref{eq:Pexpanded}.
By substitution of this expression into \eqref{eq:Hexpanded}, we find that in terms of $v$, the Hamiltonian becomes
\begin{multline}\label{eq:Hv}
\mathcal{H}(v) = H_0 + h_i \Omega^{ij} v_j \\
  + \frac{1}{2} \left(- \Omega^{ik}A_{kl}\Omega^{lj} - {K^{ik}}_l h_k \Omega^{lj} \right) v_i v_j + \Or({\| v \|}^3)\text{.}
\end{multline}
Here we have used that $v = \Or(\| x \|)$: the fact that $v$ lies in the image space of $\Omega$ guarantees $\Omega^{ij} v_j \neq 0$ in \equaref{eq:paramx}.
If $x=0$ is an equilibrium point, the linear term in \equaref{eq:Hv} vanishes ($\Omega^{ij}h_j=0$).
The matrix of the quadratic term in \equaref{eq:Hv}, which is identical to the Hessian matrix of the Hamiltonian $\mathcal H$ restricted to the symplectic leaf, is identical to $\langle M\Omega \rangle$ \eqref{eq:MJexplicit}.
(We must write the angular brackets $\langle \cdot \rangle$ here because $v$ was assumed to lie in the image space of $\Omega$.)
Thus, if $x=0$ is a local minimum of $\mathcal H$ on the symplectic leaf that contains the point $x=0$, the matrix $\langle M\Omega \rangle$ is positive semidefinite and the method presented in \secref{sec:redHDGEP} can be used.

\begin{acknowledgments}
We thank A. Secchi, J. H. Mentink, K. Y. Guslienko and O. Eriksson for useful discussions.

This work is part of the research programme of the Foundation for Fundamental Research on Matter (FOM), which is part of the Netherlands Organisation for Scientific Research (NWO).
\end{acknowledgments}

\end{document}